\documentclass[12pt,prd, aps, showpacs, groupedaddress, superscriptaddress,floatfix,preprintnumbers]{revtex4}
\usepackage{graphicx}
\usepackage{bm}
\usepackage{multirow}
\usepackage{axodraw}
\usepackage{graphicx}
\usepackage{epsfig}
\usepackage{psfrag}
\usepackage{soul}
\usepackage{color}
\usepackage{multirow}
\usepackage{mathptmx}
\usepackage{mathrsfs}
\usepackage{amsmath, amssymb}
\usepackage{slashed}
\usepackage{verbatim} 
\usepackage{subfigure}

\newcommand{\wm}{\widetilde{m}}

\newcommand{\qslash}{\slashed{q}}

\newcommand{\ident}{1}

\newcommand{\pslash}{\not{\!p}}

\newcommand{\BKRESULT}{B_K^{\overline{\mathrm{MS}}}(3\,\mathrm{GeV}) = 0.529(5)_\mathrm{stat}(15)_\chi(2)_{\mathrm{FV}}(11)_{\mathrm{NPR}}}
\newcommand{\BKRESULTTWOGEV}{B_K^{\overline{\mathrm{MS}}}(2\,\mathrm{GeV}) = 0.549(5)_\mathrm{stat}(15)_\chi(2)_{\mathrm{FV}}(21)_{\mathrm{NPR}}}

\newcommand{\BKRESULTRGI}{\hat{B}_K^{\overline{\mathrm{RGI}}} = 0.749(7)_\mathrm{stat}(21)_\chi(3)_{\mathrm{FV}}(15)_{\mathrm{NPR}}}

\newcommand{\BKRESULTRGIQUAD}{\hat{B}_K^{\overline{\mathrm{RGI}}} = 0.749(27)_{\mathrm{combined}}}

\newcommand{\NDR}{\textrm{NDR}}
\newcommand{\MSbar}{\overline{\textrm{MS}}}
\newcommand{\RIMOM}{\textrm{RI-MOM}}

\newcommand{\RISMOM}{\textrm{RI-SMOM}}

\newcommand{\RISMOMmu}{\textrm{RI-SMOM}_{\gamma_{\mu}}}

\newcommand{\A}{\textrm{A}}
\newcommand{\als}{\alpha_s}
\newcommand{\FslashA}[1]{\!\not{\hbox{\kern-2pt ${#1}$}}}
\newcommand{\FslashB}[1]{\!\not{\hbox{\kern+1pt ${#1}$}}}
\newcommand{\CF}{C_F}
\newcommand{\Nc}{N}
\newcommand{\nf}{n_f}
\newcommand{\lmsdmus}{\log{p^2\over\mu^2}}
\allowdisplaybreaks

\usepackage{dcolumn}

\newcounter{Kanji}
\ifnum\theKanji=1
\usepackage{CJK}
\fi

\newcommand\ensB{ {\bf 2~} }
\newcommand\ensA{ {\bf 1~} }
\newcommand\ensE{ {\bf e~} }

\newcommand\regensburg{Institut f\"ur Theoretische Physik, Universit\"at Regensburg, 93040 Regensburg, Germany}
\newcommand\riken{RIKEN-BNL Research Center, Brookhaven National  Laboratory, Upton, NY 11973, USA}
\newcommand\bnl{Physics Department, Brookhaven National Laboratory, Upton, NY 11973, USA}
\newcommand\edinb{SUPA, School of Physics, The University of
  Edinburgh, Edinburgh EH9 3JZ, UK}

\newcommand\cu{Physics Department, Columbia University, New York,
  NY 10027, USA}

\newcommand\uconn{Physics Department, University of Connecticut,
  Storrs, CT 06269-3046, USA}
\newcommand\soton{School of Physics and Astronomy, University of
  Southampton,  Southampton SO17 1BJ, UK}
\newcommand\kek{Institute of Particle and Nuclear Studies,  KEK, Tsukuba, 305-0801, Japan}

\newcommand\uva{Department of Physics, University of Virginia, 382 McCormick Road, Charlottesville, VA 22904-4714.}
\newcommand{\julich}{J\"ulich Supercomputing Centre, Institute for Advanced Simulation, Forschungszentrum J\"ulich GmbH, 52425 J\"ulich, Germany}
\newcommand{\indiana}{Department of Physics, Indiana University, Bloomington, IN 47405, USA}
\newcommand{\mpimunchen}{Max-Planck-Institut f{\"u}r Physik, F{\"o}hringer Ring 6, 80805 M{\"u}nchen, Germany}

\newcounter{Outline}
\newcounter{Introduction}
\newcounter{ME}
\newcounter{ChiralExtrapStrategy}
\newcounter{ChiralExtrap}
\newcounter{NPR}
\newcounter{ContPT}
\newcounter{Conclusions}
\newcounter{Acknowledgments}
\newcounter{Appendix}
\newcounter{Tables}
\newcounter{Figures}
\newcommand{\ba}{\begin{eqnarray}}
\newcommand{\ea}{\end{eqnarray}}
\newcommand{\bas}{\begin{eqnarray*}}
\newcommand{\eas}{\end{eqnarray*}}
\newcommand{\be}{\begin{equation}}
\newcommand{\ee}{\end{equation}}
\newcommand{\bes}{\begin{equation*}}
\newcommand{\ees}{\end{equation*}}
\newcommand{\bi}{\begin{itemize}}
\newcommand{\ei}{\end{itemize}}

\newcommand{\bcentre}{\begin{center}}
\newcommand{\ecentre}{\end{center}}

\font\tenmsb=msbm10 scaled\magstep1
\font\sevenmsb=msbm7 scaled\magstep1
\font\fivemsb=msbm5 scaled\magstep1
\newfam\msbfam
\textfont\msbfam=\tenmsb
\scriptfont\msbfam=\sevenmsb
\scriptscriptfont\msbfam=\fivemsb

\def\Re{\mathop{\rm Re}}		    
\def\Im{\mathop{\rm Im}}		    

\def\su3{SU(3)}

\def\tD{\mbox{D}\kern-0.65em\raise0.15ex\hbox{/}\kern0.15em} 
\def\sD{\mbox{\scriptsize D}\kern-0.5em\raise0.15ex\hbox{\scriptsize/}}
\def\ssD{\mbox{\tiny D}\kern-0.42em\raise0.15ex\hbox{\tiny/}}

\def\dslash{\hbox{\(\partial\)}\kern-0.5em\raise0.15ex\hbox{/}} 

\def\su3{SU(3)}

\def\nicefrac#1#2{\leavevmode\kern.1em\raise.5ex\hbox{\the\scriptfont0 #1}\kern-
.1em/\kern-.15em\lower.25ex\hbox{\the\scriptfont0 #2}}

\begin{document}
\bibliographystyle{apsrev}

\ifnum\theKanji=1
\begin{CJK*}{UTF8}{}
\fi

\title{Continuum Limit of $B_K$ from 2+1 Flavor Domain Wall QCD}

\author{Y.~Aoki}\affiliation{\riken}\affiliation{Present address:
Kobayashi-Maskawa Institute for the Origin of Particles and the
Universe (KMI), Nagoya University, Nagoya 464-8602, Japan}
\author{R.~Arthur}\affiliation{\edinb}
\author{T.~Blum}\affiliation{\uconn}
\author{P.A.~Boyle}\affiliation{\edinb}
\author{D.~Br\"ommel}\affiliation{\soton}\affiliation{\julich}
\author{N.H.~Christ}\affiliation{\cu}
\author{C.~Dawson}\affiliation{\uva}
\author{T.~Izubuchi}\affiliation{\riken}\affiliation{\bnl}
\author{C.~Jung}\affiliation{\bnl}
\author{C.~Kelly}\affiliation{\edinb}
\author{R.D.~Kenway}\affiliation{\edinb}
\author{M.~Lightman}\affiliation{\cu}
\author{R.D.~Mawhinney}\affiliation{\cu}
\ifnum\theKanji=1
\CJKfamily{min}
\author{Shigemi Ohta (太田滋生)} 
\fi
\ifnum\theKanji=0
\author{Shigemi Ohta} 
\fi

\affiliation{\kek}
\affiliation{Department of Particle and Nuclear Physics, Sokendai Graduate University of Advanced Studies, Hayama, Kanagawa 240-0193, Japan}
\affiliation{\riken}

\author{C.T.~Sachrajda}\affiliation{\soton}
\author{E.E.~Scholz}\affiliation{\regensburg}
\author{A.~Soni}\affiliation{\bnl}
\author{C.~Sturm}\affiliation{\bnl}\affiliation{\mpimunchen}
\author{J.~Wennekers}\affiliation{\edinb}
\author{R.~Zhou}\affiliation{\uconn}\affiliation{\indiana}

\collaboration{RBC and UKQCD Collaborations}
%
%
\noaffiliation{CU-TP-1196, Edinburgh-2010/12, KEK-TH-1366, RBRC-843, SHEP-1016, MPP-2010-172}


\date{\today}
\maketitle
\ifnum\theKanji=1
\end{CJK*}
\fi

\begin{center}
In memory of Jan Wennekers
\end{center}

\centerline{ABSTRACT}

We determine the neutral kaon mixing matrix element $B_K$
in the continuum limit with 2+1 flavors of domain wall fermions,
using the Iwasaki gauge action at two different lattice spacings.  
These lattice fermions have near exact chiral symmetry and 
therefore avoid artificial lattice operator mixing.

We introduce a significant improvement to the conventional NPR method in which the 
bare matrix elements are renormalized non-perturbatively in the RI-MOM scheme 
and are then converted into the $\overline{\textrm{MS}}$ scheme using continuum 
perturbation theory. In addition to RI-MOM, we introduce and implement four non-exceptional 
intermediate momentum schemes that suppress infrared non-perturbative uncertainties 
in the renormalization procedure. We compute the conversion factors relating the matrix elements 
in this family of RI-SMOM schemes
and $\overline{\textrm{MS}}$ at one-loop order. Comparison of the results obtained using these
different intermediate schemes allows for a more reliable estimate of the unknown higher-order
contributions and hence for a correspondingly more robust estimate of the systematic error. 
We also apply a recently proposed approach in which twisted boundary conditions are used to 
control the Symanzik expansion for off-shell vertex functions 
leading to a better control of the renormalization in the continuum limit.

We control chiral extrapolation errors by considering both the NLO
SU(2) chiral effective theory, and an analytic mass expansion.
We obtain $\BKRESULT $. This corresponds to
$\BKRESULTRGI$. Adding all sources of error in quadrature we
obtain $\BKRESULTRGIQUAD$, with an overall combined error
of 3.6\%.

\ifnum\theOutline=1
\newpage
\newpage
\fi

\refstepcounter{section}
\setcounter{section}{0}


\newpage
\section{Introduction}
\label{sec:Introduction}

\ifnum\theIntroduction=1
The indirect $CP$ violation parameter of the neutral kaon system
\begin{equation}
\epsilon_K = \frac{A(K_L \to (\pi \pi)_{I=0})}{A(K_S \to (\pi \pi)_{I=0})}, 
\end{equation}
was measured first at BNL in a Nobel Prize winning experiment~\cite{Christenson:1964fg}, and is
now experimentally measured as 
$|\epsilon_K|=(2.228\pm 0.011)\,10^{-3}$~\cite{Amsler:2008zzb}.
Since CP is not an exact symmetry of the weak interations,
the eigenstates $K_L$ and $K_S$ of the mass matrix of neutral kaon system
are not eigenstates of CP. We characterise the state mixing via 
 \begin{equation}
 K_S = p K^0 - q \bar K^0 \quad\textrm{and}\quad
 K_L = p K^0 + q \bar K^0 
 \end{equation}
where $p^2 + q^2 = 1$, and $\frac{p}{q} = \frac{1 + \bar \epsilon}{1 - \bar \epsilon}$.

$\epsilon_K$ receives its dominant contribution from ``indirect" CP violation via state-mixing,
mediated by the imaginary part of the $\Delta S=2$ box graph. 
Before $\epsilon_K$ can be used to constrain the unitarity triangle and to provide information on
CKM matrix elements, we must therefore determine the QCD hadronic matrix element of the effective weak $\Delta S=2$ four quark operator
$$\langle K^0 |{\cal{O}}_{\mathrm{VV+AA}} | \overline{K}^0\rangle,$$ where
\begin{equation}
{\cal{O}}_{\mathrm{VV+AA}}
=
(\bar{s} \gamma_\mu d)
(\bar{s} \gamma_\mu d)
+
(\bar{s} \gamma_5 \gamma_\mu d)
(\bar{s} \gamma_5 \gamma_\mu d)\,.
\end{equation}
It is conventional to define the bag parameter $B_K$ from this matrix element as
\begin{equation}
B_K = \frac{\langle K^0 |{\cal{O}}_{\mathrm{VV+AA}} | \overline{K}^0\rangle}
           {\frac{8}{3} f_K^2 M_K^2}\,,
\end{equation}
where $M_K$ and $f_K$ are the mass and leptonic decay constant of the kaon.
The kaon bag parameter is thus of fundamental importance in studies of CP violation,
and as the hadronic matrix element is non-perturbative, lattice QCD is the only known framework for
its determination from first principles. 

Since the operator ${\cal{O}}_{\mathrm{VV+AA}}$ depends on the renormalization scheme and scale used in its definition, 
$B_K$ also has the same scheme and scale dependence. Therefore, for phenomenological use, it is convenient to introduce
the renormalization-group-invariant counterpart of $B_K$,
$$\hat{B}_K = \omega^{-1}_{A}(\mu,\nf) B_K^{A}(\mu,\nf),$$
where the Wilson coefficient, $\omega^{-1}_{A}(\mu,\nf)$,
for the various schemes $A$ used in this
paper are given in Equations (\ref{eq:omega_a}) through (\ref{eq:jRIMOM}), 
and we use the numerical values for the 2+1 flavour theory 
in our conversion.

We have recently calculated $B_K$ in dynamical 2+1 flavored simulations~\cite{Antonio:2007pb,Allton:2008pn} with a
total error of about 5.5\%. It was observed by Buras and Guadagnoli~\cite{Buras:2008nn}, 
that our result~\cite{Antonio:2007pb} was sufficiently accurate that
additional care needs to be taken in relating it to the measured value of $\epsilon_K$.
Previously ignored subdominant effects of direct $CP$ violation arising from the $\Delta S=1$
Hamiltonian amount to a few percent and must now be incorporated.

The short distance contribution $\bar \epsilon_K$~\cite{Blum:2001xb,Winstein:1992sx} differs 
from $\epsilon_K$, predominantly due to direct CP violation 

\begin{equation}
\epsilon_K = \bar \epsilon_K + i \frac{\Im A_0}{\Re A_0}.
\end{equation}
Here $A_0$ is the $K^0 \to \pi \pi$ amplitude for the isospin 0 final state defined via
  \begin{equation}
 A(K^0 \to \pi \pi(I)) = A_I \exp{i \delta_I}\quad\textrm{and}\quad
 A(\bar K^0 \to \pi \pi(I)) = A_I^* \exp{i \delta_I}
 \end{equation}
 and $\delta_I$ is the $\pi \pi$ phase shift in the $I=0$ or $I=2$ final state.

Reliable calculation of $A_0$ amplitudes remains a challenging project
to which our collaboration is devoting a considerable effort
\cite{Lightman:2010xx,Lightman:2009cu,Li:2008kc,Kim:2009fe,Kim:2005gka,Kim:2003xt}.
Using the measured value $Re \frac{\epsilon^\prime_K}{\epsilon_K} = (1.65 \pm 0.26)\times
10^{-3}$~\cite{Amsler:2008zzb}, assuming the Standard Model is correct and making plausible assumptions
in estimating the somewhat less difficult ratio $\frac{{\rm Im} A_2}{{\rm Re} A_2}$, the subdominant contribution to $\epsilon_K$ can
be effectively incorporated into a correction factor $\kappa_{\epsilon_K}$~\cite{Buras:2008nn}:
\begin{equation}\label{eq:bk_epsilonk}
\epsilon_K =
\kappa_{\epsilon_K} \hat{B}_K
\frac{G_F^2 f_K^2 M_K M_W^2}{6\sqrt{2} \pi^2 \Delta M_K}\Im(\lambda_t )  e^{i\frac{\pi}{4}}
    \,\Big\{
        \Re(\lambda_c) \left[\eta_1 S_0(x_c) - \eta_3 S_0(x_c,x_t)\right]
    -   \Re(\lambda_t) \eta_2 S_0(x_t)
    \Big\},
\end{equation}
where $\lambda_x = V_{xd}V_{xs}^\ast$ contain the entries of the CKM
matrix $V_{x y}$, $\eta_i$ are perturbative QCD corrections
\cite{Buras:1998raa} and the $S_0$ are Inami-Lim functions of mass
ratios $x_q = \frac{m_q^2}{m_W^2}$. In References~\cite{Buras:2008nn,Buras:2010pz} the correction factor was estimated to
 be $\kappa_{\epsilon_K} \approx 0.94 \pm 0.02$, and here the fractional error on this small correction is large (0.02 in a correction of size 0.06) 
and model dependent.

The correction factor also includes an estimate of long distance contributions 
corresponding to two insertions of the $\Delta S
=1$ Hamiltonian, with two pions propagating long distances between them
\cite{Buras:2010pz}. The results of our present work are
sufficiently precise that it has become necessary to determine as many
contributions as possible using lattice gauge methods; 
efforts in RBC-UKQCD are underway in this direction
\cite{christLat2010,kaonsatlattice2010}.

In this paper we improve on our 
earlier calculations~\cite{Antonio:2007pb,Allton:2008pn} in three major ways. First
of all, we simulate at a second value of the 
lattice spacing which allows us to perform a continuum 
extrapolation. Secondly, we refine our approach to non-perturbative renormalization to implement intermediate schemes
defined with no exceptional momentum channels and thereby reduce the infrared non-perturbative uncertainties. 
Finally, we also use twisted boundary conditions to remove the requirement to use
the Fourier modes of our lattice for our renormalization of off-shell amplitudes:
this gives complete freedom of choice of the momentum at each lattice spacing and
enables a more reliable continuum extrapolation of the renormalized operator.

Our final result for $B_K$ from the present analysis is obtained using
an off-shell momentum scheme renormalization. When converted to $\overline{\rm MS}$ with
$p^2 = \mu^2 = (3 {\rm GeV})^2$ it is:
\begin{equation}\label{eq:bkfinal}
\boxed{\ \BKRESULT\,.\ }
\end{equation}
The 3 GeV scale for our result is made accessible by our improved renormalization techniques,
and enables us to reduce perturbative error compared to a 2 GeV renormalization scale. 
For comparison to other results we also quote the standard operator normalization:
\begin{equation}\label{eq:bkrgi}
\boxed{\ \BKRESULTRGI\,.\ }
\end{equation}

The full analysis of systematic errors presented in this paper augments and finalizes an earlier conference
presentation \cite{Kelly:2009fp}. The result Equation (\ref{eq:bkfinal}) 
represents around a factor of four reduction in the 
error during the last five years or so.

The structure of the remainder of this paper is as follows.
In the next section we discuss the details of our simulations and present 
the measured values of the bare matrix elements. In Section~\ref{sec:NPR} 
we discuss the definition of several new momentum renormalization schemes and perform the non-perturbative renormalization of the bare lattice operator ${\cal{O}}_{\mathrm{VV+AA}}$ into these schemes. In this section we also perform the one-loop perturbative matching from the momentum schemes into $\overline{\textrm{MS}}$. Having obtained the matrix elements at the values of the quark masses and lattice spacing at which we perform our simulations, we present the 
simultaneous chiral and continuum extrapolations of the renormalized matrix elements
in Section~\ref{sec:ChiralExtrap}. 
We will discuss the phenomenological context of our results in
the concluding Section~\ref{sec:Conclusions} of this paper.

\fi

\section{Simulation parameters and matrix elements}
\label{sec:ME}
\ifnum\theME=1
\begin{table}
\centering
\begin{tabular}{c|ccc}
\hline
Lattice         & $m_h$  & $m_l$   & traj.(\# meas.)\\
\hline
\multirow{3}{*}{\ensA ($32^3\times 64$)} & $0.03$ & $0.004$ & 260-3250 (300)\\
                & $0.03$ & $0.006$ & 500-3610 (312)\\
                & $0.03$ & $0.008$ & 260-2770 (252)\\
\hline
\multirow{2}{*}{\ensB ($24^3\times 64$)} & $0.04$ & $0.005$ & 900-8940 (202)\\
                & $0.04$ & $0.01$  & 1460-8540 (178) 
\end{tabular}
\caption{Ensemble details. Here traj. refers to the Monte Carlo trajectories
 used in our measurements. The bracketed \# meas. refers to the number of 
measurements, separated by 
20 MD time units (10 trajectories) for the \ensA ensembles,
and
40 molecular dynamics time units (40 trajectories) for the \ensB ensembles.
To reduce the effects of
auto-correlations we block-average our data over 80 MD time units 
and use blocked measurements for the purposes of statistical analysis.
\label{tab:ensembles} 
}
\end{table}

Details of our ensembles are given in references
\cite{Allton:2008pn,thirtytwocubed},
and are summarised in Table~\ref{tab:ensembles}. We use the Iwasaki gauge 
action 
\cite{Iwasaki:1985we}
with 2+1 flavors of dynamical domain wall fermions \cite{Furman:1994ky}.
This action was chosen to balance topology change against
chirality after a careful study 
\cite{Antonio:2008zz,Antonio:2006px,Boyle:2007fn} recognising 
a general problem that topological tunneling
will vanish towards the continuum limit in any local update
due to the gauge field potential barrier
\cite{Antonio:2008zz,Boyle:2007fn,Schaefer:2009xx}.
These lattice fermions have near exact chiral symmetry and 
avoid artificial lattice operator mixing, while retaining acceptable
topology change in our region of simulation. 

We have two
lattices of similar physical volume at two lattice spacings:
\begin{enumerate}
\item[(i)] Our finer lattice has $32^3\times 64\times 16$ points and a coupling $\beta = 2.25$, which our analysis suggests corresponds to an 
inverse lattice spacing
$a^{-1} = 2.28(3)$\,GeV. We refer to the ensembles with $\beta=2.25$ as the \ensA ensemble set.
\item[(ii)] Our coarser lattice has $24^3\times 64\times 16$ points and a coupling $\beta = 2.13$, 
corresponding to $a^{-1} = 1.73(3)$\,GeV. The ensembles with $\beta=2.13$ are labeled as the \ensB ensemble set.
\end{enumerate}
For each ensemble set we use a number of valence masses to increase the amount of information in the light
mass regime. We use our standard notation for quark masses. $m_l$ and $m_h$ represent respectively the lighter and heavier of the two sea-quark masses (the sea consists of two quarks with mass $m_l$ and one with mass $m_h$). For the valence masses we use subscripts from the end of the alphabet $m_v$, $m_x$ and $m_y$ as appropriate. $m_{l,h}$ are masses in the DWF action used in the simulation whereas the valence masses appear in the corresponding partially quenched action. Because of the finite extent of the fifth dimension, small residual mass effects are present and the multiplicatively renormalizable bare quark masses are defined as $\widetilde{m}_{l,h,v,x,y}=m_{l,h,v,x,y}+m_{\textrm{res}}$, where $m_{\textrm{res}}$ is the residual mass.
The values of the valence quark masses used in our measurements are summarised in
Table~\ref{tab:partialquenched}. As in Reference~\cite{Allton:2008pn}, we will restrict our analysis, which relies on SU(2) chiral perturbation theory, to light-quark masses corresponding to pions lighter than about 420\,MeV.

\begin{table}
\centering
\begin{tabular}{c|c|c|c}
\hline
Lattice                          & $m_h$ & $\{m_l\}$           & $\{m_v\}$\\
\hline
\ensA $32^3\times 64$ 		 & $0.03$& $0.004,0.006,0.008$ & $0.002,0.004,0.006,0.008,0.025,0.03$\\
\hline
\ensB $24^3\times 64$ 		 & $0.04$& $0.005,0.01$        & $0.001,0.005,0.01,0.03,0.02,0.04$\\
\end{tabular}
\caption{Details of partially quenched valence masses $\{m_v\}$ on each ensemble. Meson correlation functions were computed for all possible pairings of 
valence masses.
\label{tab:partialquenched} }
\end{table}

We use two approaches to calculate the matrix element
$\langle K^0 |{\cal{O}}_{\mathrm{VV+AA}}| \overline{K}^0\rangle$. Both combine periodic and anti-periodic boundary conditions
in the time direction
to eliminate the leading, unwanted
\textit{around-the-world} propagation of the meson states that arise with a finite lattice in the time direction.
In both cases we use gauge-fixed wall sources to
create a $K^0$ state and annihilate a $\overline{K}^0$ state,
and form a ratio
\begin{equation}
B_K^{\rm lat} = \frac{\langle K^0(t_1) | {\cal O}_{VV+AA}(t) | \bar{K}^0(t_2) \rangle }
{\frac{8}{3}\langle K^0(t_1) | A_0(t) \rangle \langle A_0(t) | \bar{K}^0(t_2) \rangle}.
\end{equation}
For convenience we use the local axial current interpolating operators
in the denominator, and this ratio must be multiplied by a renormalization
constant 
\begin{equation}
Z_{B_K} = \frac{Z_{{\cal O}_{VV+AA}}}{Z_A^2},
\end{equation}
to obtain physically normalized matrix elements.

On our \ensA ensembles we used a single source at $t=0$ and
used the $(P+A)$ combination for the forward propagating $K$ meson.
This has the effect of creating
$(P+A)\times(P+A) = PP + AA + PA + AP$ combinations in meson
propagators, and the meson state has periodicity $2 L_T$, where $L_T=64$ is the temporal extent of the lattice.
Similarly
the $(P-A)$ combination is taken 
for the backward propagating $\overline{K}$ meson.
These Fermion boundary conditions are implemented on gauge
links crossing the toroidal wrapping plane between $t=0$
and $t=L_T-1$.
On each successive gauge configuration we selected a different
time $t_{\rm src}$ at which to insert the kaon sources. For simplicity
this was implemented by translating
the gauge configuration and redefining $t_{\rm src}$ to be zero.
The boundary condition described above is then applied.

The above approach requires half the number of propagator inversions on
each configuration (and enables us to sample more frequently at fixed cost)
compared to that taken on the \ensB ensembles.
On our \ensB ensembles we used a
source at $t=5$ and a source at $t=59$ requiring seperate
inversions for each source. For each propagator
entering a meson, we took the average of periodic (P) and anti-periodic
(A) solutions. 

The $\Delta S=2$ four-quark operator ${\cal{O}}_{\mathrm{VV+AA}}$ is inserted on all times
between the kaon creation and anti-kaon annihilation operators.
The locations of the kaon, anti-kaon and operator all receive $L^3$
volume averages, giving a low variance estimate of the correlation function.

The quality of the data can be gauged from 
Figures~\ref{fig:mpi32_0.004_unitary}
through~\ref{fig:bk24_0.005_unitary}, displaying the lightest simulated
pion, heaviest eta and a typical kaon matrix element  fit to $B_K^{\rm lat}$
for each of the two lattice spacings. More examples can be found in ref \cite{thirtytwocubed}.
Tables~\ref{tab:32cmatrixelements} and \ref{tab:24cmatrixelements} display 
the fitted values for the matrix element $B_K^{\rm lat}$ 
on each lattice. The fitted meson masses are as in 
reference~\cite{thirtytwocubed}.
\begin{figure}[tp]
\centering
\includegraphics*[width=10cm]{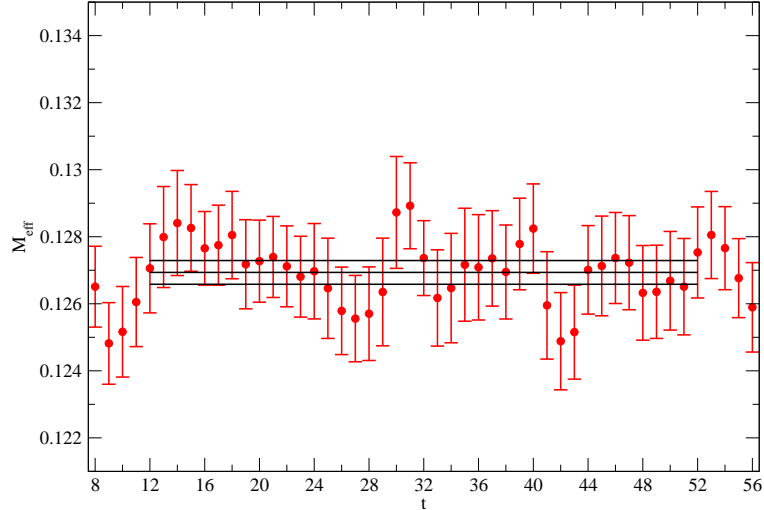}
\caption{Effective mass plateau of the lightest unitary simulated pion ($m_h=0.03$, $m_x=m_y=m_l=0.004$) on the \ensA ensembles. Here the plateau is obtained from the wall-local PP correlator, but the fit displayed is to all pseudoscalar correlators.} 
\label{fig:mpi32_0.004_unitary}
\end{figure}

\begin{figure}[tp]
\centering
\includegraphics*[width=10cm]{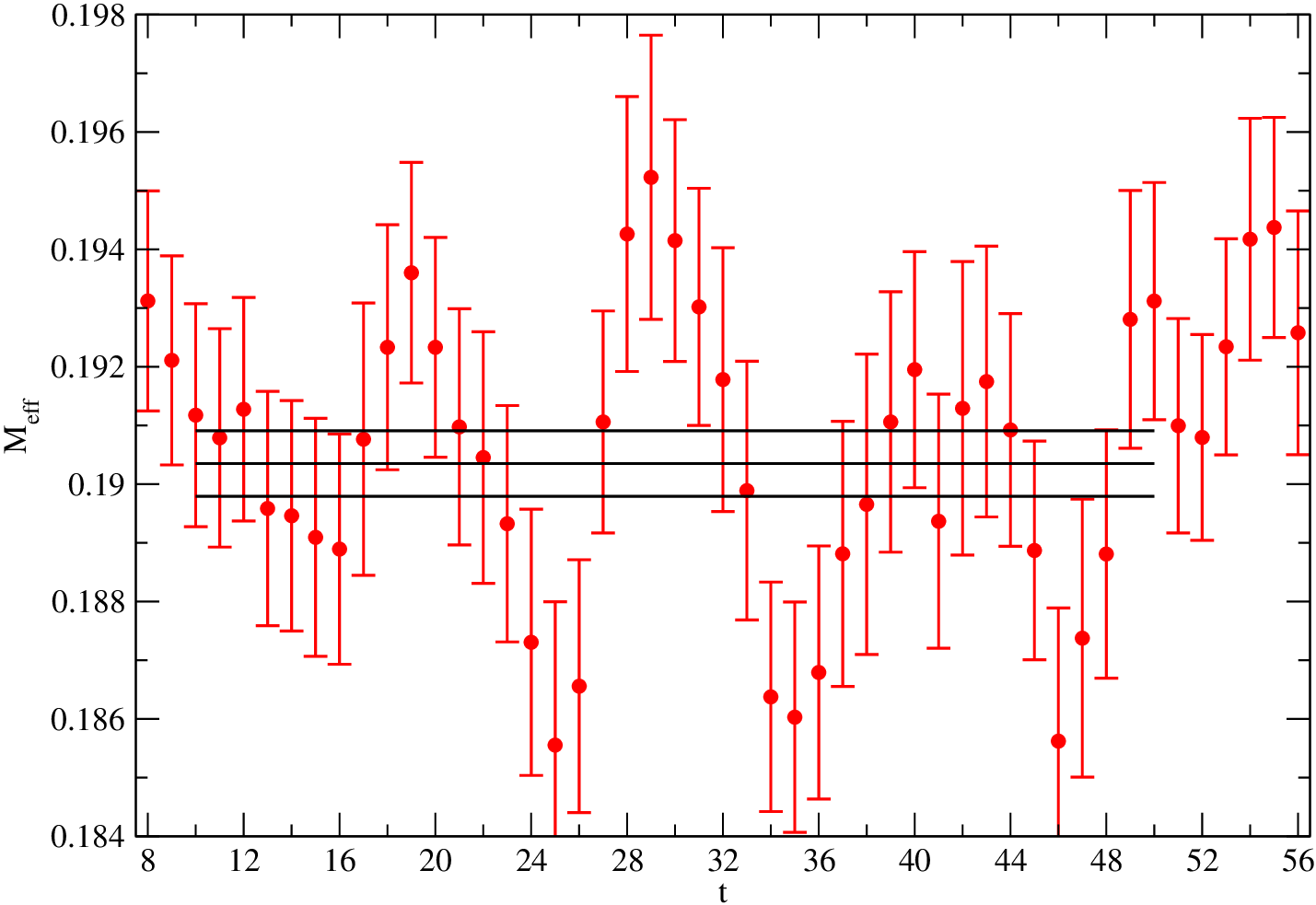}
\caption{Effective mass plateau of the lightest unitary simulated pion ($m_h=0.04$, $m_x=m_y=m_l=0.005$) on the \ensB ensembles. Here the plateau is obtained from the wall-local PP correlator, but the fit displayed is to all pseudoscalar correlators.} 
\label{fig:mpi24_0.005_unitary}
\end{figure}

\begin{figure}[tp]
\centering
\includegraphics*[width=10cm]{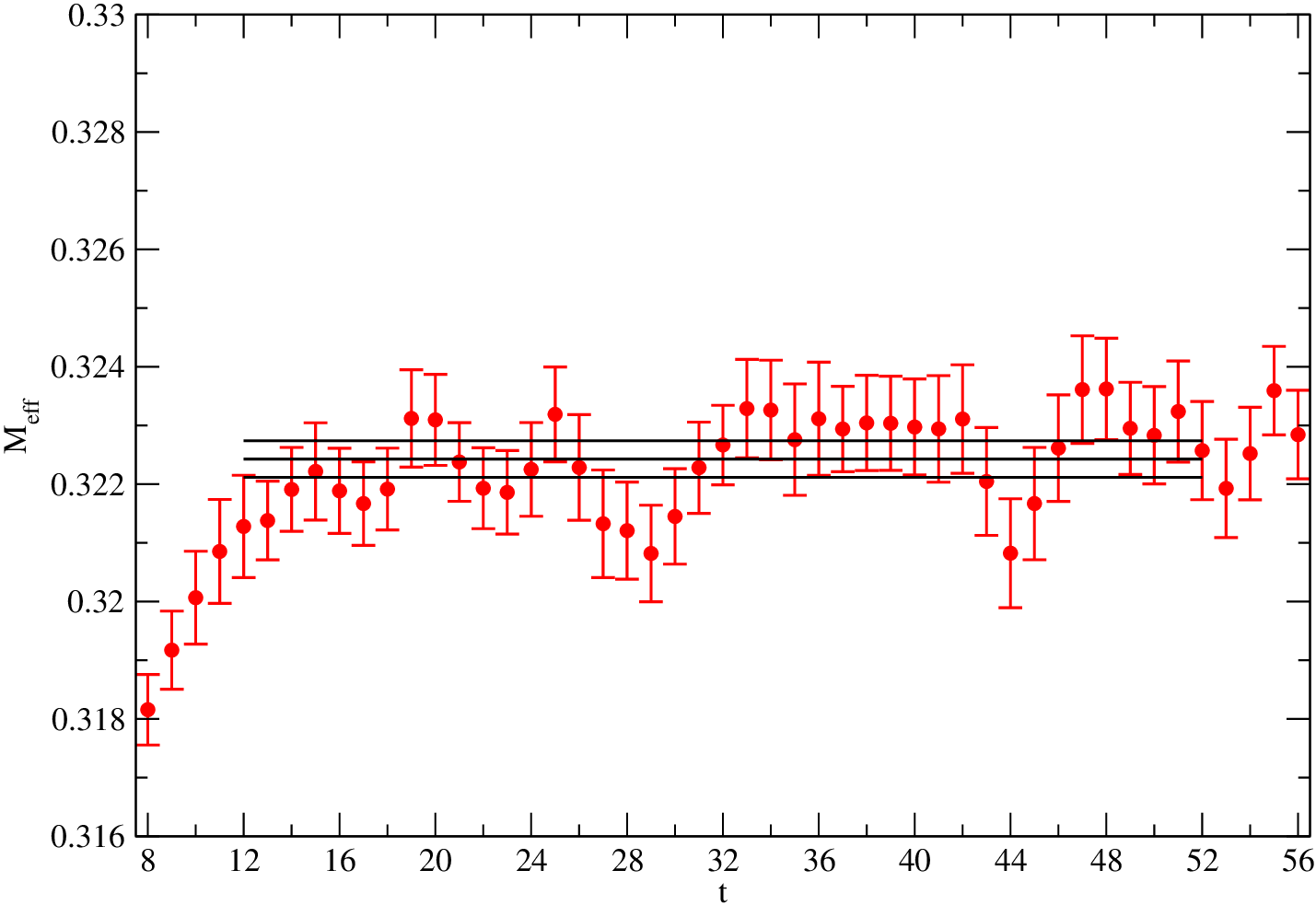}
\caption{Effective mass plateau of the heaviest simulated eta ($m_x=m_y=m_h=0.03$, $m_l=0.008$) on the \ensA ensembles. Here the plateau is obtained from the wall-local PP correlator, but the fit displayed is to all pseudoscalar correlators.} 
\label{fig:meta32_0.03_0.008}
\end{figure}

\begin{figure}[tp]
\centering
\includegraphics*[width=10cm]{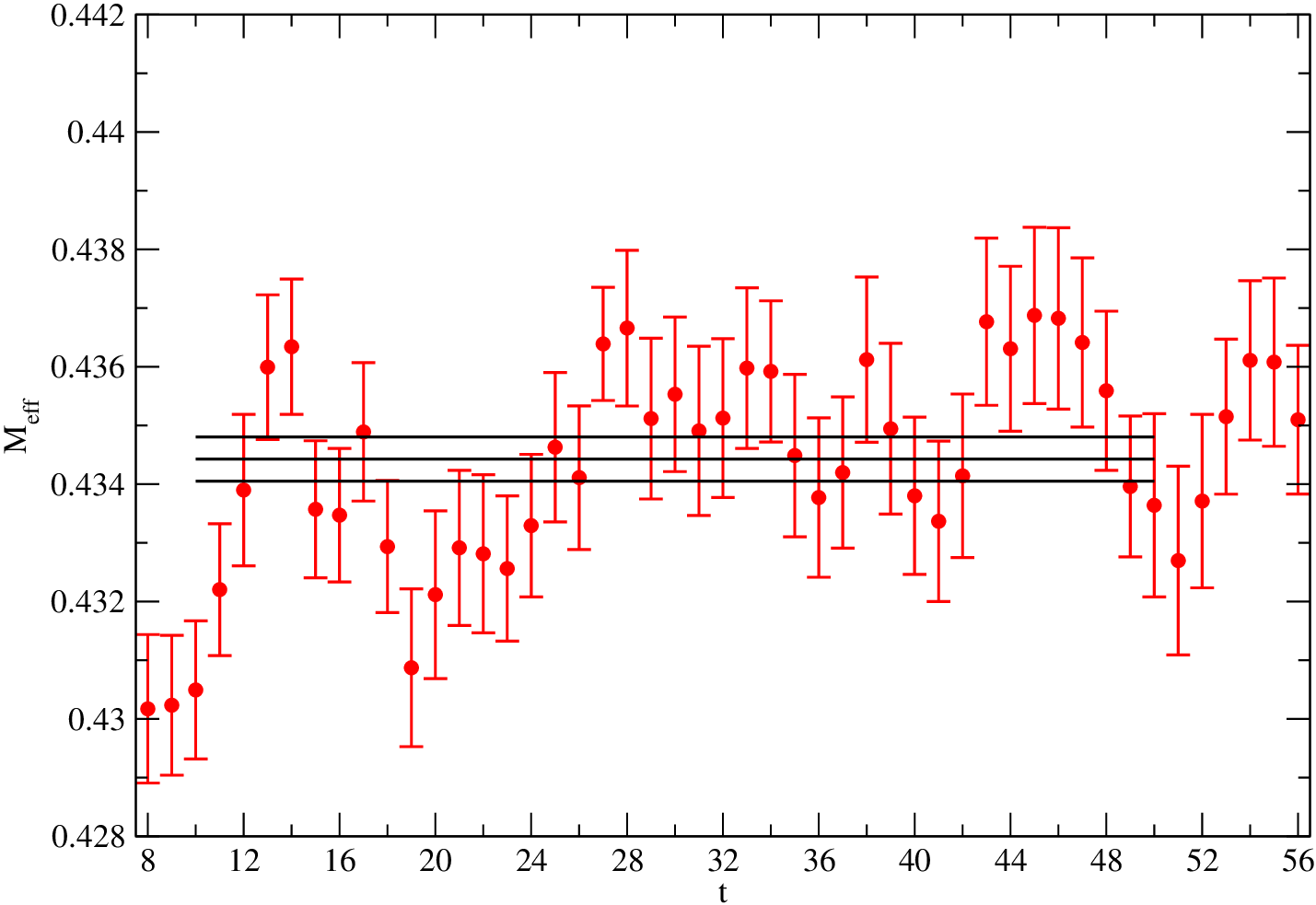}
\caption{Effective mass plateau of the heaviest simulated eta ($m_x=m_y=m_h=0.04$, $m_l=0.01$) on the \ensB ensembles. Here the plateau is obtained from the wall-local PP correlator, but the fit displayed is to all pseudoscalar correlators.} 
\label{fig:meta24_0.04_0.01}
\end{figure}

\begin{figure}[tp]
\centering
\includegraphics*[width=10cm]{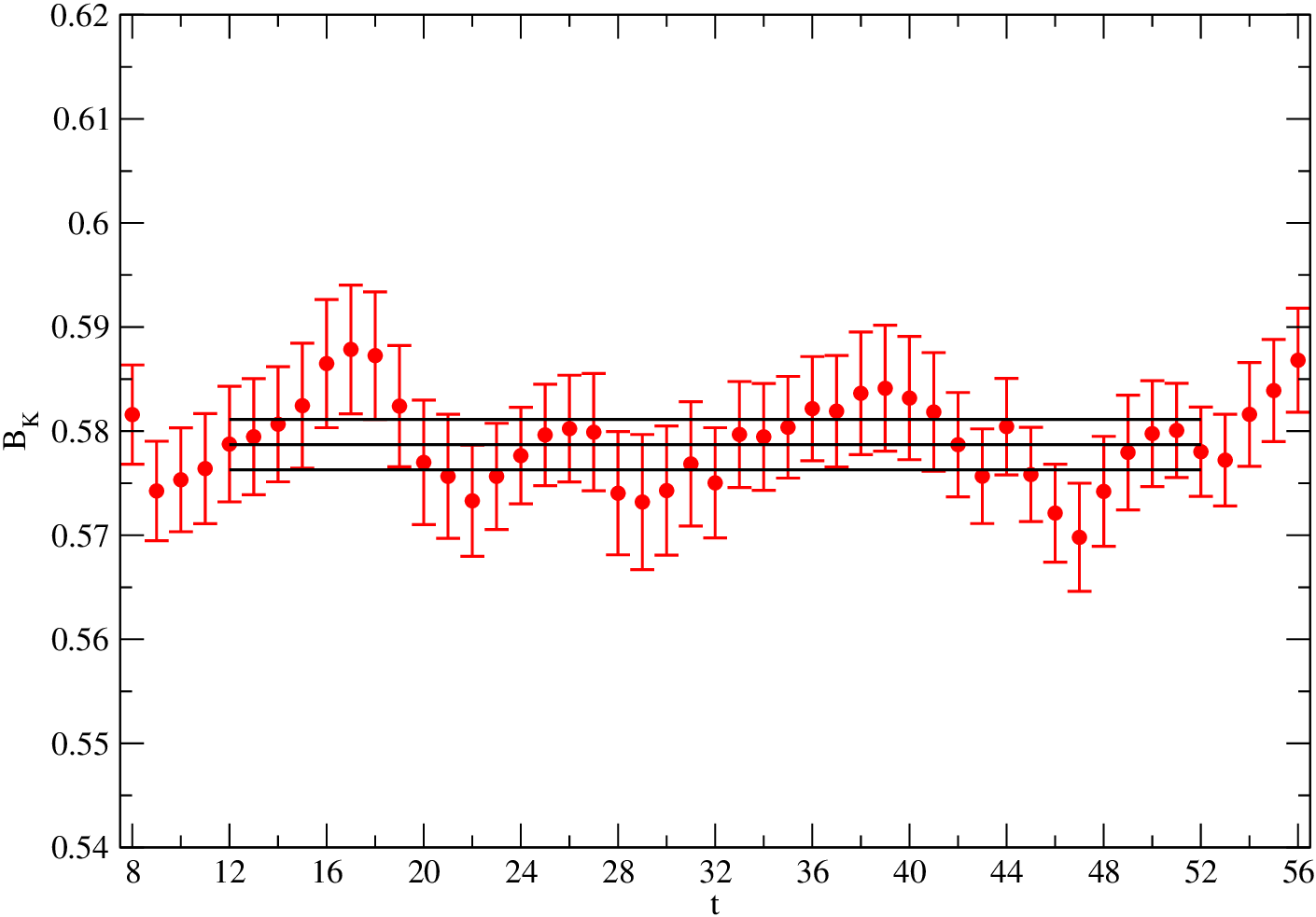}
\caption{A typical $B_K^{\rm lat}$ matrix element 
correlator ($m_y=m_h=0.03$, $m_x=m_l=0.004$) on the \ensA ensembles.} 
\label{fig:bk32_0.004_unitary}
\end{figure}

\begin{figure}[tp]
\centering
\includegraphics*[width=10cm]{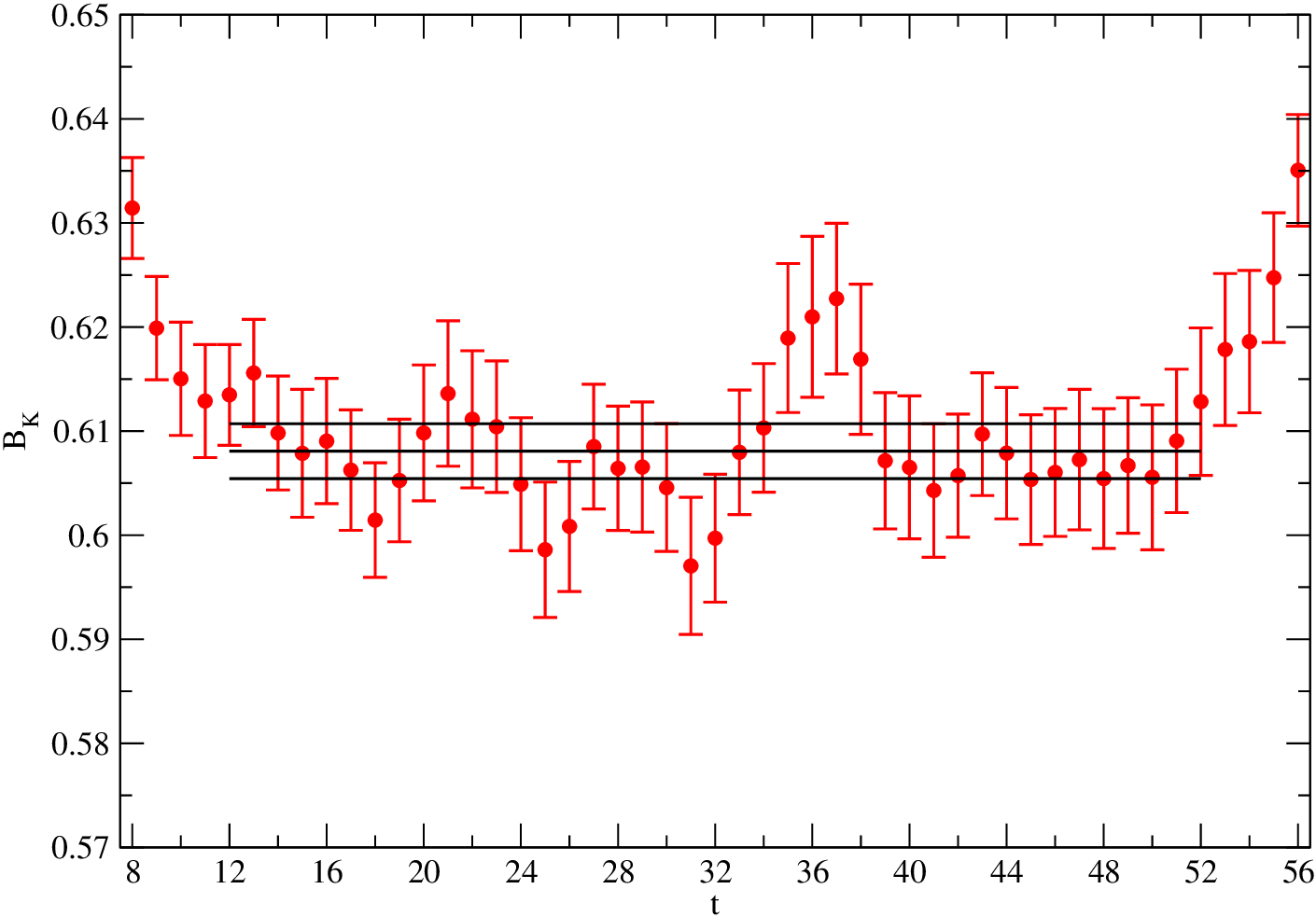}
\caption{A typical $B_K^{\rm lat}$ matrix element correlator ($m_y=m_h=0.04$, $m_x=m_l=0.005$) on the \ensB ensembles.} 
\label{fig:bk24_0.005_unitary}
\end{figure}

\begin{table}
\begin{tabular}{cc|ccc}
\hline
$m_x$ & $m_y$ & $B_{xy}(m_l=0.004)$ & $B_{xy}(m_l=0.006)$ & $B_{xy}(m_l=0.008)$ \\
\hline
0.03 & 0.03 & 0.6289(12)  & 0.6305(12) & 0.6295(12)\\
0.025 & 0.03 & 0.6199(12) & 0.6214(12) & 0.6207(12)\\
0.008 & 0.03 & 0.5862(17) & 0.5878(17) & 0.5878(19)\\
0.006 & 0.03 & 0.5823(19) & 0.5838(21) & 0.5838(22)\\
0.004 & 0.03 & 0.5787(24) & 0.5801(27) & 0.5798(28)\\
0.002 & 0.03 & 0.5767(46) & 0.5772(43) & 0.5781(50)\\
0.025 & 0.025 & 0.6100(13) & 0.6116(13) & 0.6110(13)\\
0.008 & 0.025 & 0.5725(16) & 0.5745(17) & 0.5741(18)\\
0.006 & 0.025 & 0.5679(17) & 0.5701(20) & 0.5694(21)\\
0.004 & 0.025 & 0.5634(21) & 0.5659(24) & 0.5649(26)\\
0.002 & 0.025 & 0.5601(39) & 0.5629(37) & 0.5630(43)\\
0.008 & 0.008 & 0.5135(18) & 0.5178(19) & 0.5141(20)\\
0.006 & 0.008 & 0.5047(19) & 0.5096(20) & 0.5056(22)\\
0.004 & 0.008 & 0.4951(21) & 0.5013(23) & 0.4969(25)\\
0.002 & 0.008 & 0.4852(28) & 0.4939(32) & 0.4901(34)\\
0.006 & 0.006 & 0.4949(20) & 0.5004(22) & 0.4961(24)\\
0.004 & 0.006 & 0.4842(23) & 0.4908(25) & 0.4864(27)\\
0.002 & 0.006 & 0.4727(29) & 0.4813(34) & 0.4781(35)\\
0.004 & 0.004 & 0.4721(26) & 0.4791(29) & 0.4753(31)\\
0.002 & 0.004 & 0.4584(32) & 0.4663(37) & 0.4647(39)\\
0.002 & 0.002 & 0.4408(39) & 0.4473(44) & 0.4500(48)\\
\end{tabular}
\caption{Fitted $B_K^{\rm lat}$ matrix element values on the \ensA ensembles. For heavy-light matrix elements, $m_y$ is the heavy quark mass. We chose a fit range of $t=12-52$.
\label{tab:32cmatrixelements} }
\end{table}

\begin{table}
\begin{tabular}{cc|ccc}
\hline
$m_x$ & $m_y$ & $B_{xy}(m_l=0.005)$ & $B_{xy}(m_l=0.01)$\\
\hline
0.04 & 0.04 & 0.6565(12) & 0.6562(12)\\
0.03 & 0.04 & 0.6435(14) & 0.6430(13)\\
0.02 & 0.04 & 0.6298(16) & 0.6291(14)\\
0.01 & 0.04 & 0.6154(20) & 0.6145(17)\\
0.005& 0.04 & 0.6081(26) & 0.6078(24)\\
0.001& 0.04 & 0.6017(48) & 0.6072(53)\\
0.03 & 0.03 & 0.6286(14) & 0.6280(13)\\
0.02 & 0.03 & 0.6124(16) & 0.6117(14)\\
0.01 & 0.03 & 0.5949(19) & 0.5943(16)\\
0.005& 0.03 & 0.5860(23) & 0.5860(20)\\
0.001& 0.03 & 0.5787(40) & 0.5835(40)\\
0.02 & 0.02 & 0.5929(17) & 0.5924(15)\\
0.01 & 0.02 & 0.5712(19) & 0.5711(16)\\
0.005& 0.02 & 0.5598(23) & 0.5603(19)\\
0.001& 0.02 & 0.5505(36) & 0.5547(31)\\
0.01  & 0.01 & 0.5431(22) & 0.5439(18)\\
0.005 & 0.01 & 0.5272(26) & 0.5284(21)\\
0.001 & 0.01 & 0.5134(37) & 0.5164(29)\\
0.005 & 0.005 & 0.5075(31) & 0.5085(24)\\
0.001 & 0.005 & 0.4893(42) & 0.4903(31)\\
0.001 & 0.001 & 0.4652(55) & 0.4631(40)\\
\end{tabular}
\caption{Fitted $B_K^{\rm lat}$ matrix element values on the \ensB ensembles. For heavy-light matrix elements, $m_y$ is the heavy quark mass. We chose a fit range of $t=12-52$.
\label{tab:24cmatrixelements} }
\end{table}

\subsection{Reweighting}

As explained above, at each lattice spacing we have performed the simulations using a number of light-quark 
masses but only a single sea strange-quark mass. As we can only determine the physical strange quark mass 
$m_s$ after the analysis is complete, our imperfect pre-simulation estimate of $m_s$ has been a source of 
error in previous calculations,
where we could only adjust the valence strange quark mass or use SU(3) chiral perturbation theory to 
estimate the effects of varying the unitary strange quark mass. 
We do not expect significant effects from small adjustments of the
sea strange-quark mass and \textit{reweighting} gives us a tool to demonstrate this
without doubling the cost of the simulation. 
For more discussion we
refer to our papers \cite{Allton:2008pn,thirtytwocubed}.

\begin{figure}[tp]
\centering
\includegraphics*[width=10cm]{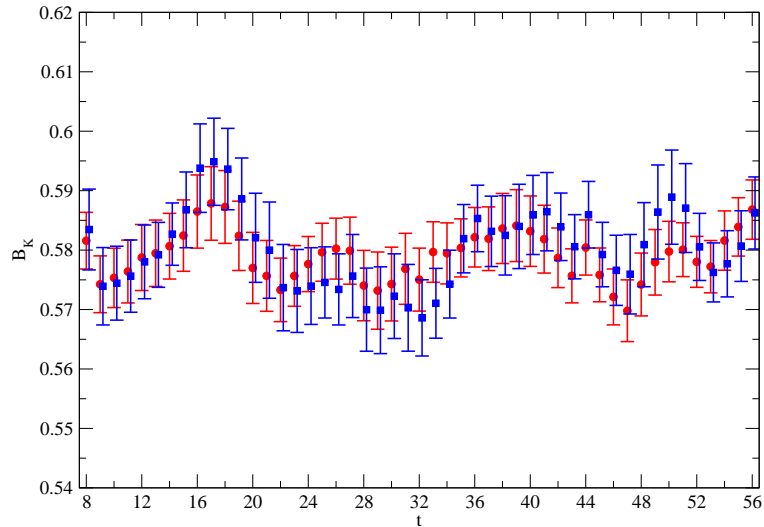}
\caption{An overlay of a typical $B_K^{\rm lat}$ matrix element 
($m_y=0.03$, $m_x=m_l=0.004$) on the \ensA ensembles at two values of the sea strange quark mass: $m_h=0.03$ (red) and $m_h=0.027$ (blue). The latter is at our closest reweight to the physical strange mass.}  
\label{fig:bk32_rwcomparison}
\end{figure}

\begin{figure}[tp]
\centering
\includegraphics*[width=10cm]{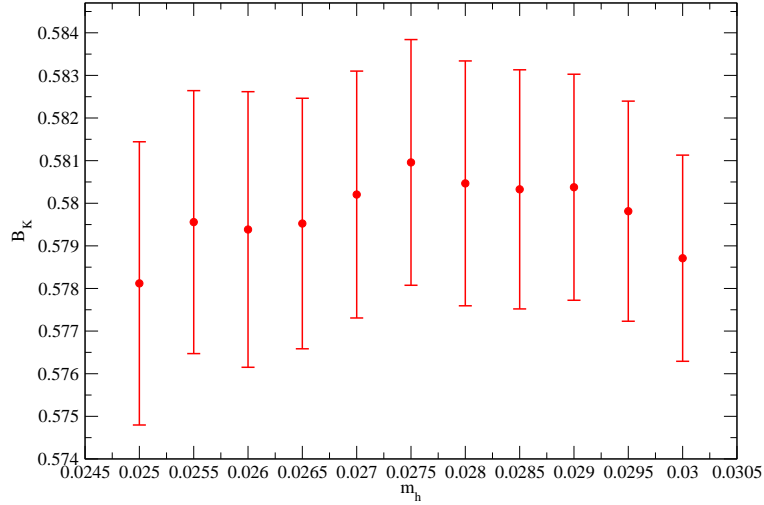}
\caption{The $m_h$ dependence of a typical $B_K^{\rm lat}$ matrix element 
 ($m_y=0.03$, $m_x=m_l=0.004$) on the \ensA ensembles.}  
\label{fig:bk32_mhdependence}
\end{figure}

Figure~\ref{fig:bk32_rwcomparison} shows an overlay of a typical kaon 
$B_K^{\rm lat}$ matrix element correlator at the
simulated sea strange-quark mass and the physical value. Figure~\ref{fig:bk32_mhdependence} shows the
dependence of the fitted value 
of the matrix element of ${\cal{O}}_{\mathrm{VV+AA}}$ 
on the sea strange-quark mass; the dependence is very small and
barely statistically significant.

For both ensemble sets, we compute the propagators at two valence strange-quark masses: $m_y=0.03$ and $0.025$ for the \ensA ensembles and $m_y=0.04$ and $0.03$ for the \ensB ensembles. When computing kaonic quantities we reweight the sea strange mass $m_h$ to both valence 
strange-quark masses $m_y$ such that $m_h = m_y$ in our observables. 
For each lattice and at each value of $m_l$ we therefore have results with two strange quark masses 
with $m_h=m_y$, one at the strange-quark mass at which we perform the simulation and the second obtained by reweighting. 
This enables us to interpolate linearly in the unitary strange quark mass to the physical point. In 
Tables \ref{tab:32cphysmhmatrixelements} and \ref{tab:24cphysmhmatrixelements} 
we give the values for the heavy-light $B_{xy}$ matrix element on each ensemble; it is to these data that we perform our simultaneous chiral fits in Section \ref{sec:ChiralExtrap}.

\begin{table}
\begin{tabular}{c|ccc}
\hline
$m_x$ & $B_{xh}(m_l=0.004)$ & $B_{xh}(m_l=0.006)$ & $B_{xh}(m_l=0.008)$ \\
\hline
0.008 & 0.5802(27) & 0.5807(29) & 0.5829(26)\\
0.006 & 0.5758(29) & 0.5764(32) & 0.5789(29)\\
0.004 & 0.5715(33) & 0.5721(38) & 0.5752(36)\\
0.002 & 0.5679(49) & 0.5680(52) & 0.5742(59)\\
\end{tabular}
\caption{Heavy-light $B_K^{\rm lat}$
matrix element values on the \ensA ensembles at the physical $m_h=0.0273(7)$, $m_h+m_\mathrm{res}=0.0278(7)$ obtained from the NLO PQChPT combined fits of Section~\ref{sec:ChiralExtrapRes}. These values are obtained by first reweighting to $m_h=m_y$ then linearly interpolating in the unitary strange mass.
\label{tab:32cphysmhmatrixelements} }
\end{table}

\begin{table}
\begin{tabular}{c|cc}
\hline
$m_x$ & $B_{xh}(m_l=0.005)$ & $B_{xh}(m_l=0.01)$\\
\hline
0.02 & 0.6191(32) & 0.6190(27)\\
0.01 & 0.6035(35) & 0.6029(31)\\
0.005 & 0.5959(38) & 0.5949(37)\\
0.001 & 0.5892(64) & 0.5904(65)\\
\end{tabular}
\caption{Heavy-light $B_K^{\rm lat}$
matrix element values on the \ensB ensembles at the physical $m_h= 0.035(1)$, $m_h+m_\mathrm{res} = 0.038(1)$ obtained from the NLO PQChPT combined fits of Section~\ref{sec:ChiralExtrapRes}. These values are obtained by first reweighting to $m_h=m_y$ then linearly interpolating in the unitary strange mass.
\label{tab:24cphysmhmatrixelements}
} 
\end{table}

\fi

\section{Non-perturbative renormalisation}
\label{sec:NPR}
\ifnum\theNPR=1
In this section we discuss the renormalization of the $\Delta S=2$ operator ${\cal{O}}_{\mathrm{VV+AA}}$, whose matrix elements we are computing. We start by performing non-perturbative renormalization, calculating numerically the renormalization factor which relates the bare lattice operator corresponding to our choice of the discrete QCD action to that defined in some intermediate renormalization scheme. For this to be feasible, of course, it is necessary that the intermediate scheme can be implemented numerically and we use several momentum subtraction schemes which are generalizations of the original RI-MOM scheme
\cite{Martinelli:1994ty}. In phenomenological applications, our results for the matrix element $\langle K^0 |{\cal{O}}_{\mathrm{VV+AA}}| \overline{K}^0\rangle$ have to be combined with the Wilson coefficient function which is calculated in perturbation theory, most frequently using renormalization schemes based on dimensional regularization, such as the NDR scheme. It is therefore necessary to combine the coefficient function and the operator matrix element in the same scheme. Below we present the matching factors which relate the operator renormalized in our intermediate schemes to the corresponding operator in the NDR scheme. Since dimensional regularization cannot be implemented in lattice simulations, this (continuum) matching is performed in perturbation theory (at one-loop order) and is of course independent of the lattice calculations. The procedure described above can be summarised as follows:
\begin{eqnarray*}
\textrm{Bare Lattice Operator}&\overset{\textrm{NPR}}{\to}& \textrm{Renormalized Operator in Momentum Subtraction Scheme}\\
&&\\
&\overset{\textrm{Perturbation Theory}}{\to}& \textrm{Renormalized Operator in }\overline{\textrm{MS}}\textrm{-NDR Scheme.}
\end{eqnarray*}

The momentum subtraction schemes which we use require the evaluation of the Green functions for the transition $d(p_1)\overline{s}\hspace{1pt}(p_2)\to \overline{d}(p_3)s(p_4)$ with a suitable choice of the momenta $p_i$. In the past, see in particular Reference~\,\cite{Antonio:2007pb}, the results were presented using the RI-MOM kinematic configuration in which $p_1=-p_2=p_3=-p_4$
\cite{Donini:1995xj}. 
Whilst this is correct asymptotically, i.e. when the $p_i^2$ are sufficiently large for each choice of the quark masses, it was argued in References~\cite{Aoki:2007xm,Aoki:2009ka,Sturm:2009kb} that performing the renormalization using Green functions with no exceptional channels, i.e. with no channels in which the square of the momentum $q^2$ is small, suppresses the non-asymptotic chiral symmetry breaking effects more effectively. In addition to the theoretical arguments, numerical evidence was presented demonstrating the suppression of terms which violated the chiral Ward-Takahashi identities, such as the equality of the renormalization constants of the vector and axial currents and of the scalar and pseudoscalar densities. Although the effects are small, typically of the order of a few percent, lattice calculations are becoming sufficiently precise that the reduction of this systematic error is necessary.

For $B_K$, the RI-MOM kinematics defined in the previous paragraph clearly have exceptional channels (e.g. $p_1+p_2=0$) and in this paper we generalize the non-exceptional RI-SMOM schemes of References~\cite{Aoki:2007xm,Aoki:2009ka,Sturm:2009kb} to the four-quark operator. The choice of non-exceptional kinematics is not unique of course and in this paper we choose to study the Green function
\begin{equation}\label{eq:smomkinematics}
d(p_1)\overline{s}(-p_2)\to \bar{d}(-p_1)s(p_2)\,
\end{equation}
with $p_1^2=p_2^2=(p_1-p_2)^2\equiv p^2$ for a 
variety of momenta satisfying these conditions. In our notation
below $q=p_1-p_2$.

We briefly mention that we have previously investigated 
non-exceptional (or strictly speaking \emph{less}
exceptional) momenta for four-quark operators~\cite{Aoki:2007xm};
here the operator was inserted only at a single point on the lattice and
the method was less statistically precise than our current work. 
Chirality mixing in the four-quark operator basis
arising in the infra-red $p^2$ region was found to be strongly suppressed
\cite{Aoki:2007xm},
thus revealing the true, good chiral properties of DWF. However,
the corresponding perturbative calculation to match \emph{this} kinematic
point to the continuum $\MSbar$~scheme was not available, and this was of
largely academic interest in displaying the quality of Domain Wall Fermions.

The remainder of the section is organised as follows. In the next subsection we introduce 4 RI-SMOM renormalization schemes, all of them defined with the kinematics of Equation\,(\ref{eq:smomkinematics}). In Subsection~\ref{subsec:matching} we calculate the perturbative matching factors relating ${\cal{O}}_{\mathrm{VV+AA}}$ in the 4 RI-SMOM schemes with that in the $\overline{\textrm{MS}}$-NDR renormalization scheme. We review some aspects of the non-perturbative renormalization of the lattice operator into a RI-SMOM renormalization scheme in Subsection~\ref{subsec:volume-average} and finally in Subsection~\ref{subsec:zbk} we combine the NPR computation and matching calculation to obtain the total renormalization factor relating the lattice and $\overline{\textrm{MS}}$-NDR operators.

\subsection{RI-SMOM Renormalization Schemes for ${\cal{O}}_{\mathrm{VV+AA}}$}\label{subsec:smomdef}

\label{sec:schemes}

We follow the procedure which was defined for the renormalization of the 
four-quark operators in the RI-MOM Scheme~\cite{Donini:1995xj}, but now 
with the kinematics defined in Equation (\ref{eq:smomkinematics}). We begin 
with the evaluation of the amputated four-point Green function 
$\Lambda_{\alpha\beta,\gamma\delta}^{ij,kl}$ of the operator 
${\cal{O}}_{\mathrm{VV+AA}}$, where  $\alpha$, $\beta$, $\gamma$, and 
$\delta$ are the 
spinor labels corresponding to the incoming $\overline{s}$ and $d$ quarks 
and outgoing $s$ and $\bar{d}$ quarks respectively and $i$, $j$, $k$, $l$ 
are the 
corresponding colour labels. Analogously to the definition of the RI-MOM 
scheme, we impose conditions on the amputated Green
functions at the renormalization 
scale in such a way that they are automatically satisfied by the tree-level 
Green functions.
To this end we introduce two projection operators
 $P^{ij,kl}_{(X),\,\alpha\beta,\gamma\delta}$,
with $X\in\{1,2\}$:
\begin{eqnarray}
P^{ij,kl}_{(1),\,\alpha\beta,\gamma\delta}&=&\frac{1}{256\*N\*(N+1)}
[(\gamma^\nu)_{\beta\alpha}(\gamma_\nu)_{\delta\gamma}+(\gamma^\nu\gamma^5)_{\beta\alpha}
(\gamma_\nu\gamma^5)_{\delta\gamma}]\, \delta_{ij}\delta_{kl}\,,
\label{eq:projector1}\\
P^{ij,kl}_{(2),\,\alpha\beta,\gamma\delta}&=&{1\over64\*q^2\*N\*(N+1)}
\left[(\FslashA{q})_{\beta\alpha}(\FslashA{q})_{\delta\gamma}
     +(\FslashA{q}\gamma_5)_{\beta\alpha}(\FslashA{q}\gamma_5)_{\delta\gamma}\right]
\delta_{ij}\delta_{kl}\,,\label{eq:projector2}
\end{eqnarray}
where $N=3$ is the number of colours. These projectors are constructed to give 1 when contracted with the tree-level result for $\Lambda_{\alpha\beta,\gamma\delta}^{ij,kl}$ given in Equation (\ref{eq:tree}) below.

In order to specify the renormalization condition on the operator we have to include a factor of $\sqrt{Z_q}$ for every external quark line, where $Z_q$ is the wave function renormalization factor, and here again we use two possible definitions, called RI-SMOM and RI-SMOM$_{\gamma_\mu}$ in Reference~\cite{Sturm:2009kb}.
Here, we do not reproduce the explicit definitions in terms of the renormalization of the quark propagator, but note that they are chosen to satisfy the Ward Takahashi identities when combined with the renormalization conditions on the vertex function for the (conserved) vector current using two different projectors. Specifically in the SMOM-scheme
\begin{equation}\label{eq:zqrismomdef}
Z_q^{\textrm{RI-SMOM}}=\frac{q_\mu}{12q^2}\,\textrm{Tr}[\Lambda_V^\mu\qslash]\,,
\end{equation}
where the trace is over both colour and spinor indices, $q$ is the momentum transfer at the vector current and $\Lambda_V$ is the amputated two point function with the incoming (outgoing) quark having momentum $p_1$ ($p_2$) with $q=p_1-p_2$ and with $p_1^2=p_2^2=q^2$ chosen to be the renormalization scale. For the second scheme we use the same projector as in the definition of the RI-MOM scheme, but with the non-exceptional kinematics as above,
\begin{equation}\label{eq:zqrismomgammamudef}
Z_q^{\textrm{RI-SMOM}_{\gamma_\mu}}=\frac{1}{48}\,\textrm{Tr}[\Lambda_V^\mu\gamma^\mu]\,.
\end{equation}

We label the renormalized four-quark operator in each of the four schemes by two labels $(X,Y)$ with $X=\qslash$ or $\gamma^\mu$ depending on which of the projectors Equation (\ref{eq:projector1}) or (\ref{eq:projector2}) are used for the 
vertex and similarly $Y=\qslash$ or $\gamma^\mu$ depending on which of the definitions Equations (\ref{eq:zqrismomdef}) 
or (\ref{eq:zqrismomgammamudef}) are used for the wavefunction renormalization. Thus for example,
\begin{equation}\label{eq:o11def}
{\cal O}_{R,VV+AA}^{(\gamma_\mu,\qslash)}=Z^{(\gamma_\mu,\qslash)}_{\cal O}{\cal O}_{B,VV+AA},
\end{equation}
where
\begin{equation}\label{eq:z11def}
Z^{(\gamma_\mu,\qslash)}_{\cal O}=(Z_q^{\textrm{RI-SMOM}})^2\,\frac{1}{P_{(1),\alpha\beta,\gamma\delta}^{ij,kl} \Lambda_{B,\alpha\beta,\gamma\delta}^{ij,kl}}\,.
\end{equation}
We have introduced the subscripts $R$ and $B$ in Equations (\ref{eq:o11def}) and (\ref{eq:z11def}) to denote \textit{renormalized} and \textit{bare} (or lattice) quantities respectively. The remaining renormalized operators are defined similarly:
\begin{eqnarray}\label{eq:z12def}
Z^{(\gamma_\mu,\gamma_\mu)}_{\cal O}&=&(Z_q^{\textrm{RI-SMOM}_{\gamma_\mu}})^2\,\frac{1}{P_{(1),\alpha\beta,\gamma\delta}^{ij,kl} \Lambda_{B,\alpha\beta,\gamma\delta}^{ij,kl}}\\
Z^{(\qslash,\qslash)}_{\cal O}&=&(Z_q^{\textrm{RI-SMOM}})^2\,\frac{1}{P_{(2),\alpha\beta,\gamma\delta}^{ij,kl} \Lambda_{B,\alpha\beta,\gamma\delta}^{ij,kl}}\label{eq:z21def}\\
Z^{(\qslash,\gamma_\mu)}_{\cal O}&=&(Z_q^{\textrm{RI-SMOM}_{\gamma_\mu}})^2\,\frac{1}{P_{(2),\alpha\beta,\gamma\delta}^{ij,kl} \Lambda_{B,\alpha\beta,\gamma\delta}^{ij,kl}}\label{eq:z22def}
\end{eqnarray}
and in each case ${\cal O}_{R,VV+AA}^{(X,Y)}=Z^{(X,Y)}_{\cal O}{\cal O}_{B,VV+AA}$, with $X,Y=\qslash$ or $\gamma_\mu$.

In addition to the four renormalization schemes defined above, we also use 
the standard RI-MOM scheme as the intermediate scheme in our conversion to 
$\overline{MS}$. The reason for introducing several renormalization schemes 
is that it allows us some control over the lattice and perturbative 
uncertainties. After performing the perturbative matching to the NDR scheme, 
each of these intermediate schemes should lead to the same value of the 
matrix element of ${\cal O}_{VV+AA}^{\textrm{NDR}}$. The spread of results 
obtained using the 5 schemes is therefore a measure of the uncertainties. 
In particular, since the matching coefficients from the intermediate schemes 
to the NDR scheme are currently available only at one-loop order (see 
Subsection \ref{subsec:matching}), the spread of results is an indication 
of the size of the higher-order terms.
We now turn to the evaluation of the matching coefficient at one-loop order.

\subsection{Perturbative Conversion to the $\NDR$ Scheme}\label{subsec:matching}

In this subsection, we calculate the conversion (matching) factors between
the four RI-SMOM schemes defined in Subsection \ref{subsec:smomdef} above
and the naive dimensional reduction (\NDR) scheme for the $\Delta S=2$
operator ${\cal O}_{VV+AA}=\bar{s}\gamma^\mu_L d\,\bar{s}\gamma_{\mu\,L} d$
(where $\gamma^\mu_L\equiv\gamma^\mu(1-\gamma^5)$ and we only consider the parity even component) using continuum
perturbation theory at the one-loop level.
The two-loop anomalous dimensions are also calculated
to derive  the renormalization group (RG) running of the operator
in these schemes.

We now use perturbation theory to convert the operators into the NDR schemes with the treatment of evanescent operators as in Reference~\cite{Buras:2000if}, as will be explained below. As explained above, for
$B_K$ the RI-SMOM schemes are defined in terms of projections of the amplitude $d(p_1)\overline{s}(-p_2)\to\bar{d}(-p_1)s(p_2)$, where $p_1^2=p_2^2=(p_1-p_2)^2\equiv p^2$ with $p_1\neq p_2$. For $p^2$ in the perturbative regime there is no channel with soft momenta, thus reducing infrared effects. At tree level we have the 4 diagrams in Figure~\ref{fig:fourlo}, where the circles represent the two currents $\bar{s}\gamma^\mu_Ld$, the arrows on the quark lines denote the flow of fermion number and the direction of the momenta are indicated explicitly below the corresponding momentum. Even though both momenta $p_1$ are ingoing and both momenta $p_2$ are outgoing, it is convenient to introduce the minus signs and to think of the process as $d(p_1) \bar{s}(-p_2)\to\bar{d}(-p_1)s(p_2)$ because then the signs also implicitly keep track of the spinor and colour labels (see Figure~\ref{fig:locombined}).
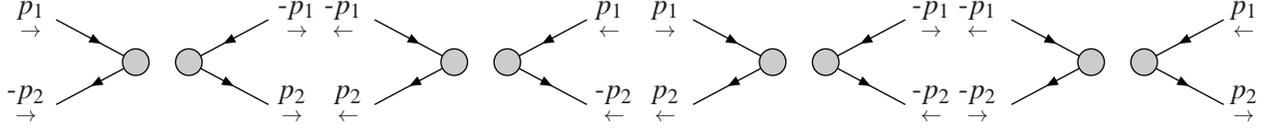
\begin{figure}[t]
\begin{center}
\begin{picture}(500,50)(45,30)
\ArrowLine(60,66)(90,50)\ArrowLine(90,50)(60,34)\ArrowLine(140,66)(110,50)\ArrowLine(110,50)(140,34)
\GCirc(90,50){5}{0.8}\GCirc(110,50){5}{0.8}
\Text(56,66)[r]{$\underset{\rightarrow}{p_1}$}\Text(56,34)[r]{$\underset{\rightarrow}{\textrm{-}p_2}$}
\Text(144,66)[l]{$\underset{\rightarrow}{\textrm{-}p_1}$}\Text(144,34)[l]{$\underset{\rightarrow}{p_2}$}
\ArrowLine(180,66)(210,50)\ArrowLine(210,50)(180,34)\ArrowLine(260,66)(230,50)\ArrowLine(230,50)(260,34)
\GCirc(210,50){5}{0.8}\GCirc(230,50){5}{0.8}
\Text(176,66)[r]{$\underset{\leftarrow}{\textrm{-}{p_1}}$}\Text(176,34)[r]{$\underset{\leftarrow}{p_2}$}
\Text(264,66)[l]{$\underset{\leftarrow}{p_1}$}\Text(264,34)[l]{$\underset{\leftarrow}{\textrm{-}p_2}$}
\ArrowLine(300,66)(330,50)\ArrowLine(330,50)(300,34)\ArrowLine(380,66)(350,50)\ArrowLine(350,50)(380,34)
\GCirc(330,50){5}{0.8}\GCirc(350,50){5}{0.8}
\Text(296,66)[r]{$\underset{\rightarrow}{p_1}$}\Text(296,34)[r]{$\underset{\leftarrow}{p_2}$}
\Text(384,66)[l]{$\underset{\rightarrow}{\textrm{-}p_1}$}\Text(384,34)[l]{$\underset{\leftarrow}{\textrm{-}p_2}$}
\ArrowLine(420,66)(450,50)\ArrowLine(450,50)(420,34)\ArrowLine(500,66)(470,50)\ArrowLine(470,50)(500,34)
\GCirc(450,50){5}{0.8}\GCirc(470,50){5}{0.8}
\Text(416,66)[r]{$\underset{\leftarrow}{\textrm{-}{p_1}}$}\Text(504,66)[l]{$\underset{\leftarrow}{p_1}$}
\Text(416,34)[r]{$\underset{\rightarrow}{\textrm{-}p_2}$}\Text(504,34)[l]{$\underset{\rightarrow}{p_2}$}
\end{picture}
\caption{The four lowest order diagrams. Each circle represents the insertion of the current $\bar{s}\gamma^\mu_L d$. The $d$ or $\bar{d}$ ($s$ or $\bar{s}$) quarks have momenta $\pm p_1$ ($\pm p_2$) and the flow of fermion number is denoted by the arrow.\label{fig:fourlo}}
\end{center}
\end{figure}
Since the two currents commute, the first two diagrams are clearly equal as are the second two; thus we can rewrite the four diagrams in Figure~\ref{fig:fourlo} in terms of the two diagrams in Figure~\ref{fig:locombined}, where the spinor (Greek letters) and colour (Latin letters) indices have now been indicated explicitly.
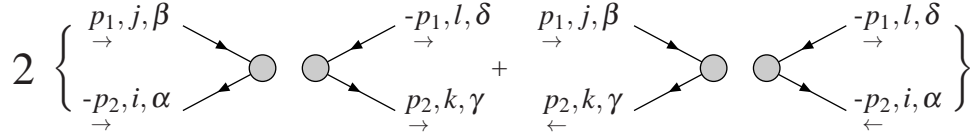
\begin{figure}[t]\begin{center}
\begin{picture}(300,40)(30,40)
\ArrowLine(60,66)(90,50)\ArrowLine(90,50)(60,34)\ArrowLine(140,66)(110,50)\ArrowLine(110,50)(140,34)
\GCirc(90,50){5}{0.8}\GCirc(110,50){5}{0.8}
\Text(56,66)[r]{$\underset{\rightarrow}{p_1},j,\beta$}\Text(56,34)[r]{$\underset{\rightarrow}{\textrm{-}p_2},i,\alpha$}
\Text(144,66)[l]{{\small $\underset{\rightarrow}{\textrm{-}p_1},l,\delta$}}\Text(144,34)[l]{{\small $\underset{\rightarrow}{p_2},k,\gamma$}}
\Text(180,50)[c]{+}
\ArrowLine(230,66)(260,50)\ArrowLine(260,50)(230,34)\ArrowLine(310,66)(280,50)\ArrowLine(280,50)(310,34)
\GCirc(260,50){5}{0.8}\GCirc(280,50){5}{0.8}
\Text(226,66)[r]{{\small $\underset{\rightarrow}{p_1},j,\beta$}}\Text(226,34)[r]{{\small $\underset{\leftarrow}{p_2},k,\gamma$}}
\Text(314,66)[l]{{\small $\underset{\rightarrow}{\textrm{-}p_1},l,\delta$}}
\Text(314,34)[l]{{\small $\underset{\leftarrow}{\textrm{-}p_2},i,\alpha$}}
\Text(0,50)[c]{\Large{2}}
\Text(15,50)[c]{\Bigg\{}\Text(355,50)[c]{\Bigg\}}
\end{picture}\end{center}
\caption{The lowest order diagrams, with spinor and colour labels exhibited. The notation is as in Figure~\ref{fig:fourlo}.\label{fig:locombined}}
\end{figure}
The mathematical expression corresponding to the diagrams in Figure~\ref{fig:locombined} is:
\begin{equation}
2\{(\gamma^\mu_L)_{\alpha\beta}\,(\gamma_{\mu\,L})_{\gamma\delta}\,\delta_{ij}\delta_{kl}- (\gamma^\mu_L)_{\gamma\beta}\,(\gamma_{\mu\,L})_{\alpha\delta}\,\delta_{il}\delta_{kj}\},
\end{equation}
where the minus sign between the terms arises from fermion statistics. The Fierz identity (for the parity even component)
\begin{equation}\label{eq:fierz}
(\gamma^\mu_L)_{\alpha\beta}\,(\gamma_{\mu\,L})_{\gamma\delta}= -(\gamma^\mu_L)_{\gamma\beta}\,(\gamma_{\mu\,L})_{\alpha\delta}\end{equation}
allows us to write the lowest order result as
\begin{equation}\label{eq:tree}
2(\gamma^\mu_L)_{\alpha\beta}\,(\gamma_{\mu\,L})_{\gamma\delta}\,\{\delta_{ij}\delta_{kl}+\delta_{il}\delta_{kj}\}\,.\ \end{equation}
Writing the result in this way, the spinor structure is just that of the first of the four diagrams in Figure
\ref{fig:fourlo}, but the colour factor is different. It will be convenient in defining the projectors to take a trace in colour space, i.e. to multiply the expression in Equation (\ref{eq:tree}) by $\delta_{ij}\delta_{kl}$ and sum over the repeated indices. This gives a colour factor at lowest order of $N^2+N$, where $N=3$ is the number of colours.

We presented the above arguments explicitly because they generalize to the one-loop calculations below. Consider for example the 4 diagrams obtained by adding a gluon between the quarks with momenta labeled $p_1$ and $-p_2$ in Figure~\ref{fig:fourlo}. Each of these four diagrams can be Fierz-transformed into each other. It is therefore sufficient to calculate any one of the diagrams, but care needs to be taken in order to evaluate the colour factor correctly.

Fierz identities are four dimensional relations whereas in $\NDR$ one works in $D=4+2\varepsilon$ dimensions. This is the origin of the so called evanescent operators such as
\begin{equation}\label{eq:e1vlldef}
E_1=(\bar{s}^{\,i}\gamma^\mu_Ld^j) (\bar{s}^{\,j}\gamma_{\mu\,L}d^i)-
(\bar{s}{\,i}\gamma^\mu_Ld^i) (\bar{s}^{\,j}\gamma_{\mu\,L}d^j)
\end{equation}
which vanish in 4-dimensions by the Fierz identity, Equation~(\ref{eq:fierz}). Note the relative minus sign compared to
Equation (\ref{eq:fierz}) due to the interchange of fermion fields. It is
conventional to define the $\NDR$ operators having subtracted the
evanescent operators, i.e. using the 4-dimensional Fierz identities
(analogously to subtracting the Euler constant and $\log(4\pi)$ when
defining the $\MSbar$ scheme). This is possible because the evanescent
operators vanish in 4 dimensions and are therefore proportional to
$\varepsilon$ and are only combined with the $1/\varepsilon$
divergence. Their contribution is therefore independent of momenta. The
evanescent operators are therefore removed by one-loop counterterms, and
must be included when evaluating the two-loop anomalous
dimension~\cite{Buras:2000if,Ciuchini:1997bw}. In order to compare our
result for the one-loop counterterms with Reference~\cite{Buras:2000if} we
evaluate their coefficients. We use the same basis of three operators as
in Reference~\cite{Buras:2000if}; in addition to $E_1$ defined in Equation (\ref{eq:e1vlldef}) we introduce
\begin{eqnarray}
E_2&=&(\bar{s}^i\gamma_\mu\gamma_\nu\gamma_\rho P_Ld^i)(\bar{s}^j\gamma^\mu\gamma^\nu\gamma^\rho P_Ld^j)-(16+4\epsilon)(\bar{s}^i\gamma^\mu_Ld^i) (\bar{s}^j\gamma_{\mu\,L}d^j)\label{eq:e2vlldef}\\
E_3&=&(\bar{s}^i\gamma_\mu\gamma_\nu\gamma_\rho P_Ld^j)(\bar{s}^j\gamma^\mu\gamma^\nu\gamma^\rho P_Ld^i)-(16+4\epsilon)(\bar{s}^i\gamma^\mu_Ld^i) (\bar{s}^j\gamma_{\mu\,L}d^j)\,,\label{eq:e3vlldef}
\end{eqnarray}
where $P_L=1-\gamma^5$~\footnote{In Reference~\cite{Buras:2000if} a label $VLL$ is included on $E_{1,2,3}$ to distinguish them from evanescent operators appearing in other processes. Since we are only studying $K$\,-\,$\bar{K}$ mixing here and there is no ambiguity, we omit this label for compactness of notation.}. In comparing our results with Reference~\cite{Buras:2000if} the reader should note that we use $D=4+2\epsilon$ to denote the number of dimensions whereas the authors of Reference~\cite{Buras:2000if} use $D=4-2\epsilon$.

\subsubsection{Evaluating the Diagrams}\label{subsec:diagrams}

There are two independent Feynman diagrams which have to be evaluated (see Figure~\ref{fig:oneloop}) and we now present the results for these diagrams. The results are presented before taking the traces corresponding to the projection operators which define the $\RISMOM$ schemes,
 and so contain flavour and colour indices. The expressions for the remaining diagrams can then be readily obtained from those in 
Figure~\ref{fig:oneloop} by symmetries, except for the contribution of the evanescent operators to the one-loop counterterm which we also discuss later. Leaving the indices free also provides us with the flexibility to use a variety of renormalization schemes (such as the schemes defined in Subsection\,\ref{subsec:smomdef}) which we exploit at the end of this Section.

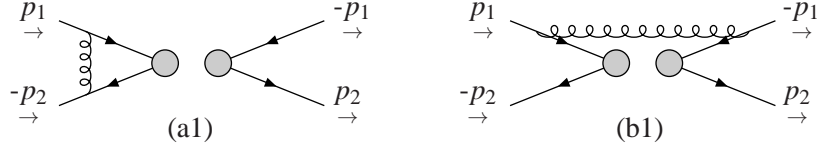
\begin{figure}[t]
\begin{center}
\begin{picture}(300,45)(40,25)
\ArrowLine(50,66)(90,50)\ArrowLine(90,50)(50,34)\ArrowLine(150,66)(110,50)\ArrowLine(110,50)(150,34)
\Gluon(60,62)(60,38){2}{4}
\GCirc(90,50){5}{0.8}\GCirc(110,50){5}{0.8}
\Text(46,66)[r]{{\small $\underset{\rightarrow}{p_1}$}}\Text(46,34)[r]{{\small $\underset{\rightarrow}{\textrm{-}p_2}$}}
\Text(154,66)[l]{{\small $\underset{\rightarrow}{\textrm{-}p_1}$}}\Text(154,34)[l]{{\small $\underset{\rightarrow}{p_2}$}}
\Text(100,25)[c]{(a1)}
\ArrowLine(220,66)(260,50)\ArrowLine(260,50)(220,34)\ArrowLine(320,66)(280,50)\ArrowLine(280,50)(320,34)
\Gluon(230,62)(310,62){-2.2}{11.5}
\GCirc(260,50){5}{0.8}\GCirc(280,50){5}{0.8}
\Text(216,66)[r]{{\small $\underset{\rightarrow}{p_1}$}}\Text(216,34)[r]{{\small $\underset{\rightarrow}{\textrm{-}p_2}$}}
\Text(324,66)[l]{{\small $\underset{\rightarrow}{\textrm{-}p_1}$}}\Text(324,34)[l]{{\small $\underset{\rightarrow}{p_2}$}}
\Text(270,25)[c]{(b1)}
\end{picture}
\caption{The two independent one-loop Feynman diagrams to be evaluated.\label{fig:oneloop}}
\end{center}
\end{figure}

Diagram (a1) gives the following result:
\begin{eqnarray}
\frac{g^2C_F}{16\pi^2}\delta_{ij}\delta_{kl}&\Big\{&-\gamma^\rho_L\otimes\gamma_{\rho\,L}\,
\bigg[\log\frac{p^2}{\mu^2}+\frac23 C_0-1\bigg]+ \frac23\,\frac{\not{\!p}_1\gamma^\rho_R\not{\!p}_1+\not{\!p}_2\gamma^\rho_R\not{\!p}_2}{p^2}\otimes\gamma_{\rho\,L}
\nonumber\\&&-\frac{1+2C_0}{3}\,\frac{\not{\!p}_1\gamma^\rho_R\not{\!p}_2}{p^2}\otimes\gamma_{\rho\,L}
-\frac13\,\frac{\not{\!p}_2\gamma^\rho_R\not{\!p}_1}{p^2}\otimes\gamma_{\rho\,L}\Big\}+\label{eq:diaga}\\
(1-\xi)\,\frac{g^2C_F}{16\pi^2}\delta_{ij}\delta_{kl}&
\Big\{&\gamma^\rho_L\otimes\gamma_{\rho\,L}\,\bigg[\log\frac{p^2}{\mu^2}+\frac{C_0-4}{3}\bigg]
+\frac{C_0-1}{3}\,\frac{\gamma^\rho_L\not{\!p}_1\not{\!p}_2+\not{\!p}_1\not{\!p}_2\gamma^\rho_L}{p^2}\otimes\gamma_{\rho\,L}
\nonumber\\
&&+\frac{C_0}{3}\,\frac{\not{\!p}_1\gamma^\rho_R\not{\!p}_2}{p^2}\otimes\gamma_{\rho\,L}
-\frac{C_0-2}{3}\,\frac{\not{\!p}_1\not{\!p}_2\gamma^\rho_L\not{\!p}_1\not{\!p}_2}{p^4}\otimes\gamma_{\rho\,L}\Big\} \nonumber\\
&\equiv&C_F\delta_{ij}\delta_{kl}\,A_{\alpha\beta,\gamma\delta}\,,\label{eq:Adef}
\end{eqnarray}
where $C_0=\frac{2}{3}\Psi^\prime(\frac13)-(\frac23\pi)^2\simeq 2.34391$ and $\Psi(x)$ is the digamma-function $\Psi(x)=\Gamma^\prime(x)/\Gamma(x)$. In Equation~(\ref{eq:diaga}), $X\otimes Y$ denotes $X_{\alpha\beta}Y_{\gamma\delta}$,
$\gamma^\mu_R=\gamma^\mu(1+\gamma^5)$ and $\xi$ is the gauge parameter defined so that $\xi=0$ corresponds to the Landau gauge and $\xi=1$ to the Feynman gauge. It will prove to be a convenient shorthand to define $A_{\alpha\beta,\gamma\delta}$ as in Equation~(\ref{eq:Adef}).

The expression for diagram (b1) is
\begin{eqnarray}
\frac{g^2}{16\pi^2}\,T^a_{ij}\,T^a_{kl}\,&\Bigg\{&\gamma^\rho_L\gamma^\nu\gamma^\mu\otimes\gamma_{\rho\,L} \gamma_\nu\gamma_\mu\,
\bigg[\frac14\log\frac{p^2}{\mu^2}-\frac{2(1-\log 2)}{3} \bigg]+\nonumber\\
&&(1-\xi)\,\gamma^\rho_L\otimes\gamma_{\rho\,L}\,\bigg[-\log\frac{p^2}{\mu^2}+\frac{4(1-\log 2)}{3}\bigg]+
\label{eq:diagb}\\ &&\hspace{-0.5in}\frac{\gamma^\rho_L\pslash_1\gamma^\mu\otimes\gamma_{\rho\,L}\pslash_1\gamma_\mu}{p^2}\,
\left(\frac{1+8\log2}{6}-(1-\xi)\frac{4\log 2-1}{6}\right)\Bigg\}\,.\nonumber
\end{eqnarray}

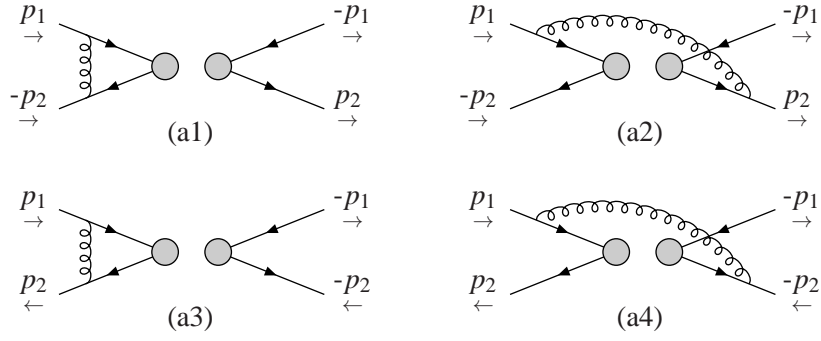
\begin{figure}[t]
\begin{center}
\begin{picture}(300,110)(40,25)
\ArrowLine(50,66)(90,50)\ArrowLine(90,50)(50,34)\ArrowLine(150,66)(110,50)\ArrowLine(110,50)(150,34)
\Gluon(60,62)(60,38){2}{4}
\GCirc(90,50){5}{0.8}\GCirc(110,50){5}{0.8}
\Text(46,66)[r]{{\small $\underset{\rightarrow}{p_1}$}}\Text(46,34)[r]{{\small $\underset{\leftarrow}{p_2}$}}
\Text(154,66)[l]{{\small $\underset{\rightarrow}{\textrm{-}p_1}$}}\Text(154,34)[l]{{\small $\underset{\leftarrow}{\textrm{-}p_2}$}}
\Text(100,25)[c]{(a3)}
\ArrowLine(220,66)(260,50)\ArrowLine(260,50)(220,34)\ArrowLine(320,66)(280,50)\ArrowLine(280,50)(320,34)
\GlueArc(255.63,2.11)(65.146,33.43,113.17){2.0}{15}
\GCirc(260,50){5}{0.8}\GCirc(280,50){5}{0.8}
\Text(216,66)[r]{{\small $\underset{\rightarrow}{p_1}$}}\Text(216,34)[r]{{\small $\underset{\leftarrow}{p_2}$}}
\Text(324,66)[l]{{\small $\underset{\rightarrow}{\textrm{-}p_1}$}}\Text(324,34)[l]{{\small $\underset{\leftarrow}{\textrm{-}p_2}$}}
\Text(270,25)[c]{(a4)}
\ArrowLine(50,136)(90,120)\ArrowLine(90,120)(50,104)\ArrowLine(150,136)(110,120)\ArrowLine(110,120)(150,104)
\Gluon(60,132)(60,108){2}{4}
\GCirc(90,120){5}{0.8}\GCirc(110,120){5}{0.8}
\Text(46,136)[r]{{\small $\underset{\rightarrow}{p_1}$}}\Text(46,104)[r]{{\small $\underset{\rightarrow}{\textrm{-}p_2}$}}
\Text(154,136)[l]{{\small $\underset{\rightarrow}{\textrm{-}p_1}$}}\Text(154,104)[l]{{\small $\underset{\rightarrow}{p_2}$}}
\Text(100,95)[c]{(a1)}
\ArrowLine(220,136)(260,120)\ArrowLine(260,120)(220,104)\ArrowLine(320,136)(280,120)\ArrowLine(280,120)(320,104)
\GlueArc(255.63,72.11)(65.146,33.43,113.17){2.0}{15}
\GCirc(260,120){5}{0.8}\GCirc(280,120){5}{0.8}
\Text(216,136)[r]{{\small $\underset{\rightarrow}{p_1}$}}\Text(216,104)[r]{{\small $\underset{\rightarrow}{\textrm{-}p_2}$}}
\Text(324,136)[l]{{\small $\underset{\rightarrow}{\textrm{-}p_1}$}}\Text(324,104)[l]{{\small $\underset{\rightarrow}{p_2}$}}
\Text(270,95)[c]{(a2)}
\end{picture}
\caption{Four one-loop diagrams whose Feynman integrals are given by that of diagram (a1) in Figure~\ref{fig:oneloop}.\label{fig:1agen}}
\end{center}
\end{figure}

Diagrams (a1) and (b1) in Figure~\ref{fig:oneloop} are not the only ones which need to be evaluated but, apart from the subtlety associated with the evanescent operators (which we neglect for the moment but to which we return shortly), they are the only ones for which the Feynman integrals need to be evaluated. Consider first the four diagrams in Figure~\ref{fig:1agen}, in which one end of the gluon is attached to the quark labeled with momentum $p_1$ and the other to one with momentum $\pm p_2$. The results of the four diagrams in Figure~\ref{fig:1agen} can then be deduced by inspection:
\begin{equation}
\begin{matrix}
(a1)= A_{\alpha\beta,\gamma\delta}C_F\delta_{ij}\delta_{kl};&
\qquad\ \ (a2)= -A_{\gamma\beta,\alpha\delta}T^a_{ij}T^a_{kl};\\
\ \ (a3)= -A_{\gamma\beta,\alpha\delta}C_F\delta_{il}\delta_{kj};&
\qquad(a4)= A_{\alpha\beta,\gamma\delta}T^a_{ij}T^a_{kl}\,.
\end{matrix}
\label{eq:a1to4}\end{equation}
To these must be added the contributions from the four diagrams in which one end of the gluon is attached to the quark with momentum $-p_1$. These are obtained from the results in Equation~(\ref{eq:a1to4}) by making the substitutions $\alpha\leftrightarrow\gamma,\beta\leftrightarrow\delta,i\leftrightarrow k,j\leftrightarrow l$, and the sum of the eight diagrams is to be multiplied by 2 to include the diagrams obtained by interchanging the two currents. In this way we obtain a total answer for the 16 diagrams in which a gluon is attached to quarks of different flavour
\begin{eqnarray}
C_a&=&2(A_{\alpha\beta,\gamma\delta}+A_{\gamma\delta,\alpha\beta})(C_F\delta_{ij}\delta_{kl}+T^a_{il}T^a_{kj})- \label{eq:ca}\\ &&\hspace{-0.55in}2(A_{\gamma\beta,\alpha\delta}+A_{\alpha\delta,\gamma\beta})(C_F\delta_{il}\delta_{kj}+T^a_{ij}T^a_{kl})+
\frac{g^2}{16\pi^2}\,\frac{1}{\epsilon}\,\left[\frac14\left(E_3^{\textrm{tree}}-\frac{1}{N}E_2^{\textrm{tree}}\right) -(4+\xi)E_1^{\textrm{tree}}\right]\,.\nonumber
\end{eqnarray}
The last term contains the contribution from the evanescent operators which we have ignored up to now in this discussion. They arise because in rewriting the divergent terms in terms of the spinor structure $(\gamma^\rho_L)_{\alpha\beta}(\gamma_{\rho\, L})_{\gamma\delta}$ or $(\gamma^\rho_L)_{\alpha\beta}(\gamma_{\rho\, L})_{\gamma\delta}$ we have used the spinor Fierz identities which are not valid in $D=4+2\epsilon$ dimensions. These contributions only arise in the presence of the $\epsilon$ ultraviolet divergence and are hence straightforward to identify. When evaluating the conversion factor between the $\RISMOM$ and $\NDR$ schemes, we will use projection operators which have some symmetry in the indices and which effectively simplify the expression in Equation~(\ref{eq:ca})\,.

\begin{figure}[t]
\begin{center}
\begin{picture}(300,115)(40,-45)
\ArrowLine(50,66)(90,50)\ArrowLine(90,50)(50,34)\ArrowLine(150,66)(110,50)\ArrowLine(110,50)(150,34)
\Gluon(60,62)(140,62){-2.2}{11.5}
\GCirc(90,50){5}{0.8}\GCirc(110,50){5}{0.8}
\Text(46,66)[r]{{\small $\underset{\rightarrow}{p_1}$}}\Text(46,34)[r]{{\small $\underset{\rightarrow}{\textrm{-}p_2}$}}
\Text(154,66)[l]{{\small $\underset{\rightarrow}{\textrm{-}p_1}$}}\Text(154,34)[l]{{\small $\underset{\rightarrow}{p_2}$}}
\Text(100,25)[c]{(b1)}
\ArrowLine(220,66)(260,50)\ArrowLine(260,50)(220,34)\ArrowLine(320,66)(280,50)\ArrowLine(280,50)(320,34)
\Gluon(230,62)(310,62){-2.2}{11.5}
\GCirc(260,50){5}{0.8}\GCirc(280,50){5}{0.8}
\Text(216,66)[r]{{\small $\underset{\rightarrow}{p_1}$}}\Text(216,34)[r]{{\small $\underset{\leftarrow}{p_2}$}}
\Text(324,66)[l]{{\small $\underset{\rightarrow}{\textrm{-}p_1}$}}\Text(324,34)[l]{{\small $\underset{\leftarrow}{\textrm{-}p_2}$}}
\Text(270,25)[c]{(b2)}
\ArrowLine(50,-4)(90,-20)\ArrowLine(90,-20)(50,-36)\ArrowLine(150,-4)(110,-20)\ArrowLine(110,-20)(150,-36)
\Gluon(140,-32)(60,-32){-2.2}{11.5}
\GCirc(90,-20){5}{0.8}\GCirc(110,-20){5}{0.8}
\Text(46,-4)[r]{{\small $\underset{\rightarrow}{p_1}$}}\Text(46,-36)[r]{{\small $\underset{\rightarrow}{\textrm{-}p_2}$}}
\Text(154,-4)[l]{{\small $\underset{\rightarrow}{\textrm{-}p_1}$}}\Text(154,-36)[l]{{\small $\underset{\rightarrow}{p_2}$}}
\Text(100,-45)[t]{(b3)}
\ArrowLine(220,-4)(260,-20)\ArrowLine(260,-20)(220,-36)\ArrowLine(320,-4)(280,-20)\ArrowLine(280,-20)(320,-36)
\Gluon(310,-32)(230,-32){-2.2}{11.5}
\GCirc(260,-20){5}{0.8}\GCirc(280,-20){5}{0.8}
\Text(216,-4)[r]{{\small $\underset{\rightarrow}{p_1}$}}\Text(216,-36)[r]{{\small $\underset{\leftarrow}{p_2}$}}
\Text(324,-4)[l]{{\small $\underset{\rightarrow}{\textrm{-}p_1}$}}\Text(324,-36)[l]{{\small $\underset{\leftarrow}{\textrm{-}p_2}$}}
\Text(270,-45)[t]{(b4)}
\end{picture}
\caption{Four one-loop diagrams whose Feynman integrals are related to that of diagram (b1) in Figure~\ref{fig:oneloop}.\label{fig:b1gen}}
\end{center}
\end{figure}
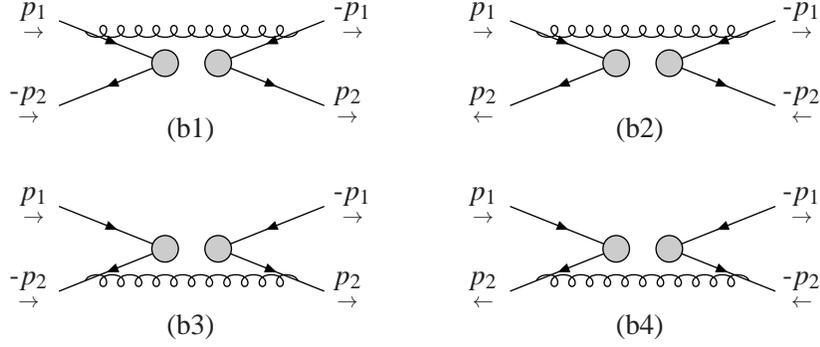

Next we consider the 8 diagrams whose Feynman integral is given by the expression in Equation~(\ref{eq:diagb}). Four of these are shown in 
Figure~\ref{fig:b1gen} and the remaining 4 are obtained by switching the two currents (and are equal to those in Figure~\ref{fig:b1gen}). The result for each of the diagrams (b2)--(b4) can be deduced by inspection from that for (b1) given in Equation~(\ref{eq:diagb}) and for the total contribution from the 8 diagrams we find:
\begin{eqnarray}
C_b&=&\frac{g^2}{16\pi^2}\,\frac{N-1}{N}\,O_{\Delta S=2}^{\textrm{tree}}\left\{1+(3+\xi)\log\frac{p^2}{\mu^2}-\frac{4(1-\log 2)}{3}(7+\xi)\right\}\nonumber\\
&&+\frac{g^2}{16\pi^2}\,\frac{2X^b_{\alpha\beta\gamma\delta,ijkl}}{p^2}\,\left\{\frac{1+8\log 2}{6}-(1-\xi)\frac{4\log2-1}{6}\right\}\label{eq:cb}\\
&&+\frac{g^2}{16\pi^2}\,\left\{\frac{1}{4\epsilon}\left(E_3-\frac1NE_2\right)- \frac{1-\xi}{\epsilon}E_1 \right\}\,,\nonumber
\end{eqnarray}
where
\begin{eqnarray}
X^b_{\alpha\beta\gamma\delta,ijkl}&=&\{(\gamma^\rho_L\pslash_1\gamma^\mu)_{\alpha\beta}\, (\gamma_{\rho\,L}\pslash_1\gamma_\mu)_{\gamma\delta}+(\gamma^\mu\pslash_2\gamma^\rho_L)_{\alpha\beta}\, (\gamma_\mu\pslash_2\gamma_{\rho\,L})_{\gamma\delta}\}T^a_{ij}T^a_{kl}-\nonumber\\
&&\hspace{-0.5in}\{(\gamma^\rho_L\pslash_1\gamma^\mu)_{\gamma\beta}\, (\gamma_{\rho\,L}\pslash_1\gamma_\mu)_{\alpha\delta}+(\gamma^\mu\pslash_2\gamma^\rho_L)_{\gamma\beta}\, (\gamma_\mu\pslash_2\gamma_{\rho\,L})_{\alpha\delta}\}T^a_{kj}T^a_{il}\,.\label{eq:xbdef}
\end{eqnarray}

\subsubsection{The Conversion Factor}\label{subsec:conversion}

Having kept the external colour and spinor indices uncontracted in Subsection\,\ref{subsec:diagrams}, we are in a position to determine the conversion factors relating the $\Delta S=2$ four-quark operator defined in the four RI-SMOM schemes to that in the NRD scheme. The conversion factors, $C_{B_K}^{(X,Y)}$, are defined by
\begin{equation}\label{eq:cbkdef}
{\cal O}_{VV+AA}^{\NDR}(\mu)=C_{B_K}^{(X,Y)}(p^2/\mu^2)\, {\cal O}_{VV+AA}^{(X,Y)}(p),
\end{equation}
where for convenience at this stage we keep $p$ as the
renormalization scale in the $\textrm{RI-SMOM}(X,Y)$ schemes and $\mu$ as the
renormalization scale in the $\NDR$ scheme. Since in this subsection we are only concerned with renormalized quantities we drop the subscript $R$ denoting \textit{renormalized}. From the definition of the RI-SMOM renormalization schemes given in Equations\,(\ref{eq:z11def})\,--\,(\ref{eq:z22def}) we see that the conversion factors can be obtained from the equations
\begin{equation}
{\left(C^{(Y)}_q\right)^2\over C^{(X,Y)}_{B_K}}\,P^{ij,kl}_{(X),\,\alpha\beta,\gamma\delta}\*\Lambda^{\NDR,\;ij,kl}_{\alpha\beta,\gamma\delta}
=1,
\label{eq:BK cfactor def}
\end{equation}
where, as throughout this paper, $\Lambda$ represents the amputated Green function. $C^{(Y)}_q$ are the conversion factors relating the wave-function renormalization factors in the $\overline{\textrm{MS}}$ scheme and that in the RI-SMOM scheme labeled by $Y$,
$C^{(Y)}_q=Z^{\MSbar}_q/Z^{(Y)}_q$. At one-loop order these
were already obtained in Reference~\cite{Sturm:2009kb},
\begin{eqnarray}
\label{eq:Cpsi RISMOM}
C^\text{\RISMOM}_q&=&1+\frac{g^2}{16\pi^2}\,\CF\,\xi\left[\log \frac{p^2}{\mu^2}-1\right]+\mathcal{O}(g^4)\,,\\
\label{eq:wav2}
C^{\RISMOMmu}_{q}&=&1+{g^2\over16\pi^2}\CF\left[
 1 - {\xi\over2}\*\left(3 - 2\*\log{p^2\over\mu^2} - C_0\right)
\right]+\mathcal{O}(g^4)
\end{eqnarray}
where $\CF$ denotes the Casimir operator in the fundamental
representation of SU($\Nc$). These results have recently been extended to two loops
\cite{Almeida:2010ns,Gorbahn:2010bf}.

We now sketch the calculation of the conversion factor for the RI-SMOM($\gamma_\mu,\qslash$) scheme and then present the results for the other three RI-SMOM schemes. The renormalization condition in Equation~(\ref{eq:BK cfactor def}) with
the projector of Equation~(\ref{eq:projector1}) in the
$(\gamma_\mu,{\FslashA{q}})$ scheme can therefore be written in the form
\begin{equation}\label{eq:rismomvertex}
\left.(C^{\RISMOM}_{q})^2
\,P^{ij,kl}_{(1),\,\alpha\beta,\gamma\delta}\,
\Lambda^{\NDR,\;ij,kl}_{\alpha\beta,\gamma\delta}\right|_{\mbox{\scriptsize{non-except.}}}\!\!\!=
C^{(\gamma_\mu,\FslashB{q})}_{B_K}.
\end{equation}
From Equation~(\ref{eq:rismomvertex}), together with the expressions in
Equation~(\ref{eq:ca}), (\ref{eq:cb}) and (\ref{eq:Cpsi RISMOM}) we can
evaluate the conversion factor between the $(\gamma_\mu,\FslashA{q})$ and the $\NDR$
scheme.

There are 3 contributions to the conversion factor:
\begin{enumerate}
\item The total contribution from diagrams such as those in Figure~\ref{fig:1agen} above, in which the gluon is exchanged between a strange quark or antiquark and a down quark or antiquark, is:
\begin{eqnarray}
D_a&=&\frac{g^2}{16\pi^2}\frac{(N-1)(N+2)}{N}\left\{-\xi\log\frac{p^2}{\mu^2}-1+\frac{3-C_0}{2}\xi
\right\}O^{(\gamma_\mu,\qslash)}_{VV+AA}(p)
\nonumber\\ &&+\frac{g^2}{16\pi^2}\frac{1}{2\epsilon}\,\left[-(8+2\xi)\,E_1 -\frac{1}{2N}E_2+\frac12E_3\right]\label{eq:d1def} \end{eqnarray}
where $N=3$ is the number of colours ($(3-C_0)/2\simeq 0.328046$).
\item The corresponding contribution from diagrams, such as those in Figure~\ref{fig:b1gen} above, in which a gluon is exchanged between quarks of the same flavour (i.e. the two strange quarks or the two down quarks), is:
\begin{eqnarray}
D_b&=&\frac{g^2}{16\pi^2}\,\frac{N-1}{N}\left\{(3+\xi)\log\frac{p^2}{\mu^2}+12\log2-7+2\xi(2\log2-1)
\right\}O^{(\gamma_\mu,\qslash)}_{VV+AA}(p)\nonumber\\
&&+\frac{g^2}{16\pi^2}\left(\frac{1}{4\epsilon}(E_3-\frac{1}{N}E_2)-\frac{1}{\epsilon}(1-\xi) E_1\right)\,.\label{eq:d2def}
\end{eqnarray}
\item Finally we have the contribution from the quark wave-function renormalization:
\begin{equation}
D_c=\frac{g^2C_F}{16\pi^2}\,2\xi\bigg\{\log\frac{p^2}{\mu^2}-1\bigg\}O^{(\gamma_\mu,\qslash)}_{VV+AA}(p) \,.\label{eq:d3def}\end{equation}
\end{enumerate}

Before presenting the final result we make two observations:
\begin{enumerate}
\item The total term with evanescent operators is
\begin{equation}\frac{g^2}{16\pi^2}\,\frac{1}{\epsilon}\,\left(\frac12(E_3-\frac{1}{N}E_2)-5E_1\right)\,.\end{equation}
This term is eliminated by introducing counterterms which are equal and opposite to this. The result agrees with (2.15) and (2.22) of Reference~\cite{Buras:2000if} (recall again that we are using $D=4+2\epsilon$ and the authors of \cite{Buras:2000if} are using $D=4-2\epsilon$).
\item
The total logarithmic term is
\begin{equation}
\frac{g^2}{16\pi^2}(3-\frac{3}{N})\,\log\frac{p^2}{\mu^2}\,,
\end{equation}
which agrees with the known anomalous dimension.
\end{enumerate}

The final result for the conversion factor
${C^{(\gamma_\mu,\FslashB{q})}_{B_K}}$ is given by
\begin{eqnarray}
C^{(\gamma_\mu,\FslashB{q})}_{B_K}&=& 1+{g^2\over16\*\pi^2}\left[
          {1\over\Nc}\*\left(9 - 3\*\log{p^2\over\mu^2} - 12\*\log{2}\right)
        - 8
        + 12\*\log{2}
        + 3\*\log{p^2\over\mu^2}
        - \Nc
          \right.\nonumber\\&&\quad\;\,
         +\left.
          \xi\*\left( {1\over\Nc}\*\left(C_0 - 4\*\log{2}\right)
                    - {1\over2}
                    - {C_0\over2}
                    + 4\*\log{2}
                    + {\Nc\over2}\*\left(1 - C_0\right)
              \right)
                               \right]+\mathcal{O}(g^4)\nonumber\\
                       &\stackrel{\Nc=3}{=}&
     1+{g^2\over16\*\pi^2}\left[
          2\*\log{p^2\over\mu^2}
        + 8\*\log{2}
        - 8
        + \xi\*\left(   1
                     - {5\over3}\*C_0
                     + {8\over3}\*\log{2}
              \right)
                               \right]+\mathcal{O}(g^4)\nonumber\\
                       &\simeq&\rule{0mm}{1cm}
     1+\frac{g^2}{16\pi^2}\left[
          2\log\frac{p^2}{\mu^2}
        - 2.45482
        - \xi\,1.05812
                              \right]+\mathcal{O}(g^4)\,.
\label{eq:conversion}
\end{eqnarray}
The remaining three conversion factors are obtained from equations Equations~(\ref{eq:ca}), (\ref{eq:cb}) and (\ref{eq:Cpsi RISMOM}) or (\ref{eq:wav2}) in a similar way and we only present the final results. For the $(\gamma_\mu,\gamma_\mu)$ scheme we find
\begin{eqnarray}
C^{(\gamma_\mu,\gamma_\mu)}_{B_K}&=&
               1+{g^2\over16\*\pi^2}\left[
       {1\over\Nc}\*\left( 8 - 12\*\log{2} - 3\*\lmsdmus \right)
     - 8
     + 12\*\log{2}
     + 3\*\lmsdmus
          \right.\nonumber\\&&\quad\;\,
         +\left.
   \xi\*\left(  {1\over2\*\Nc}\*(1 + C_0 - 8\*\log{2})
              - {1\over2}
              - {C_0\over2}
              +  4\*\log{2}\right)
                 \right]+\mathcal{O}(g^4)\nonumber\\
                     &\stackrel{\Nc=3}{=}&
     1+{g^2\over16\*\pi^2}\left[
       2\*\lmsdmus
     + 8\*\log{2}
     - {16\over3}
     - \xi\*\left( {1\over3} + {1\over3}\*C_0 - {8\over3}\*\log{2} \right)
          \right]+\mathcal{O}(g^4)\nonumber\\
                     &\simeq&\rule{0mm}{1cm}
     1+{g^2\over16\*\pi^2}\left[
          2\*\lmsdmus + 0.211844 + \xi\*\,0.733757
          \right]+\mathcal{O}(g^4)\,.
\end{eqnarray}
For the remaining two schemes we use the
second projector in Equation~(\ref{eq:projector2}) and impose
\begin{equation}
\label{eq:cond2}
\left.(C^{(Y)}_q)^2{1\over64q^2N(N+1)}
\,P^{ij,kl}_{(2),\,\alpha\beta,\gamma\delta}\,
\Lambda^{\NDR,\;ij,kl}_{\alpha\beta,\gamma\delta}
\right|_{\mbox{\scriptsize{non-except.}}}\!\!\!=C^{(\FslashB{q},Y)}_{B_K}
\end{equation}
again with $q=p_1-p_2$ and $p_1^2=p^2_2 = q^2=p^2$. The
conversion factors are
\begin{eqnarray}
C^{(\FslashB{q},\FslashB{q})}_{B_K}&=&
     1+{g^2\over16\*\pi^2}\left[
       {1\over\Nc}\*\left(9 - 3\*\lmsdmus - 12\*\log{2}\right)
     + 12\*\log{2}
     - 9
     + 3\*\lmsdmus
          \right.\nonumber\\&&\quad\;\,
         +\left.
      \xi\*\left( {1\over\Nc}\*\left(  C_0
                                     - 4\*\log{2}\right)
                                     - C_0
                                     + 4\*\log{2}
                              \right)
          \right]+\mathcal{O}(g^4)\nonumber\\
                     &\stackrel{\Nc=3}{=}&
               1+{g^2\over16\*\pi^2}\left[
         2\*\lmsdmus
       + 8\*\log{2}
       - 6
       + \xi\*\left({8\over3}\*\log{2} - {2\over3}\*C_0 \right)
                 \right]+\mathcal{O}(g^4)\nonumber\\
                   &\simeq&\rule{0mm}{1cm}
              1+{g^2\over16\*\pi^2}\left[
        2\*\lmsdmus
      - 0.454823
      + \xi\*\,0.285788
                   \right]+\mathcal{O}(g^4)\,.
\end{eqnarray}
and
\begin{eqnarray}
C^{(\FslashB{q},\gamma_\mu)}_{B_K}&=&
         1+{g^2\over16\*\pi^2}\left[
         {1\over\Nc}\*\left(8 - 12\*\log{2} - 3\*\lmsdmus\right)
       + 12\*\log{2}
       - 9
       + 3\*\lmsdmus
       + \Nc
          \right.\nonumber\\&&\quad\;\,
         +\left.
    \xi\*\left(
           {1\over2\*\Nc}\*( 1 + C_0 - 8\*\log{2})
         - C_0
         + 4\*\log{2}
         + {\Nc\over2}\*\left( C_0 - 1 \right)
         \right)
                  \right]+\mathcal{O}(g^4)\nonumber\\
                     &\stackrel{\Nc=3}{=}&
          1+{g^2\over16\*\pi^2}\left[
        2\*\log{p^2\over\mu^2}
      + 8\*\log{2}
      - {10\over3}
      + \xi\*\left({8\over3}\*\log{2} + {2\over3}\*C_0 - {4\over3} \right)
                  \right]+\mathcal{O}(g^4)\nonumber\\
          &\simeq&\rule{0mm}{1cm}
           1+{g^2\over16\*\pi^2}\left[
          2\*\log{p^2\over\mu^2}
        + 2.211844
        + \xi\*\,2.077664
          \right]+\mathcal{O}(g^4)\,.
\end{eqnarray}
The results for the four conversion factors for the RI-SMOM schemes together with that for RI-MOM are summarized in Table~\ref{tab:projBK}.

\begin{table}
\begin{tabular}{|c|c|}
\hline
Scheme for  & \multirow{2}{*}{$C_{B_K}$ for $\xi=0$}\\[-0.3cm]
four quark operator & \\
\hline
RI-MOM   & {$1+{\als\over4\*\pi}\*(0.87851...)+\mathcal{O}(\als^2)$}\\
\hline
$(\gamma_\mu, \FslashA{q})$ & {$1+{\als\over4\*\pi}\*(-2.45482...)+\mathcal{O}(\als^2)$}\\
$(\gamma_\mu, \gamma_\mu)$ & {$1+{\als\over4\*\pi}\*(0.21184...)+\mathcal{O}(\als^2)$}\\
$(\FslashA{q}, \FslashA{q})$ & {$1+{\als\over4\*\pi}\*(-0.45482...)+\mathcal{O}(\als^2)$}\\
$(\FslashA{q}, \gamma_\mu)$ & {$1+{\als\over4\*\pi}\*(2.21184...)+\mathcal{O}(\als^2)$}\\
\hline
\end{tabular}
 \caption{\label{tab:projBK} Summary of the conversion factors (in the Landau gauge) of the four quark operator from the RI-(S)MOM schemes to the $\overline{\text{MS}}$[\NDR] scheme.}
\end{table}


\subsubsection{Two-Loop Anomalous Dimension}

We follow the conventions of Reference~\cite{Buras:2000if} and define the anomalous dimension $\gamma$ of the renormalized operator $O$ by
\begin{equation}
\mu\frac{dO(\mu)}{d\mu}=-\gamma(\mu)\,O(\mu)\,,
\end{equation}
where $\mu$ is the renormalization scale.
Expanding $\gamma$ as a perturbation series
\begin{equation}
\gamma(\mu)=\frac{g^2(\mu)}{16\pi^2}\gamma^{(0)}+\frac{g^4(\mu)}{(16\pi^2)^2}\gamma^{(1)}
+ \mathcal{O}\left(\frac{g^2(\mu)}{16\pi^2}\right)^3\,,
\end{equation}
the one and two-loop coefficients in the $\MSbar$-$\NDR$
scheme (called $\NDR$ in the following) are~\cite{Buras:1989xd}
\begin{eqnarray}
\gamma^{(0)\,\NDR} &=& 6 - {6\over\Nc} \stackrel{\Nc=3}{=} 4
\qquad\textrm{and}\\
\gamma^{(1)\,\NDR} &=&
- {22\over3} - {57\over2\*\Nc^2} + {39\over\Nc} - {19\over6}\*\Nc
+ \nf\*\left({2\over3} - {2\over3\*\Nc}\right)
\stackrel{\Nc=3}{=}-7+{4\over9}\*\nf\,,
\end{eqnarray}
where $\nf=3$ is the number of flavours contributing to the running in the region of interest.

Now let the conversion factor between the $\NDR$ scheme and a scheme A which is defined in the Landau gauge so that the gauge parameter is not renormalized be given by
\begin{equation}
O^{\NDR}(\mu)=\left(1+\frac{g^2(\mu)}{16\pi^2}\,\Delta r_{\A\to\NDR} + \mathcal{O}\left(\frac{g^2(\mu)}{16\pi^2}\right)^2\right)\,O^{\A}(\mu)\,.
\end{equation}
In the following we consider for the 5 schemes $\A\in\{\RIMOM$,
$(\gamma_\mu,\qslash)$, $(\gamma_\mu,\gamma_\mu)$, $(\qslash,\qslash)$, $(\qslash,\gamma_\mu)\}$.

From Equation~(\ref{eq:conversion}) we see that $\Delta
r_{\RISMOM\to\NDR}\simeq-2.45482$ and from Section 5 of
Reference~\cite{Buras:2000if} we read
\begin{equation}
\Delta r_{\RIMOM\to\NDR}=
-7 + {7\over\Nc} + 12\*\left(1 - {1\over\Nc}\right)\*\log{2}
\stackrel{\Nc=3}{\simeq}0.878511\,.
\end{equation}
For the one-loop anomalous dimensions the equation
$\gamma^{(0)\,\A}=\gamma^{(0)\,\NDR}$ holds and the relations between the
two-loop anomalous dimensions are given by
\begin{equation}
\gamma^{(1)\,\A}=\gamma^{(1)\,\NDR}-2\beta_0\Delta r_{\A\to\NDR}\,,
\end{equation}
where $\beta_0$ is the one-loop coefficient of the QCD $\beta$-function
which is defined by
\begin{equation}
\beta={\partial{\alpha_s(\mu)/(4\*\pi)}\over\partial\log(\mu^2)}=-\beta_0\*\left({\alpha_s(\mu)\over4\*\pi}\right)^2-\beta_1\*\left({\alpha_s(\mu)\over4\*\pi}\right)^3+\mathcal{O}(\alpha_s^4)
\end{equation}
with
\begin{eqnarray}
\beta_0&=&{11\over3}\*\Nc-{2\over3}\*\nf\,,\\
\beta_1&=&{34\over3}\*\Nc^2 + \left({1\over\Nc} - {13\over3}\*\Nc\right)\*\nf\,,
\end{eqnarray}
and $\alpha_s(\mu)=g^2(\mu)/(4\*\pi)$ is the strong coupling constant. In this way we obtain in the Landau gauge
\begin{eqnarray}
\gamma^{(1)\,\NDR}&=&
 - {57\over2\*\Nc^2}
 + {39\over\Nc}
 - {22\over3}
 - {19\over6}\*\Nc
 - \nf\*{2\over3}\*\left[{1\over\Nc} - 1\right]
\underset{\nf=3}{\stackrel{\Nc=3}{=}}-\frac{17}{3}\,,\\
\gamma^{(1)\,\RIMOM}&=&
 - {57\over2\*\Nc^2}
 + {39\over\Nc}
 - {176\over3} + 88\*\log{2}
 + \Nc\*\left({289\over6} - 88\*\log{2}\right)\nonumber\\
&+& \nf\*\left[   {1\over\Nc}\*\left({26\over3} - 16\*\log{2}\right)
                - {26\over3} + 16\*\log{2}
        \right]
\underset{\nf=3}{\stackrel{\Nc=3}{\simeq}}-21.4799\,,\\
\gamma^{(1)\,(\gamma_\mu,\qslash)}&=&
 - {57\over2\*\Nc^2}
 + {39\over\Nc}
 - {220\over3} + 88\*\log{2}
 + \Nc\*\left({111\over2} - 88\*\log{2}\right)
 + {22\over3}\*\Nc^2 \nonumber\\
&+&\nf\*\left[ {1\over\Nc}\*\left({34\over3} - 16\*\log{2}\right)
              - 10 + 16\*\log{2}
              - {4\over3}\*\Nc
       \right]
\underset{\nf=3}{\stackrel{\Nc=3}{\simeq}}38.5201\,,\\
\gamma^{(1)\,(\gamma_\mu,\gamma_\mu)}&=&
 - {57\over2\*\Nc^2}
 + {39\over\Nc}
 - 66 + 88\*\log{2}
 + \Nc\*\left({111\over2} - 88\*\log{2}\right)\nonumber\\
&+&\nf\*\left[{1\over\Nc}\*\left(10 - 16\*\log{2}\right) + 16\*\log{2} - 10 \right]
\underset{\nf=3}{\stackrel{\Nc=3}{\simeq}}-9.47986\,,\\
\gamma^{(1)\,(\qslash,\qslash)}&=&
 - {57\over2\*\Nc^2}
 + {39\over\Nc}
 - {220\over3} + 88\*\log{2}
 + \Nc\*\left({377\over6} - 88\*\log{2}\right) \nonumber\\
&+& \nf\*\left[ {1\over\Nc}\*\left({34\over3} - 16\*\log{2}\right)
              - {34\over3} + 16\*\log{2}
        \right]
\underset{\nf=3}{\stackrel{\Nc=3}{\simeq}}2.52014\,,\\
\gamma^{(1)\,(\qslash,\gamma_\mu)}&=&
 - {57\over2\*\Nc^2}
 + {39\over\Nc}
 - 66 + 88\*\log{2}
 + \Nc\*\left({377\over6} - 88\*\log{2}\right)
 - {22\over3}\*\Nc^2\nonumber\\
&+&\nf\*\left[  {1\over\Nc}\*\left(10 - 16\*\log{2}\right)
              - {34\over3} + 16\*\log{2}
              + {4\over3}\*\Nc\right]
\underset{\nf=3}{\stackrel{\Nc=3}{\simeq}}-45.4799\,.
\end{eqnarray}
In Reference~\cite{Ciuchini:1995cd,Ciuchini:1997bw} a factor has been
introduced to convert the results to the renormalization group
independent (scale invariant) value defined by
\begin{equation}
\label{eq:omega_a}
Z^{\mbox{\scriptsize{RGI}}}_{B_{K}}(\nf)=\omega^{-1}_{A}(\mu,\nf)\*Z^{A}_{B_{K}}(\mu,\nf)\,,
\end{equation}
where $\A$ again labels the scheme. At next-to-leading order the
contribution to the evolution of the operator is written in terms of a
quantity called $J_{\A}^{(\nf)}$
\begin{equation}
\label{eq:omega_aJ}
\omega^{-1}_{A}(\mu,\nf)=\als(\mu)^{-\gamma^{(0)}/(2\*\beta_0)}\*\left[1+{\als(\mu)\over4\*\pi}\*J^{(\nf)}_{A}\right]\,,
\end{equation}
as defined in Appendix D of Reference~\cite{Aoki:2007xm}. In the notation 
used here it is given by
\begin{equation}\label{eq:jdef}
J_{\A}^{(\nf)}=-\left(\frac{\gamma^{(1)}}{2\beta_0}-\frac{\gamma^{(0)}\beta_1}{2\beta_0^2}\right)\,.
\end{equation}
With $\Nc=3$ we find in the Landau gauge
\begin{eqnarray}
\label{eq:jNDR}
J_{\NDR}^{(3)}&=&{13095 - 1626\*\nf + 8\*\nf^2\over
                  6\*(2\*\nf - 33)^2}
\underset{\nf=3}{\simeq}1.89506\,,\\
\label{eq:jRIMOM}
J_{\RIMOM}^{(3)}&=&-{17397 - 2070\*\nf + 104\*\nf^2\over
                     6\*(2\*\nf - 33)^2} + 8\*\log{2}
\underset{\nf=3}{\simeq}2.77357\,,\\
J_{(\gamma_\mu,\qslash)}^{(3)}&=&-{39177 - 4710\*\nf + 184\*\nf^2\over
                      6\*(2\*\nf - 33)^2} + 8\*\log{2}
\underset{\nf=3}{\simeq}-0.55976\,,\\
J_{(\gamma_\mu,\gamma_\mu)}^{(3)}&=&-{7251 - 866\*\nf + 40\*\nf^2\over
                        2\*(2\*\nf - 33)^2} + 8\*\log{2}
\underset{\nf=3}{\simeq}2.10691\,,\\
J_{(\qslash,\qslash)}^{(3)}&=&-{26109 - 3126\*\nf + 136\*\nf^2\over
                         6\*(2\*\nf - 33)^2} + 8\*\log{2}
\underset{\nf=3}{\simeq}1.44024\,,\\
J_{(\qslash,\gamma_\mu)}^{(3)}&=&-{2895 - 338\*\nf + 24\*\nf^2\over
                           2\*(2\*\nf - 33)^2} + 8\*\log{2}
\underset{\nf=3}{\simeq}4.10691\,.
\end{eqnarray}
The first two results in Equations~(\ref{eq:jNDR}) and (\ref{eq:jRIMOM}) can be
taken from Reference~\cite{Ciuchini:1997bw} and agree with (D4) and (D3)
respectively in Reference~\cite{Aoki:2007xm}.

.

\subsection{Volume averaged vertex functions}\label{subsec:volume-average}

In contrast to earlier RBC-UKQCD publications~\cite{Aoki:2007xm}, in the present study we have developed volume-source 
NPR for four quark operators with a generalised momentum configuration. 
As will be demonstrated below, this volume averaging greatly improves the statistical precision. The technique is similar in style
to previous analyses introduced for bilinear operators
by the QCDSF collaboration \cite{Gockeler:1998ye}. The advantage of the method arises from the 
fact that the amputated vertex functions are evaluated  with the operator insertion averaged over
all $L^4$ lattice sites, as opposed to the single-point source operator
insertion. The resulting statistical errors are tiny and systematic effects like
${\cal{O}}_4$ breaking lattice artefacts dominate. These must be included in
the error analysis or removed using, for example, the techniques of \cite{Arthur:2010ht} (which we also do in this study).

We define the four momentum source, used on a Landau gauge-fixed configuration, as
\begin{equation}
\eta_p(x) = e^{i p_\mu x^\mu} \delta_{ij} \delta_{\alpha\beta}\,,
\end{equation}
where $i$, $j$ and $\alpha$, $\beta$
are color and spinor labels respectively and the momenta take the values
\begin{equation}
p_\mu = n_\mu \frac{ 2\pi}{L},
\end{equation}
where $n$ is a four-vector of integers.

On a given gauge field $U_\mu(x)$ we solve the equation
\begin{equation}\label{eq:momsrcdirac}
M(x,y) G_p(y) = \eta_p(x),
\end{equation}
and $M$ is the domain wall fermion matrix with $(5-M_5)\ident$ on the
site diagonal portion.

In performing the NPR, as explained above, we select two momenta $p_1$ and $p_2$ satisfying $p_1^2 = p_2^2 = (p_1-p_2)^2$. 
In order to reduce the artefacts arising from the breaking of ${\cal O}_4$ symmetry, 
we selected values for $p_1^2 = p_2^2 = (p_1-p_2)^2$, such that while still satisfying the Fourier constraints
we best minimise $\sum_i p_i^4$ as documented in Table~\ref{tab:momenta}.
Alternatively, following ref,\,\cite{Arthur:2010ht}, 
we may impose twisted boundary conditions \cite{Boyle:2003ui,Bedaque:2004kc,deDivitiis:2004kq,Sachrajda:2004mi,Flynn:2005in,Boyle:2007wg} on the quark fields
\begin{equation}
q(x+L) = e^{iBx} q(x)\quad\textrm{where}\quad
B_\mu = \frac{\theta \pi}{L_\mu}
\end{equation}
Equation (\ref{eq:momsrcdirac}) is then modified to
\begin{equation}
M(x,y) \tilde{G}_p(y) = \eta_p(x)\quad\textrm{where}\quad
\tilde{G}(y,p) = e^{-iBy} G_{p+B}(y)
\end{equation}
Thus by varying the twist angle $\theta$ we can vary the magnitude of the momentum
without changing the direction. Our choices of $p$ and $B$ are documented in Table
\ref{tab:tw_momenta}. The particular choices here are the non-exceptional directions that minimise $\sum_i p_i^4$.
We choose the components of $B$ equal and always in the same direction as $p$: for example if $p = (0,1,1,0)$ then $B = \frac{\pi}{L}(0,\theta,\theta,0)$\,.

\begin{table}[t]
\begin{tabular}{ccc|ccc}
\hline
$24^3\times 64$ & $p_1$ & $p_2$ &
$32^3\times 64$ & $p_1$ & $p_2$ \\
\hline
& (0,4,4,0) & (4,0,4,0) &
 & (3,2,2,2) & (3,2,-1,-4)  \\
& (1,2,2,8) & (-2,-1,2,8) &
 & (4,2,2,0) & (4,0,-2,4) \\
& (1,4,2,8) & (2,-1,4,8) &
 & (4,4,3,2) & (4,3,-1,-8) \\
& (2,2,4,0) & (4,-2,2,0) &
 & (4,-5,0,-6)& (4,0,-5,-6) \\
& (2,3,2,8) & (3,-2,2,8) &
 & (-4,-1,-4,2)& (-4,-4,1,2) \\
& (-3,1,1,8) & (1,1,3,8) &
& & \\
\hline
\end{tabular}
\caption{
Non-exceptional discrete momenta used for the evaluation of
amputated Green's functions in our NPR analysis. The momenta here
are listed in $(x,y,z,t)$ order for our $24^3\times 64$ and $32^3\times 64$
lattices. The integer Fourier mode numbers $\{n_i\}$ are given and the lattice momenta
are related via $a p_i = \frac{n_i 2\pi}{L_i}$. The exceptional momenta used
correspond to $p_2=p_1$ for the same set of momenta.
\label{tab:momenta}
}
\end{table}

\begin{table}[t]
\begin{tabular}{c|cc|c}
\hline
$24^3\times 64$ & $p_1 $ & $p_2$ & $\theta$  \\
\hline
& (-3,0,3,0) & (0,3,3,0)  &  $\frac{3}{16} n$ : $n = \{-2,1...,12\}$\\
& (-4,0,4,0) & (0,4,4,0)  &  $\frac{3}{2}$ \\
\hline
$32^3\times 64$ & $p_1 $ & $p_2$ & $\theta$  \\
\hline
& (-3,0,3,0) & (0,3,3,0)  &  $\frac{1}{4}$ \\
& (-4,0,4,0) & (0,4,4,0)  &  $-\frac{3}{4}$ , $\frac{3}{8}$\\
& (-5,0,5,0) & (0,5,5,0)  &  $-\frac{5}{8}$ , $\frac{3}{8}$ \\
\end{tabular}
\caption{
Non-exceptional momenta and twist angles used for the evaluation of
amputated twisted Green's functions in our NPR analysis. The momenta here
are listed in $(x,y,z,t)$ order for our $24^3\times 64$ and $32^3\times 64$
lattices. The integer Fourier mode numbers $\{n_i\}$ 
are related to the lattice momenta
via $a p_i = \frac{n_i 2\pi}{L_i}$. The momentum added by the twist, $B$, is determined by the
twist angle $\theta$ giving $a p_i = \frac{ (2 n_i  +  \theta) \pi}{L_i}$. The exceptional momenta used
correspond to $p_2=p_1$ for the same set of momenta.
\label{tab:tw_momenta}
}
\end{table}

We now form phased propagators
\begin{equation}
G^\prime_{p}(x) = G_{p}(x) e^{-ip\cdot x}
                  = \sum_{y} M^{-1}(x,y)e^{ip\cdot (y-x)}\,.
\end{equation}
With twisted boundary conditions this equation is generalized to
\begin{equation}
\tilde{G}_{p}(x) e^{-ip\cdot x} = G_{p+B}(x) e^{-i(p+B)\cdot x}
                  = \sum_{y} M^{-1}(x,y)e^{i(p+B)\cdot (y-x)}
		  = G^\prime_{p+B}(x)\,,
\end{equation}
so that the phases are properly accounted for and the following discussion holds for both
twisted or untwisted propagators. For each configuration we form unamputated
bilinear and four quark vertex functions
for generic Dirac structure $\Gamma$:
\begin{equation}
\left[ \sum_x
\gamma_5 (G^\prime_{p_1}(x))^\dagger \gamma_5
\Gamma
G^\prime_{p_2}(x)\right]_{ij,\alpha\beta} ,
\end{equation}
and
\begin{equation}
\sum_x \left(
\gamma_5 (G^\prime_{p_1}(x))^\dagger \gamma_5
\Gamma
G^\prime_{p_2}(x)
\right)_{ij,\alpha\beta}
 \left(
\gamma_5 (G^\prime_{p_1}(x))^\dagger \gamma_5
\Gamma
G^\prime_{p_2}(x)
\right)_{kl,\gamma\delta}.
\end{equation}
Here, external colour and spin indices are left free
for later amputation.
We use the kinematics explained in Section~\ref{subsec:matching} in which the four-point functions have two legs with incoming momentum $p_1$ and two with outgoing momentum $p_2$.

A single $12\times 12$ object is written out for each configuration
and momentum point for the bilinear vertex functions, and a $12\times 12 \times 12 \times 12$ object for the four quark operator. For convenience, we use a single 12 valued
index below to represent both color and spin. These building blocks enable the accumulation of the following ensemble averages
\begin{eqnarray}
\left(\overline{G}^\prime_p\right)_{ab} &=&
\sum_x \langle \left( G^\prime_{p}(x)\right)_{ab} \rangle,\\
\left(V_\Gamma(p_1,p_2)\right)_{ab} &=&
\langle\sum_x
\left(
\gamma_5 (G^\prime_{p_1})^\dagger(x) \gamma_5
\Gamma
G^\prime_{p_2} (x)
\right)_{ab}
\rangle,\\
W^{stuv}_\Gamma(p_1,p_2) &=&
\langle
\sum_x
\left(
\gamma_5 (G^\prime_{p_1})^\dagger(x) \gamma_5
\Gamma
G^\prime_{p_2} (x)
\right)_{su}
 \left(
\gamma_5 (G^\prime_{p_1})^\dagger(x) \gamma_5
\Gamma
G^\prime_{p_2} (x)
\right)_{tv}
\rangle.
\end{eqnarray}
These ensemble averages are then used to construct the
amputated vertex functions for bilinears
\begin{equation}
\Lambda^{\mathrm{bilinear}}_\Gamma =
\gamma_5(\overline{G}^\prime_{p_1})^{-\dagger}\gamma_5
V_\Gamma(p_1,p_2)
(\overline{G}^\prime_{p_2})^{-1}\,,
\end{equation}
where $\Gamma \in \{ A,V,S,P,T \}$ and for four quark operators
\begin{equation}
\label{eq:lambda}
\Lambda^{\mathrm{4q}}_\Gamma =
\left(\gamma_5(\overline{G}^\prime_{p_2})^{-\dagger}\gamma_5\right)_{as}
\left(\gamma_5(\overline{G}^\prime_{p_2})^{-\dagger}\gamma_5\right)_{bt}
W^{stuv}_\Gamma(p_1,p_2)
(\overline{G}^\prime_{p_1})^{-1}_{uc}
(\overline{G}^\prime_{p_1})^{-1}_{vd}
\end{equation}
where $\Gamma \in \{ VV\pm AA, SS\pm PP, TT\}$\,.

Finally the $\Lambda^{\mathrm{4q}}_\Gamma $ are contracted with
the projectors defined in Equations (\ref{eq:projector1}) and (\ref{eq:projector2}).

\subsection{Lattice Results for the Renormalization of $B_K$}
\label{subsec:zbk}

While the methods summarized in the previous section can be directly
applied to the case at hand, it is important to adopt a strategy which
depends on amplitudes which can be accurately determined.  For example,
it is useful to directly calculate the ratio of renormalization factors
in the scheme $S$, $Z^S_{{\cal O}_{VV+AA}}/Z^2_A$, which is needed for 
the ratio of the four quark matrix element to $f_K^2$ which enters the 
actual definition of $B_K$ because the common factor of $Z_q^2$ 
appearing in the lattice calculation of $Z^S_{{\cal O}_{VV+AA}}$ and 
$Z^2_A$ cancels in this ratio.  (Here $Z_q$ is renormalization factor 
for the domain wall quark field which is central to the RI-MOM 
approach but may introduce large systematic errors if it is identified 
as the coefficient of a momentum-dependent term in the lattice quark 
propagator.)

Thus, we transform our lattice-normalized result for $B_K$ to one
normalized in the scheme $S$ by multiplying by the ratio
\begin{align}
\label{eq:zbk}
  Z_{B_K}^\text{S}= \frac{Z_{{\cal O}_{VV+AA}}^\text{S}}{Z_V^2}
= \left(\frac{\Gamma_V^2}{\Gamma_{{\cal O}_{VV+AA}}}\right)^\text{S}_{m\to0},
\end{align}
where $\Gamma_{{\cal O}_{VV+AA}}$ is the projection of the amputated Green 
function, $\Lambda^{\mathrm{4q}}_\Gamma$, with a projector from 
Equations~(\ref{eq:projector1}) and (\ref{eq:projector2}) corresponding to 
the renormalization scheme S, and $\Gamma_V = \frac{Z_q}{Z_V}$ is the 
appropriate projection of the amputated vertex function of the local 
vector current $\Lambda_V$.  Here either the local vector or axial 
current can be used since their difference is expected to be
of order $m_{\rm res}^2$.

We compute $Z_{B_K}$  in each scheme using Equation (\ref{eq:zbk}). The twisted momenta are
given in Table~\ref{tab:tw_momenta}. For the \ensA ensembles the lattice momenta approximately span
the physical range $4.0\,\text{GeV}^2<p^2<11.0\,\text{GeV}^2$. On the \ensB ensembles the momenta span
$3.25\,\text{GeV}^2<p^2<9.0\,\text{GeV}^2$. The overlap region, $4.0\,\text{GeV}^2<p^2<9.0\,\text{GeV}^2$, will
be used for continuum extrapolations.

We perform a linear extrapolation of the results to the massless limit using data
with quark masses corresponding to the dynamical light-quark masses
$m_l$. We do not observe any statistically relevant mass dependence in
$Z_{B_K}$. Since we are restricted to a single sea strange quark mass in our computation, we cannot perform a
chiral extrapolation for the third active flavour. This mismatch
between the mass-independent renormalization schemes and the finite
sea strange quark mass is included in our error budget.

The lattice data in the chiral limit is converted to the NDR scheme at the renormalization scale
$\mu=2\,\text{GeV}$ or $\mu=3\,\text{GeV}$ using the perturbative results from Section III B.

Several additional inputs are required: we define the
three flavor coupling $\alpha_s$ from the PDG 2010 central values 
$\alpha_s(M_Z) = 0.1184(7)$, 
$m_b^{\rm \overline{MS}}=4.19^{+18}_{-6} $ GeV and
$m_c^{\rm \overline{MS}}=1.27^{+7}_{-9} $ GeV by using the
four-loop running down to our renormalization scale and matching across
flavor thresholds. We combine this four-loop and 2+1 flavour 
$\alpha_s$ with the two-loop anomalous dimensions to obtain the
Wilson coefficients for both scheme change to $\overline{\rm MS}$,
and to obtain the 2+1 flavour RGI operator.
%

The perturbative contribution to the momentum
scale dependence is divided out, and the data for $Z_{B_K}^\text{S}$ is displayed in Figure~\ref{fig:zbkdata_24} and \ref{fig:zbkdata_32}.
The remaining $p^2$ dependence is a source of
systematic error and is discussed in detail in Section \ref{sec:zbksyserr}.

\begin{figure}[ht]
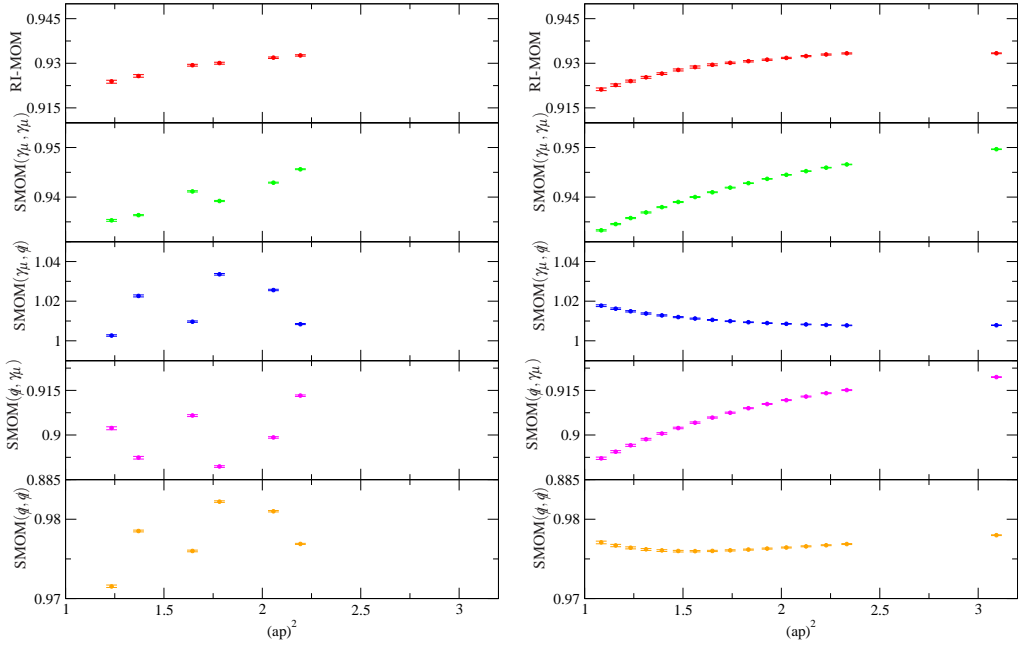

\psfrag{RI-MOM}{\tiny RI-MOM}
\psfrag{SMOMqq}{\tiny SMOM$(\slashed{q},\slashed{q})$}
\psfrag{SMOMqg}{\tiny SMOM$(\slashed{q},\gamma_\mu)$}
\psfrag{SMOMgg}{\tiny SMOM$(\gamma_\mu,\gamma_\mu)$}
\psfrag{SMOMgq}{\tiny SMOM$(\gamma_\mu,\slashed{q})$}
\begin{center}

    \subfigure{\includegraphics*[width=0.4\textwidth]{fig/ZVVpAA_24_untw.eps} }\,\,
    \subfigure{\includegraphics*[width=0.4\textwidth]{fig/ZVVpAA_24_tw.eps}}

    \caption{
We can use the 
the perturbative running to convert the chiral limit of
the ratio (\ref{eq:zbk}) 
to $\overline{MS}$ at $2$ GeV for each $p^2$
using 
 $
Z_{B_K}^\text{S}(p^2) \times \frac{
\omega_{{\rm NDR}}(\mu=2 {\rm GeV},~n_f=3)
}{
\omega_{{\rm S}}(\mu^2=p^2,~n_f=3)
}
$.
This is displayed for all five intermediate MOM
    schemes $S$ on the \ensB ensemble set ($24^3$, $a^{-1}=1.73,\text{GeV}$ lattice). The
    top two panels correspond to the original RI-MOM as the intermediate scheme and the other four rows correspond to the schemes of Section~\ref{subsec:matching}. The left-hand panels show the data with the momenta of Table~\ref{tab:momenta} and the right-hand panels show the data using the momenta in Table~\ref{tab:tw_momenta} accessible with the use of twisted boundary conditions. The scatter due to the $O(4)$ symmetry breaking in the    left hand panels is absent in the right-hand panels. For this reason we use the data with twisted boundary conditions for our analysis.
    \label{fig:zbkdata_24}}
  \end{center}
\end{figure}

\begin{figure}[ht]

\psfrag{RI-MOM}{\tiny RI-MOM}
\psfrag{SMOMqq}{\tiny SMOM$(\slashed{q},\slashed{q})$}
\psfrag{SMOMqg}{\tiny SMOM$(\slashed{q},\gamma_\mu)$}
\psfrag{SMOMgg}{\tiny SMOM$(\gamma_\mu,\gamma_\mu)$}
\psfrag{SMOMgq}{\tiny SMOM$(\gamma_\mu,\slashed{q})$}

\begin{center}

    \subfigure{\includegraphics*[width=0.4\textwidth]{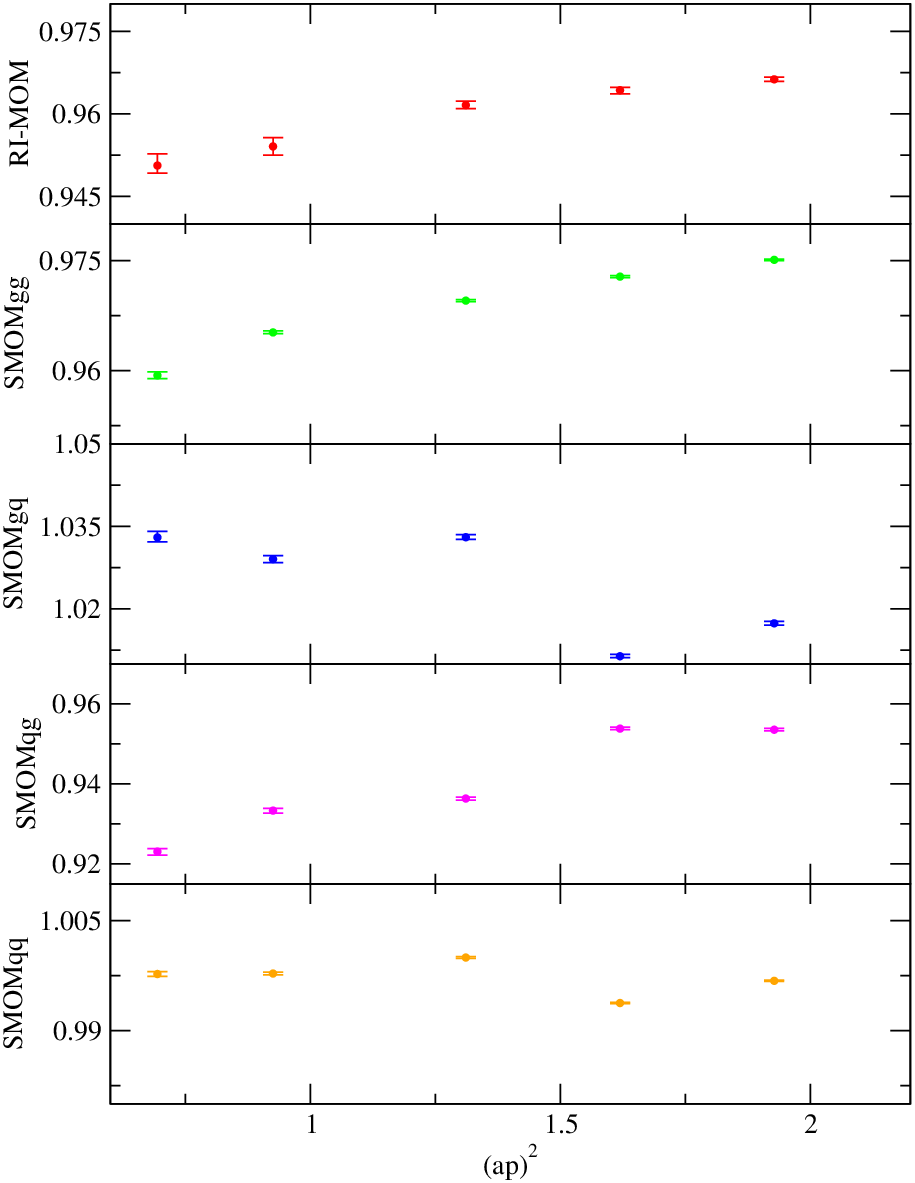}}\,\,
    \subfigure{\includegraphics*[width=0.4\textwidth]{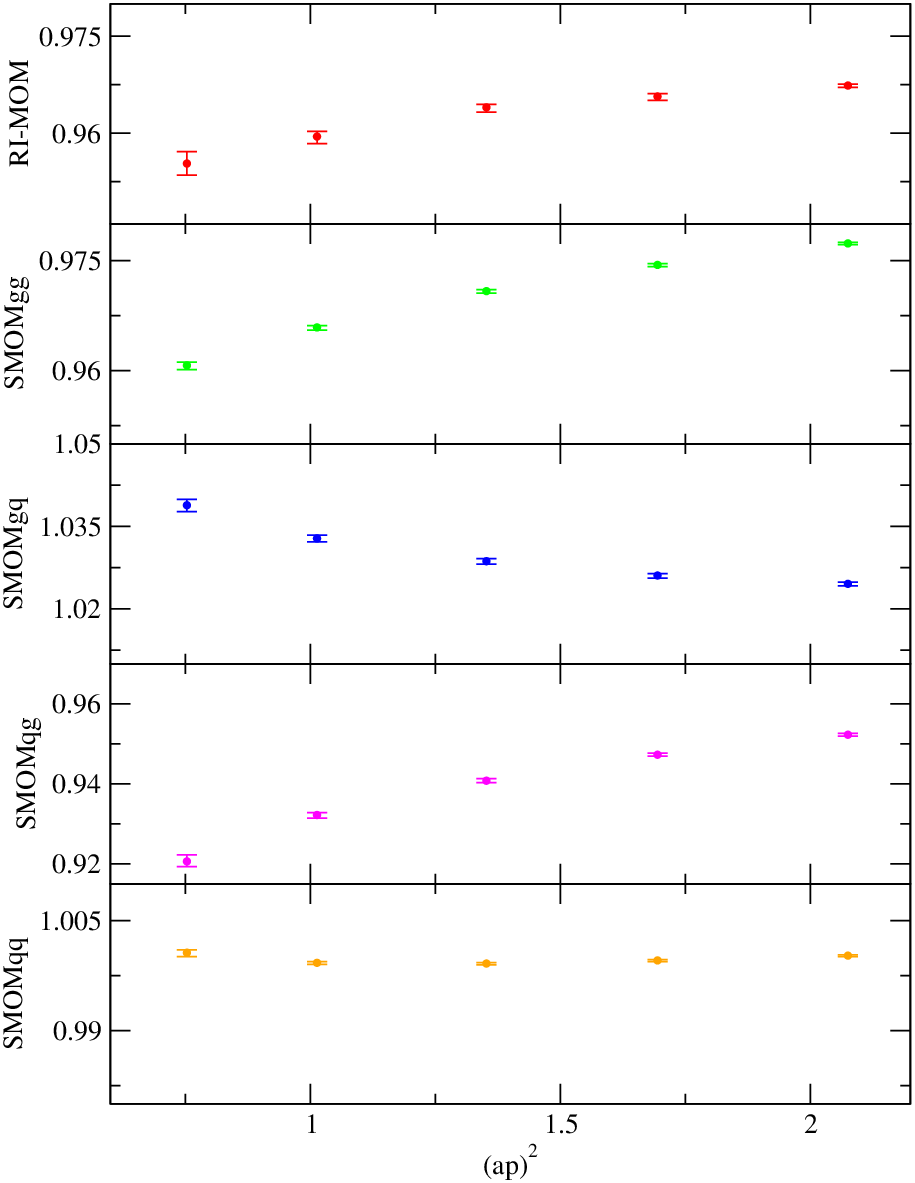} }

    \caption{
We can use the 
the perturbative running to convert the chiral limit of
the ratio (\ref{eq:zbk}) 
to $\overline{MS}$ at $2$ GeV for each $p^2$
using 
 $
Z_{B_K}^\text{S}(p^2) \times \frac{
\omega_{{\rm NDR}}(\mu=2 {\rm GeV},~n_f=3)
}{
\omega_{{\rm S}}(\mu^2=p^2,~n_f=3)
}
$.
This is displayed for all five intermediate MOM
    schemes on the \ensA ensemble set ($32^3$, $a^{-1}=2.28\,\text{GeV}$ lattice). The
    top two panels correspond to the original RI-MOM as the intermediate scheme and the other four rows correspond to the schemes of Section~\ref{subsec:matching}. The left-hand panels show the data with the momenta of Table~\ref{tab:momenta} and the right-hand panels show the data using the momenta in Table~\ref{tab:tw_momenta} accessible with the use of twisted boundary conditions. The scatter due to the breaking of $O(4)$ symmetry  is smaller on this finer lattice.
    \label{fig:zbkdata_32}}
  \end{center}
\end{figure}

\subsubsection{Systematic errors due to renormalization}
\label{sec:zbksyserr}

\begin{table}[ht]
  \begin{center}
 \begin{tabular}{c|c|c|c|c|c}
    scheme & MOM & SMOM $(\gamma_\mu,\gamma_\mu)$ & SMOM $(\gamma_\mu,\slashed{q})$ &
	SMOM $(\slashed{q},\gamma_\mu)$  & SMOM $(\slashed{q},\slashed{q})$\\ \hline
$Z_{B_K}^\text{NDR}(2\,\text{GeV})$ & 0.95541 & 0.96089 & 1.03838 & 0.92164 & 1.00028 \\
\hline
Stat & 0.00151 & 0.00046 & 0.00093 & 0.00104 & 0.00036 \\
$a^{-1}$  & 0.00045 & 0.00052 & 0.00211 & 0.00030 & 0.00129 \\
$m_s$  & 0.00846 & 0.00221 & 0.00386 & 0.00174 & 0.00151 \\
$V-A$ & 0.00551 & 0.00014 & 0.00013 & 0.00010 & 0.00014 \\
\hline
Total & 0.01022 & 0.00232 & 0.00450 & 0.00205 & 0.00202 \\
\end{tabular}
\caption{Error budget, without the perturbative truncation (PT) error, for $Z_{B_K}^\text{NDR}\,$(2 GeV) 
on the \ensA ensemble set ($\beta=2.25$, $32^3$ lattices.)\label{tab_32_2}}
  \end{center}
\end{table}

\begin{table}[ht]
  \begin{center}
 \begin{tabular}{c|c|c|c|c|c}
    scheme & MOM & SMOM $(\gamma_\mu,\gamma_\mu)$ & SMOM $(\gamma_\mu,\slashed{q})$ &
	SMOM $(\slashed{q},\gamma_\mu)$  & SMOM $(\slashed{q},\slashed{q})$\\ \hline
$Z_{B_K}^\text{NDR}(3\,\text{GeV})$ & 0.93453 & 0.94284 & 0.99252 & 0.91681 & 0.96698 \\
\hline
Stat & 0.00030 & 0.00017 & 0.00034 & 0.00038 & 0.00013 \\
$a^{-1}$  & 0.00058 & 0.00049 & 0.00137 & 0.00004 & 0.00086 \\
$m_s$  & 0.00181 & 0.00048 & 0.00039 & 0.00024 & 0.00009 \\
$V-A$ & 0.00188 & 0.00002 & 0.00002 & 0.00002 & 0.00002 \\
\hline
Total & 0.00269 & 0.00070 & 0.00147 & 0.00046 & 0.00088 \\

\end{tabular}
  \caption{Error budget without PT error for $Z_{B_K}^\text{NDR}\,\text(3 GeV)$ at
  $\beta=2.25$
  ($32^3$ lattices).\label{tab_32_3}}
  \end{center}
\end{table}

\begin{table}[hbt]
  \begin{center}
  \begin{tabular}{c|c|c|c|c|c}
    scheme & MOM & SMOM $(\gamma_\mu,\gamma_\mu)$ & SMOM $(\gamma_\mu,\slashed{q})$ &
	SMOM $(\slashed{q},\gamma_\mu)$  & SMOM $(\slashed{q},\slashed{q})$\\ \hline
$Z_{B_K}^\text{NDR}(2\,\text{GeV})$ & 0.92578 & 0.93731 & 1.01350 & 0.89936 & 0.97621 \\
\hline
Stat & 0.00028 & 0.00010 & 0.00032 & 0.00027 & 0.00011 \\
$a^{-1}$  & 0.00049 & 0.00064 & 0.00225 & 0.00013 & 0.00140 \\
$m_s$  & 0.00757 & 0.00393 & 0.00445 & 0.00054 & 0.00180 \\
$V-A$ & 0.00750 & 0.00021 & 0.00026 & 0.00021 & 0.00026 \\
\hline
Total & 0.01067 & 0.00399 & 0.00500 & 0.00065 & 0.00230 \\

  \end{tabular}
  \caption{Error budget without PT error for $Z_{B_K}^\text{NDR}\,\text(2 GeV)$ on the \ensB ensemble set
  ($\beta=2.13$, $24^3$ lattices).\label{tab_24_2}}
  \end{center}
\end{table}

\begin{table}[hbt]
  \begin{center}
  \begin{tabular}{c|c|c|c|c|c}
    scheme & MOM & SMOM $(\gamma_\mu,\gamma_\mu)$ & SMOM $(\gamma_\mu,\slashed{q})$ &
	SMOM $(\slashed{q},\gamma_\mu)$  & SMOM $(\slashed{q},\slashed{q})$\\ \hline
$Z_{B_K}^\text{NDR}(\,\text{GeV})$ & 0.90444 & 0.91983 & 0.97455 & 0.89147 & 0.94672 \\
\hline
Stat & 0.00066 & 0.00010 & 0.00029 & 0.00027 & 0.00011 \\
$a^{-1}$  & 0.00076 & 0.00051 & 0.00131 & 0.00007 & 0.00084 \\
$m_s$  & 0.00347 & 0.00181 & 0.00164 & 0.00148 & 0.00063 \\
$V-A$ & 0.00203 & 0.00003 & 0.00012 & 0.00009 & 0.00012 \\
\hline
Total & 0.00415 & 0.00188 & 0.00213 & 0.00151 & 0.00106 \\

  \end{tabular}
  \caption{Error budget without PT error for $Z_{B_K}^\text{NDR}\,\text(3 GeV)$ on the \ensB ensemble set
  ($\beta=2.13$, $24^3$ lattices).\label{tab_24_3}}
  \end{center}
\end{table}

In Tables~\ref{tab_32_2} , \ref{tab_32_3} and \ref{tab_24_2} , \ref{tab_24_3} we summarize the results and the error
budget for the schemes described in Section \ref{sec:schemes}.
There are six main contributions to the total error 
\begin{enumerate}
\item Statistical errors. These are denoted by the label ``stat" in Tables~\ref{tab_32_2}--\ref{tab_24_3}.
\item Errors due to the breaking of ${\cal O}_4$ symmetry. As explained below we eliminate these errors by evaluating the Green functions using momenta which are made accessible by the implementation of twisted boundary conditions. These are therefore absent in Tables~\ref{tab_32_2}--\ref{tab_24_3}.
\item Uncertainty in the values of the lattice spacing. We denote these by $a^{-1}$ in Tables~\ref{tab_32_2}--\ref{tab_24_3}.
\item Uncertainties due to infrared chiral symmetry breaking effects. These are only significant in the RI-MOM scheme where one manifestation is the difference in the values of $\Lambda_V$ and $\Lambda_A$. We therefore label these effects by $V-A$ in Tables~\ref{tab_32_2}--\ref{tab_24_3}.
\item Errors due to the fixed sea strange-quark mass when defining mass-independent renormalization schemes. We label this by $m_s$ in Tables~\ref{tab_32_2}--\ref{tab_24_3}.
\item Error due to the truncation of the perturbation series in the matching. We label this by PT. Since we estimate this error by comparing the results obtained in different schemes, it is absent in Tables~\ref{tab_32_2}--\ref{tab_24_3} where errors in individual schemes are presented separately.
\end{enumerate}

We define the central value for $Z_{B_K}$ through a linear interpolation
in $(ap)^2$ to the same physical scale $p^2=\mu^2$ on both
ensemble sets, and this is our chosen $\overline{\rm MS}$
renormalization scale $\mu$. We take the continuum limit of the
renormalized matrix element, removing the
lattice artefacts. This approach differs from
earlier work in our collaboration \cite{Aoki:2007xm}
where the values of the renormalization constants extrapolated to $p^2=0$ were used.

We now consider the sources of systematic error in more detail:

\textbf{${\cal O}_4$ breaking:}\\
The use of volume sources leads to tiny statistical errors and as a result
the scatter of the points around a smooth curve in $(ap)^2$
becomes a prominent source of uncertainty. This is illustrated by a comparison of the left and right-hand
plots of Figures~\ref{fig:zbkdata_24} and \ref{fig:zbkdata_32}.
The scatter in the left-hand plots, which correspond to Fourier momenta given in Table~\ref{tab:momenta}, can be attributed to artefacts 
which appear due to the breaking of rotational symmetries on the lattice. In previous studies they have been hidden
due to the statistical noise and the averaging over all degenerate $p^2$.
In a recent paper~\cite{Arthur:2010ht} it has been shown how this scatter can be avoided
using twisted boundary conditions. Instead of using the Fourier modes, we introduce twisted boundary conditions
and use momenta which are equivalent under the hypercubic group
on each lattice spacing. This eliminates the spread due to the breaking of ${\cal O}_4$ invariance.
 This expectation is confirmed in the right-hand plots in Figures~\ref{fig:zbkdata_24} and \ref{fig:zbkdata_32}, where we use the twisting angles specified in Table~\ref{tab:tw_momenta} and we therefore use the twisted data exclusively in this analysis. 
Of course, the $O(a^2)$ errors still remain -- we have simply chosen a single orientation for the lattice momentum. 
The twisting allows us to deal with these discretisation errors by taking the continuum limit of a fixed
observable with a controlled Symanzik expansion.

\textbf{Uncertainty in the lattice spacing:}\\  In order to obtain the renormalization constants
at a given physical scale we use our measured values of the lattice spacings
$a^{-1}_{24} = 1.73(3)$ and $a^{-1}_{32} = 2.28(3)$~\cite{thirtytwocubed}.
The central values quoted above for the renormalization constants are obtained using the central values for $a^{-1}$ and the errors are estimated by recalculating $Z_{B_K}$ using $a^{-1} + \Delta a^{-1}$, where $\Delta a^{-1}$ is the error in the inverse lattice spacing, and taking the difference for the estimated uncertainty.

\textbf{Infrared chiral symmetry breaking effects:}\\ 
In the original RI-MOM scheme the
difference between the bilinear vertex
functions of the vector and the axial vector current is significant
\cite{Aoki:2007xm}. We perform separate analyses using $\Lambda_V$ or
$\frac{1}{2}(\Lambda_V+\Lambda_A)$ in $Z_{B_K}$,
as these differ for the original RI-MOM kinematics due to infrared chiral symmetry effects. We
include the difference as a systematic error and take the ratio with $\Lambda_V$ as the central value. This was estimated to be
one of the largest sources error in our previous RI-mom work, 
but we now find that there is no measurable difference between the two cases for the new SMOM schemes. 

{${\mathbf m_s}$:}\\ 
We associate an error due to our treatment of data with
sea strange quarks near their physical mass while using a mass-independent scheme
when converting to $\overline{MS}$. This can be estimated by measuring the
slope of the data with respect to the simulated light-quark masses in the
chiral extrapolation of vertex functions. We take one half of this slope,
as there is now a single flavour, and multiply by the simulated
strange quark mass to obtain the systematic error. This error is rather small
for the non-exceptional momentum schemes which have a mild mass
dependence. 

{\bf Perturbative truncation:}

For each scheme a perturbative truncation
error arises because we only know the perturbative running to some fixed order.
Estimating this error is necessarily subjective as a rigorous estimate would require
us to know the unknown higher order terms.

At fixed order there are two possible approaches that may be advocated as being reasonable 
estimates of this error. Firstly, notional convergence of the perturbative series could allow
one to estimate the error as either the last term in the series, or perhaps $\alpha_s^n$, where $n$ is the order of the first unknown term, or even $\left(\frac{\alpha_s}{4\pi}\right)^n$ according to subjective taste. These differ greatly, however for
our preferred scheme ${\rm SMOM}(\slashed{q},\slashed{q})$ the last term is around 0.8\%.

Another approach is to compare the results obtained using different schemes to the order at which
we know the results, and consider that any discrepancies between the schemes after the 
well-controlled continuum limit has been taken are 
indicative of the residual perturbative uncertainty. Here again some subjectivity enters through an assessment of which and how many schemes should be considered, however this is a promising approach which we adopt.

In Reference~\cite{Arthur:2010ht} it was found that the ${\rm SMOM}(\slashed{q},\slashed{q})$ scheme was
better described by two-loop perturbative running than the other schemes. Here we also find that the residual $p^2$
dependence for the ${\rm SMOM}(\slashed{q},\slashed{q})$ scheme
is the smallest, and in Section \ref{sec:zbkstep} confirm the analysis of~\cite{Arthur:2010ht}
on our ensembles with a larger volume. This indicates that in the continuum limit,
the ${\rm SMOM}(\slashed{q},\slashed{q})$ scheme is best described by the
perturbative running, and we take the result in this scheme as
our central value. 
We note that of our schemes  $J_{(\qslash,\qslash)}^{(3)}$ was closest to $J_{\rm NDR}^{(3)}$, and this
is therefore consistent with the small size of the perturbative correction needed to change scheme.
For the error, we take the difference between the two schemes that are best
described by perturbation theory in Section \ref{sec:zbkstep}, namely the difference
between the SMOM$(\slashed{q},\slashed{q})$ and SMOM$(\gamma_\mu,\gamma_\mu)$ schemes.

We examined alternate strategies involving a weighted average of the results in all the schemes.
This selects the schemes best described by perturbation theory,
and deweights those poorly described by perturbation theory. Here the relative
weight might be determined by the slope of each scheme after removing perturbative
running. We find that in this case the overall error is slightly smaller than that obtained from the
difference of the results in the SMOM$(\slashed{q},\slashed{q})$ and SMOM$(\gamma_\mu,\gamma_\mu)$ schemes, and
so we adopt the latter as the more conservative error.

We also note from our tables that at the higher scale the difference between schemes is smaller.
For example on our finer lattice, i.e. closer to the continuum limit, we find that the rms error
between the different schemes is reduced from around $0.04$ to $0.03$ as we go from $2$ to $3$ GeV. 
At a sufficiently high scale and in the continuum limit all schemes should give the same result. Since the difference between schemes is
a major systematic error and we believe we have good control over lattice artefacts by taking the
continuum limit, we prefer to compute $Z_{B_K}$ at the higher scale of $3$ GeV. 
The non-perturbative conversion factor to go from $2$ to $3$~GeV in a variety of schemes will 
be presented in a later section.

Finally, as a result of using a formulation of lattice QCD with good chiral properties we have no systematic error associated with operator mixing, as we explicitly demonstrate in the following subsection.

\subsubsection{Operator mixing}

The four-fermion operator $O_{VV+AA}$ renormalizes
multiplicatively when chiral symmetry is preserved. This holds, for example, for lattice
regularizations which preserve chiral symmetry and mass-independent renormalization schemes. In 
Reference~\cite{Aoki:2007xm} it
was shown that the original RI-MOM procedure, with four identical momenta
in the four-point vertex function, does not lead to vanishing mixing
with the remaining elements of the basis of dimension six operators.
Already in Reference~\cite{Aoki:2007xm} it was pointed out that schemes with non-exceptional momentum configurations
$p_1^2=p_2^2=(p_1-p_2)^2$ give mixings consistent with zero.  The
application of momentum sources to this problem dramatically decreases the statistical
error on the mixing coefficients. Therefore we are able to give
more stringent bounds on the residual mixing
which is expected to be of $O(am_\text{res}^2)$ for domain wall fermions. In Figure~\ref{fig:zbkmixing} we present results for the mixing coefficient
$Z_{VV+AA,X}$, where $X=VV-AA,SS-PP,SS+PP$ or $TT$ in the SMOM-$(\gamma_\mu, \gamma_\mu)$ scheme.
\begin{figure}[ht]
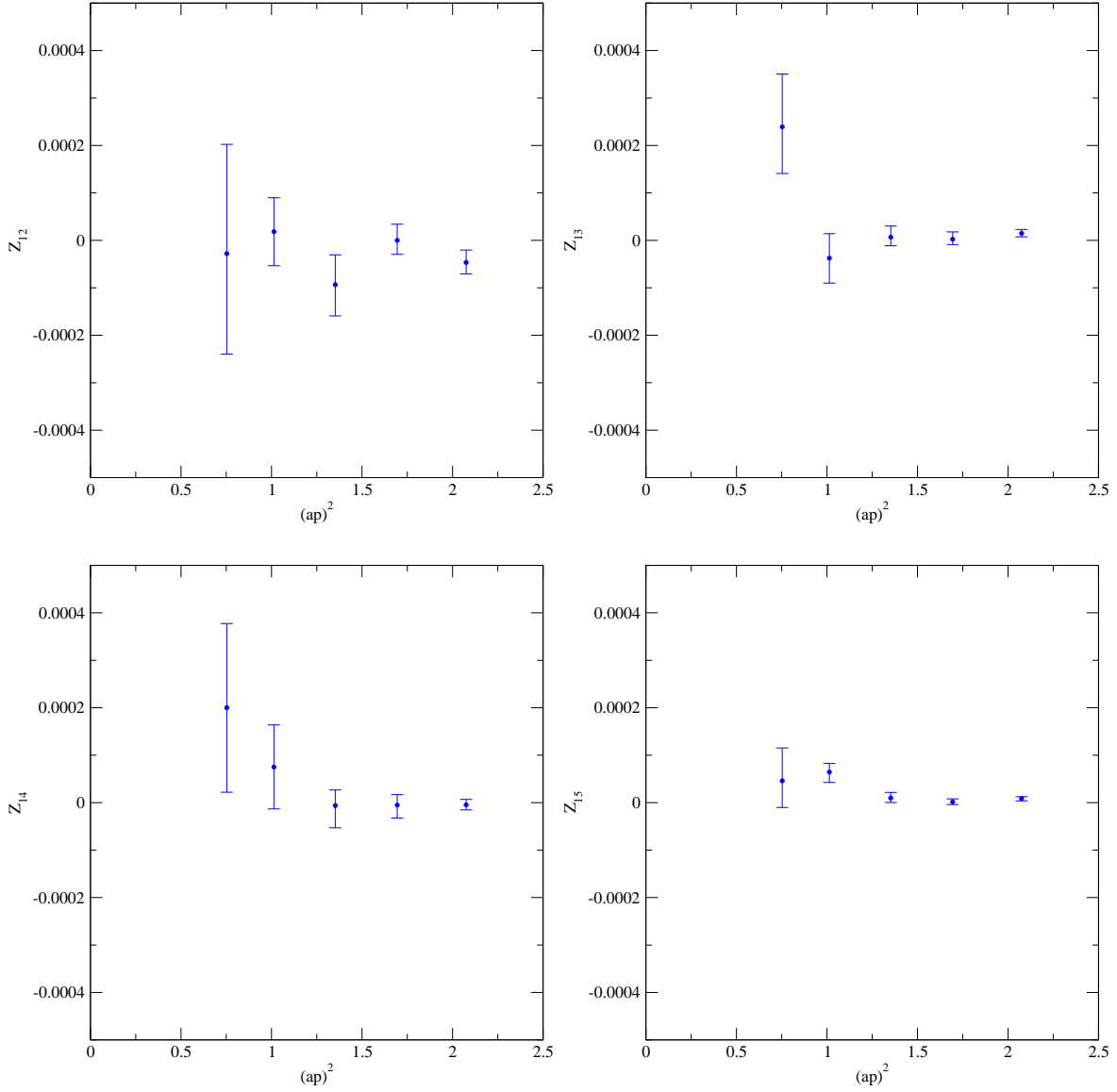

  \begin{center}
    \includegraphics*[width=.47\textwidth]{fig/01lambda_psq_fit.eps}
    \includegraphics*[width=.47\textwidth]{fig/02lambda_psq_fit.eps}\\[5mm]
    \includegraphics*[width=.47\textwidth]{fig/03lambda_psq_fit.eps}
    \includegraphics*[width=.47\textwidth]{fig/04lambda_psq_fit.eps}
   \caption{Mixing coefficient at $\beta=2.25$ for $O_1=O_{VV+AA}$ and
   the operators $O_2=O_{VV-AA}$,\ $O_3=O_{SS-PP}$,\ $O_4=O_{SS+PP}$ and
	$O_5=O_{TT}$. The data shown has been extrapolated to the chiral limit.
   \label{fig:zbkmixing}}
  \end{center}
\end{figure}
The other SMOM schemes also show similarly small mixing coefficients, while the
mixing is artificially enhanced through the pion pole contribution in the RI-MOM
scheme.
Since the mixing coefficients are found to be at least four orders of magnitude
smaller than the multiplicative factor $Z_{11}$, we conclude that the mixing
can be safely neglected even at the high statistical accuracy reached in our
computation. In the following we define the renormalization factor for $B_K$
as the multiplicative Z factor only.

\subsubsection{Step scaling functions}
\label{sec:zbkstep}

Following Reference~\cite{Arthur:2010ht} we can compute the step scaling functions $\sigma_{B_K}$. In this reference a comparison of the continuum non-perturbative step scaling
functions with the perturbative results was proposed as a means to identify the ``best'' scheme
for conversion to $\overline{\textrm{MS}}$. It was observed that the SMOM$(\slashed{q},\slashed{q})$
scheme agreed very well with the perturbative running. We also find here that this
scheme has the smallest residual slope in $p^2$ after removing the perturbative running.

Details of the step scaling scheme can be found in \cite{Arthur:2010ht}, we briefly summarize them here.
Using Equation (\ref{eq:zbk}) in the chiral limit
on each ensemble we have calculated $Z_{B_K}(p,a)$ for $p$ in the range $2.0\,\text{GeV}<p<3.0\,\text{GeV}$.
Because of our twisted boundary conditions we have been able to choose the same momentum direction consistently.
Thus renormalization constants at the same physical scale on both lattices have the same Symanzik expansion
and we can perform the continuum extrapolation of the ratio,
\begin{equation}
\Sigma_{B_K}(p,sp,a) = \frac{Z_{B_K}(sp_0,a)}{Z_{B_K}(p_0,a)}
\end{equation}
where $s$ is a scale factor between $1$ and $1.5$ and $p_0 = 2\,\text{GeV}$ to obtain
\begin{equation}
\lim_{a \to 0}\Sigma_{B_K}(p,sp,a) = \sigma_{B_K}(p,sp) = \frac{Z_{B_K}(sp_0)}{Z_{B_K}(p_0)}\,.
\end{equation}

The present calculation marks an improvement over Reference~\cite{Arthur:2010ht} where the determination of the 
lattice spacing was performed using fits to the static potential and was a large source of statistical and systematic error. Here we use the well determined values of the lattice spacing~\cite{thirtytwocubed} on these ensembles, which significantly reduces the error. Figure~\ref{fig:zbkstepscaling} shows the
step scaling functions for all four SMOM schemes, and we confirm that the SMOM$(\slashed{q},\slashed{q})$
is very well described by perturbation theory. This  motivates us to use it as our central value. In these plots we use the opposite convention to \cite{Arthur:2010ht} and plot $\frac{Z(3s \,\text{GeV})}{Z(3 \,\text{GeV})}$ where s varies between $\frac{2}{3}$ and $1$. The values of $\sigma_{B_K}(2\,\text{GeV}, 3\,\text{GeV})$ and the corresponding error budgets are presented in Table~\ref{tab:bk23step}.

\begin{figure}[htb]
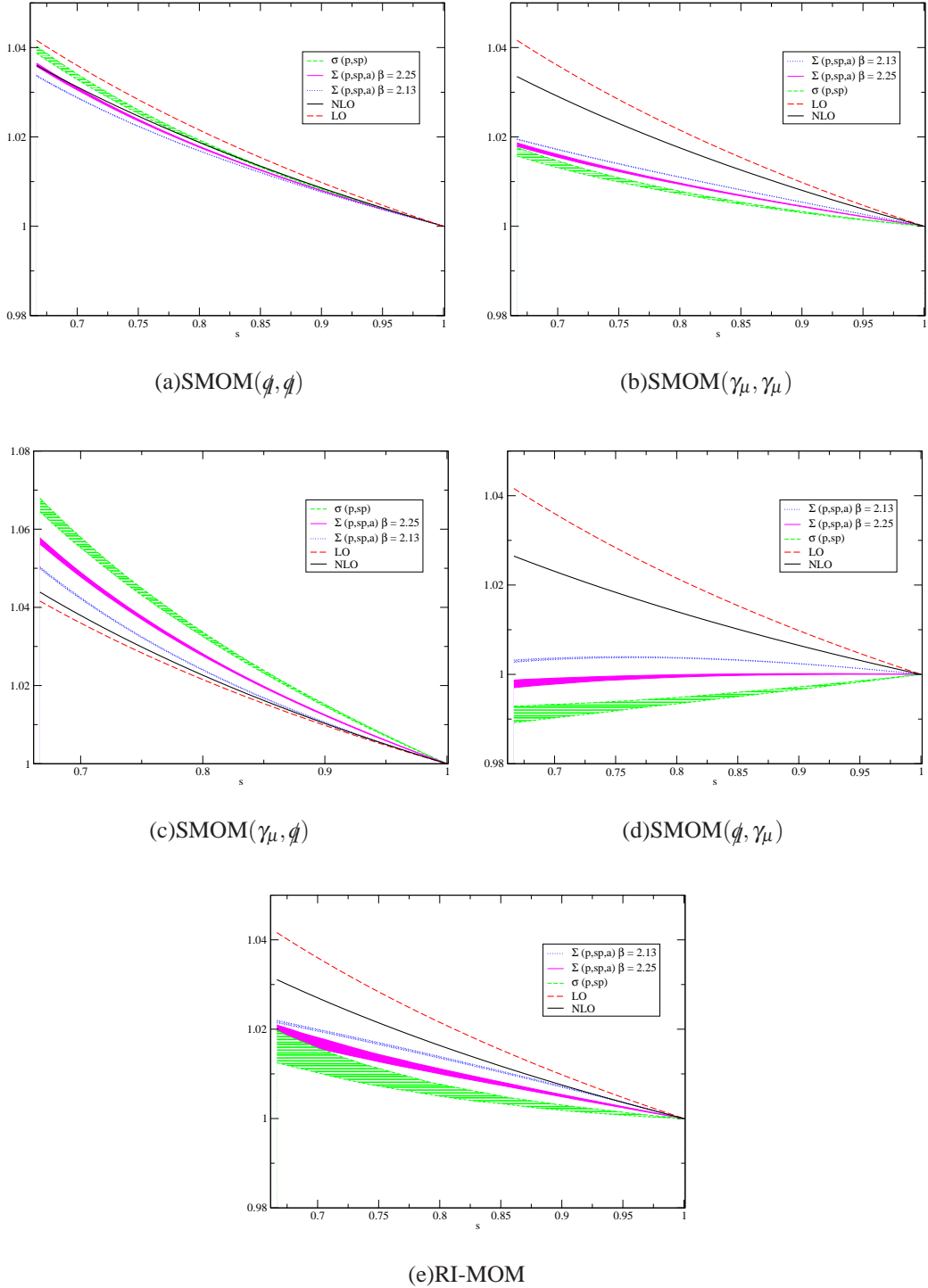

  \begin{center}
    \subfigure[SMOM$(\slashed{q},\slashed{q})$]{\includegraphics*[width=.4\textwidth]{fig/NEVVpAA_RGI.eps} }\,\,\,\,\,\,
    \subfigure[SMOM$(\gamma_\mu,\gamma_\mu)$]{\includegraphics*[width=.4\textwidth]{fig/NEVVpAA_gamma_RGI.eps}}\\[5mm]
    \subfigure[SMOM$(\gamma_\mu,\slashed{q})$]{\includegraphics*[width=.4\textwidth]{fig/NEVVpAA_gamma_q_RGI.eps}}\,\,\,\,\,\,
    \subfigure[SMOM$(\slashed{q},\gamma_\mu)$]{\includegraphics*[width=.4\textwidth]{fig/NEVVpAA_q_gamma_RGI.eps}}\\[5mm]
\subfigure[RI-MOM]{\includegraphics*[width=.4\textwidth]{fig/VVpAA_RGI.eps}}
   \caption{Continuum limit step scaling functions for all four SMOM schemes (blue) compared with one-loop perturbation theory (black). The continuum limit is a simple linear extrapolation in $a^2$. The right, $s = 1$, point corresponds to $3\,\text{GeV}$
   \label{fig:zbkstepscaling}}
  \end{center}
\end{figure}

\begin{table}[hbt]
  \begin{center}
  \begin{tabular}{c|c|c|c|c|c}
    scheme & MOM & SMOM $(\gamma_\mu,\gamma_\mu)$ & SMOM $(\gamma_\mu,\slashed{q})$ &
	SMOM $(\slashed{q},\gamma_\mu)$  & SMOM $(\slashed{q},\slashed{q})$\\ \hline
$\sigma_{B_K}(2\,\text{GeV}, 3\,\text{GeV})$ & 0.98457 & 0.98346 & 0.93783 & 1.00893 & 0.96189 \\
\hline
Stat & 0.00352 & 0.00091 & 0.00154 & 0.00186 & 0.00073 \\
$m_s$  & 0.01041 & 0.00075 & 0.00382 & 0.00056 & 0.00012 \\
$V-A$ & 0.00068 & 0.00066 & 0.00008 & 0.00042 & 0.00007 \\
\hline
Total & 0.01101 & 0.00135 & 0.00412 & 0.00199 & 0.00075 \\
\end{tabular}
  \caption{\label{tab:bk23step}Scaling factor $\sigma_{B_K}(2\,\text{GeV},~3\,\text{GeV})$ from $2$ to $3$\,GeV for each scheme. The values are the reciprocal of the left most point in Figure \ref{fig:zbkstepscaling}. The error from the uncertainty in the lattice spacing is now folded into the statistical error.}
  \end{center}
\end{table}

\fi

\section{Chiral-continuum extrapolation strategy}
\label{sec:ChiralExtrap}
\ifnum\theChiralExtrapStrategy=1
In Reference~\cite{thirtytwocubed} we perform a combined chiral-continuum fit simultaneously to 
our \ensA and \ensB ensemble sets, allowing us to 
extract the lattice spacing and physical quark masses characterising 
each ensemble set. An ensemble set is a group of ensembles with
the same value of $\beta$.   When extrapolated to physical up/down and strange
quark masses, determined via two constraints, 
we determined the lattice spacing of each ensemble set using a third
constraint. Thus, with two ensemble sets, a total of six constraints
are required, and the relation of these constraints between the different
ensemble sets determines our chosen scaling trajectory to the continuum limit:
in principle we are free to choose three quantities or ratios as having
no $a^2$ corrections in \emph{defining} our scaling trajectory.

We summarise the chiral-continuum fit
procedure and the subsequent determination of the lattice scales and
physical quark masses below.
Throughout we denote masses implicitly shifted by $m_{res}$ with a tilde
as in $\wm_l$; 
these are analogous to a PCAC mass, but as we have good chiral symmetry
the adjustment is rather small.

\subsection{Overview of method}

In Reference~\cite{thirtytwocubed} we simultaneously performed a chiral-continuum fit
of the following five quantities: 
$m_\pi$, $m_K$, $m_\Omega$, $f_\pi$ and $f_K$. 
After summarising these global fits to obtain
lattice spacings and quark masses, we will then perform
a separate chiral-continuum fit for $B_K$.
We explore two alternate sets of fit forms: 
\begin{itemize}
 \item The first form is obtained through a joint chiral and $a^2$ 
expansion at next-to-leading order in $SU(2)$ chiral perturbation theory (ChPT)
and in $a^2$. 
Throughout our analyses we use $\Lambda_\chi = 1$~GeV as the chiral scale. 
For heavy-light quantities such as $B_K$, $m_K$ and $f_K$, we use $SU(2)$ PQChPT 
to which the kaon is coupled into the theory at leading 
order in the non-relativistic expansion \cite{Allton:2008pn}.
\item The second form is obtained from a leading-order analytic expansion about 
a non-zero unphysical pion mass as advocated by Lellouch 
\cite{Lellouch:2009fg}, and including $a^2$ corrections.
The fit forms are linear in the quark masses.
By using this approach we lose the ability to take the chiral 
limit and only extrapolate to the non-zero physical point. 
\end{itemize}

\subsection{Ideal trajectory to continuum limit}

We must use six quantities to determine the scale, 
strange mass and the (degenerate) up/down mass
for each of the two lattice spacings. The discussion can be simplified if we first consider 
an ideal case where we were able to simulate at any quark mass.
In this case we would tune the input quark masses on both 
lattices until we obtain $m_\pi/m_\Omega$ and $m_K/m_\Omega$ 
simultaneously equal to their experimentally observed values.

This would define a non-perturbative, hadronic mass dependent 
renormalization condition, and the freedom we hold in defining
the trajectory to the continuum would be absorbed by defining
these quantities to be artefact free.  

\subsection{Matching at unphysical quark mass}

In practice, we are not yet able to simulate with
the physical quark masses and getting to the physical masses
involves some degree of interpolation or extrapolation.
However, the above strategy can 
be modified
{to identify the mass parameters for each ensemble which
lie on the particular scaling trajectory by requiring that a pair
of mass ratios take on convenient unphysical values} rather than
``real world'' observed ratios.  

{For example, we can require that the ratios $m_{ll}/m_{hhh}$ 
and $m_{hl}/m_{hhh}$ take the values given by one pair of input 
quark masses that were used when generating a particular ensemble.
Here the masses $m_{ll}$, $m_{hl}$ and $m_{hhh}$ are the unphysical
analogues of $m_\pi$, $m_K$ and $m_\Omega$ for our unphysical choice
of $m_l$ and $m_h$.}
{Then the pair of matching light and heavy quark masses, 
($m_l$, $m_h$), for a second ensemble set with a different value of 
$\beta$ can be obtained by interpolation in the light quark mass 
$m_l$. We also require {a matching value of $m_h$ on this
second ensemble}. As we 
only used one mass value for the strange sea quark we apply reweighting 
to assign the heavy sea quark mass the value $m_h$.} 
This self-consistent heavy quark mass reweighting and interpolation
to an equal valence mass will be {performed} iteratively.

We formulate our approach to deal with arbitrarily many 
{$\beta$ values} with ensemble set index {\bf e}.
We may then define a lattice spacing 
ratio for each ensemble set {\ensE} to the primary ensemble set
{\ensA} from the ratio of {$hhh$ baryon} masses:
\begin{equation}
\label{eq:CemRa}
R_a^{\rm \ensE} = \frac{(m_{hhh})^{\rm \ensA} }
           {(m_{hhh})^{\rm \ensE} } 
= \frac{a^{\ensA}}{a^{\rm \ensE} },
\end{equation}
where this ratio is naturally $1$ for ${\ensE}={\ensA}$.

For the quark masses that 
yielded matched pseudoscalar and {$hhh$ baryon}
masses we {characterize} the {additional logarithmic 
dependence on $a$ by defining the factors $Z_l^{\rm \ensE}$ and 
$Z_h^{\rm \ensE}$:
\begin{eqnarray}
\label{eq:CemZl}
Z_l^{\rm \ensE} &=& \frac{(\wm_l)^{\rm \ensA} }
           {R_a^{\rm \ensE} (\wm_l)^{\rm \ensE} } \\
\label{eq:CemZh}
Z_h^{\rm \ensE} &=& \frac{(\wm_h)^{\rm \ensA} }
           {R_a^{\rm \ensE} (\wm_h)^{\rm \ensE} }.
\end{eqnarray} }

As we approach the continuum limit, standard renormalized
perturbation theory implies that physically equivalent 
light and heavy quark masses will be related between two 
$\beta$ values by the same renormalization factor.  
However, for non-zero lattice spacing we expect 
$Z_l^{\rm \ensE} \ne Z_h^{\rm \ensE}$.  Further as 
$a^{\rm \ensE} \rightarrow a^{\ensA}$ these
factors each approach unity.  This implies~\cite{thirtytwocubed} 
that:
\begin{equation}
Z_h^{\rm \ensE} = Z_l^{\rm \ensE}
         \left(1+c_m \left[(a^{\rm \ensA})^2 - (a^{\rm \ensE})^2\right]\right). 
\end{equation}

While the coefficient $c_m$ must vanish as $m_l \rightarrow m_h$, 
we have not written it as proportional to $ m_h - m_l $ because the 
low energy matrix elements of the dimension 6 operators which give 
rise to these $O(a^2)$ corrections will contain the more complex 
infra-red quark mass dependence of low energy QCD.
In fact the difference between these two factors is at 
or below the 1\% level and, as can be seen from Table \ref{tab:zlzhzacomparison}, 
they were numerically indistinguishable in our study~\cite{thirtytwocubed}.  Never-the-less
we treat them as two independent quantities in our fits.

When performing an extrapolation in quark mass using both of the 
available ensembles, it is convenient to employ a mass renormalization
scheme which is closely related to the mass parameters used in those
simulations.  Thus, for any simulated {quark} mass on any ensemble set 
{\bf e}, we introduce an equivalent, matched {quark} mass $m^{\ensA}_f$, 
{expressed in lattice units} on our \ensA~ensemble set:
\begin{equation}
\label{eq:CemMmatch}
m^{\ensA}_f \equiv Z_f^{\rm \ensE} R_a^{\rm \ensE} m^{\rm \ensE}_f
\quad \mbox{for } f=l\mbox{ or }h.
\end{equation}
This $m^\ensA_f$ represents a convenient but unconventional 
renormalization scheme where $Z_m$ is defined to be unity for our 
finest lattice spacing.  This non-canonical choice of renormalization 
scheme can of course be transformed to $\overline{\rm MS}$ at a later stage.

The matching prescription ensures that 
the trajectory to the continuum is
defined such that the masses of certain simulated 
pion-like, kaon-like, and $\Omega$-like particles
are lattice artefact free. In principle, these states are \emph{only} 
lattice artefact free at the {specific simulated masses $m_l$ and
$m_h$ used to define the fixed factors $Z_l$ and $Z_h$ in Equation~(\ref{eq:CemMmatch}). 
However
in some neighbourhood $(\delta_{m_l},\delta_{m_h})$ of this simulation 
point the variations in the factors $Z_l$ and $Z_h$ will be sufficiently
small to be neglected.  Since $Z_l$ and $Z_h$ are already themselves
indistinguishable, we can safely neglect the variations in $Z_l$ as $m_l$
varies between zero and any of the (0.005, 0.01) and (0.004, 0.006, 0.008) 
quark mass values in our two ensembles.  Likewise, we will treat $Z_h$ as constant
for $\delta_{m_h}$ within 20\% of $m_h$. Thus, by taking a simulated 
pion-like object to be artefact free for one of these values of $m_l$ we
can view artefacts in all pions to be small, even in the chiral limit.}

\subsection{SU(2) power-counting}

As in
\cite{thirtytwocubed} 
we view the {light quark} mass and $a^2$ expansions as a double power series, and
work only to NLO in this double series. 
We choose the quark masses on each ensemble set such that {the ratios of} some
reference pseudoscalar masses {to the $hhh$ baryon mass} remain fixed. 
Consider the continuum SU(2) expression for the pion mass: 
\begin{equation}
\label{eq:CemNLOmll}
m_{\rm ll}^2 = \chi_l
+ \chi_l \left\{ \frac{16}{f^2} 
\left( (2 L_8^{(2)}- L_5^{(2)}) + 2  ( 2L_6^{(2)} - L_4^{(2)}) \right)
+ \frac{1}{16\pi^2f^2 \chi_l \log\frac{\chi_l}{\Lambda_\chi^2}}
\right\},
\end{equation}
where all quantities are expressed in physical units and 
\begin{equation}
\label{eq:chi_def}
\chi_l = 2 B \wm_l
\end{equation}
depends on the definition of the light quark mass $m_l$.
When we consider this in an expansion at non-zero lattice 
spacing, we represent $B$ and $\wm_l$ in our
matched lattice scheme as 
{
\begin{equation}
\chi_l = \frac{2 B^{\ensA} \wm_l^\ensA}{(a^{\ensA})^2}.
\end{equation} }

As the {LEC $B$ is scheme dependent} we have used our freedom to define a scheme where it simply multiplies
the matched bare quark mass on our \ensA ensemble.
Our matching at non-zero quark mass can be introduced to the fit directly with no further $a^2$counter terms
as the leading order $a^2$ dependence away from our match point 
{has been argued above to be small.}
For $B$ and $\wm$ expressed in this scheme there are also no order $a^2$ counter terms.

In fact, we note that if we were to apply Equation~(\ref{eq:CemNLOmll}) in independent 
fits to dimensionless masses on each ensemble set, and \emph{if} the NLO LEC's turned out to be the same
(something that our combined fit constrains to be the case), then our scaling trajectory would
require $\chi_l$ to be matched in the \emph{same} way as our earlier 
matching strategy,
{that is, $\chi_l^\ensE (a^\ensE/m^\ensE_{hhh})^2$ would be required to be unchanged 
along the trajectory}.

These constraints of identical NLO LEC's on both ensembles and fitting our data at
the (simulated) match point would induce the same relation between bare $B$'s on each ensemble that arises
naturally in our matching approach:
{
\begin{equation}
\chi_l = (a^{\rm \ensA})^{-2} B^{\rm \ensA} \wm_l^{\rm \ensA} = 
(a^{\rm \ensE})^{-2} B^{\rm \ensE} \wm_l^{\rm \ensE} 
\end{equation}
and thus
\begin{equation}
B^{\rm \ensA} = 
B^{\rm \ensE} \frac{R_a^{\rm \ensE}} {Z_l^{\rm \ensE}}.
\end{equation} }

Quantities not used to set quark masses and lattice
scales acquire $a^2$ dependence
at leading order but keep only the 
continuum portions of next-to-leading order mass-expansion
terms. For example, the SU(2), partially quenched, light pseudoscalar 
decay constant for a meson composed of quarks with masses
$m_l$ and $m_x$ is given by
\begin{equation}
\label{eq:CemNLOfll}
f^{\rm \ensE}_{ll} =  f \left\{ 1 + c_{f_\pi} (a^{\rm \ensE})^2
- \frac{2(\chi_x+\chi_l)}{(32\pi^2 f^2)}
\log\left(\frac{\chi_x+\chi_l}{2\Lambda_\chi^2}\right) 
+ \frac{16}{f^2}L_4 \chi_l + \frac{4}{f^2}L_5\chi_x \right\}.
\end{equation}
At fixed {heavy quark} mass, we take the partially quenched light 
quark mass dependence of the kaon mass and decay constant as:
\begin{equation}
\label{eq:CemNLOmxh}
m^2_{xh} = B^{(K)}(\wm_h) \wm_h 
\left\{
1+
\frac{\lambda_1(\wm_h)}{f^2}\chi_l
+
\frac{\lambda_1(\wm_h)}{f^2}\chi_x
\right\}
\end{equation}
and
\begin{equation}
\label{eq:CemNLOfxh}
\begin{array}{ccl}
f_{xh} &=& 
f^{(K)}(\wm_h) 
\left\{
1+C_{f^{(K)}} a^2
\right\}\\
&&+
f^{(K)}(\wm_h) 
\left\{
+\frac{\lambda_3(\wm_h)}{f^2}\chi_l
+\frac{\lambda_4(\wm_h)}{f^2}\chi_x
-\frac{1}{4\pi f^2} \left[ 
\frac{\chi_x+\chi_l}{2}\log\frac{\chi_x+\chi_l}{2\Lambda_\chi^2}
+\frac{\chi_l-2\chi_x}{4}\log\frac{\chi_x}{\Lambda_\chi^2}
\right]
\right\}.
\end{array}
\end{equation}
{These formula have validity once the lattice results have
been reweighted so that both valence and sea heavy quark masses 
take the value $m_h$.}

For the kaon bag parameter we use:
\begin{equation}
\label{eq:CemNLOBKxh}
B_K^{xh} = B^0_K \Big{[}\,1 + c_a a^2 + \frac{c_0\chi_l}{f^2} + \frac{\chi_x c_1}{f^2} - \frac{\chi_l}{32\pi^2 f^2}\log\left(\frac{\chi_x}{\Lambda_\chi ^2}\right)\,\Big{]}\,.
\end{equation}

\subsection{Analytic expansions}

We also consider first order Taylor expansions about
a non-zero quark mass $\wm^m$, in the style of
\cite{Lellouch:2009fg}.
By using this approach we lose the ability to take the chiral limit 
and only extrapolate to the non-zero physical point. In fact
our ansatz for $m_\pi$ has a (small when fitted) constant term that requires
some form of chiral curvature (at smaller masses) to satisfy Goldstone's theorem.
Again, we apply a power counting rule in a double expansion
in $\delta_m$ and $a^2$. 

For the mass of the pion composed of valence quarks with masses $m_x, m_y$ 
and as a function of light sea quark mass $m_l$ and fixed sea strange
mass we write the average valence mass in a meson
as $\wm_v = \frac{\wm_x+\wm_y}{2}$ and use the ansatz
\begin{equation}
\label{eq:CemANmllEX}
m_{ll}^2 = C_0^{m_\pi} + C_1^{m_\pi}(\wm_v-\wm^m) + C_2^{m_\pi}(\wm_l-\wm^m).
\end{equation}
There is no $O(a^2)$ term at the match point and so no correction
to $C_0^{m_\pi}$. Thus within our power counting we could equivalently use
\begin{equation}
\label{eq:CemANmll}
m_{ll}^2 = C_0^{m_\pi} + C_1^{m_\pi} \wm_v + C_2^{m_\pi} \wm_l,
\end{equation}
{where for convenience we redefine $C_0^{m_\pi}$ between 
Equations~(\ref{eq:CemANmllEX}) and (\ref{eq:CemANmll}).}
For decay constants, which do not vanish in the chiral limit, the 
$O(a^2)$ term is not sensitive to the choice of expansion point:
\begin{eqnarray}
\label{eq:CemANfEX}
f_{ll} &=& C_0^{f_\pi} [1+C_f a^2] + C_1^{f_\pi}(\wm_v - \wm^m) + C_2^{f_\pi}(\wm_l-\wm^m)\\
\label{eq:CemANfll}
&\equiv& C_0^{f_\pi} [1+C_f a^2] + C_1^{f_\pi}\wm_v + C_2^{f_\pi} \wm_l,
\end{eqnarray}
{where again $C_0^{f_\pi}$ has been redefined between 
Equations~(\ref{eq:CemANfEX}) and (\ref{eq:CemANfll}).}
At fixed valence and sea strange mass $m_y=m_h=m_s$, 
we take the dependence  on the light valence quark mass $m_x$ and light
sea quark mass $m_l$ of the kaon mass, kaon decay constant, 
and kaon bag parameter as
\begin{eqnarray}
\label{eq:CemANmxh}
m_{xh}^2 &=& C_0^{m_K} + C_1^{m_K}(\wm_x-\wm^m) + C_2^{m_K}(\wm_l-\wm^m)\\
 &\equiv&  C_0^{m_K} + C_1^{m_K} \wm_x + C_2^{m_K} \wm_l,
\end{eqnarray}
\begin{eqnarray}
\label{eq:CemANfxh}
f_{xh} &=& C_0^{f_K} [1+C_{f_K} a^2] + C_1^{f_K}(\wm_x-\wm^m) + C_2^{f_K}(\wm_l-\wm^m)\nonumber\\
  &\equiv&  C_0^{f_K} [1+C_{f_K} a^2] + C_1^{f_K} \wm_x + C_2^{f_K} \wm_l\,,
\end{eqnarray}
\begin{eqnarray}
\label{eq:CemANBKxh}
B_K^{xh} &=& c_0(1 + c_a a^2) + c_l (\wm_l-\wm^m) + c_v (\wm_x-\wm^m) \nonumber\\
    &\equiv& c_0(1 + c_a a^2) + c_l \wm_l + c_v \wm_x\,,
\end{eqnarray}
where again the parameters $C_0^{m_K}$, $C_0^{f_K}$ and $c_0$ have
been redefined between each pair of equations, and
implicitly depend on the strange quark mass.

\fi

\section{Chiral-continuum extrapolation results}
\label{sec:ChiralExtrapRes}
\ifnum\theChiralExtrap=1
In this section we present the joint chiral-continuum extrapolation of our data.

\subsection{Fitting procedure}

In References~\cite{Allton:2008pn,thirtytwocubed} we performed correlated fits where the 
correlation matrix is obtained by taking increasing numbers of the 
leading eigenvectors. We find no significant difference over uncorrelated 
fit results within our limited ability to estimate the correlation matrix. 
Hence for this analysis and those in References~\cite{Allton:2008pn,thirtytwocubed} we use 
uncorrelated fits.

In order to perform our fits, which include forms valid only for
fixed strange mass, we are faced with the problem that the 
physical strange mass is an output of our calculation. Thus
the combined chiral-continuum fit procedure is necessarily iterative.
The details of the procedure are documented in Reference \cite{thirtytwocubed}, 
and it suffices to note here that the iterative
process terminates when the fixed strange mass forms produce a
prediction for $m_s$ that is consistent with the guess $m_s$ to
which our data was interpolated. When doing this
we use reweighting to adjust all pionic observables 
to the current strange mass guess for each ensemble. For kaon and $\Omega$ 
observables a linear interpolation between the (unreweighted)
unitary measurement, and a second valence strange
(reweighted-to-be-unitary) measurement suffices to obtain that
observable for $\wm_y = \wm_h = \wm_s^{\rm guess}$.

\subsection{Scaling analysis}

As discussed in Section \ref{sec:ChiralExtrap}, we match our lattice data using ratios of hadronic masses $\frac{m_\pi}{m_\Omega}$ 
and $\frac{m_K}{m_\Omega}$. We choose a specific simulated value of $(\wm_l,\wm_h)^{\bf M}$ on the ensemble set ${\bf M}$ to which the other ensemble sets are matched. We refer to this as the match point. The choice of the match point defines a particular trajectory along which we approach the continuum limit. Although the physical predictions do not depend upon the particular trajectory, certain match points are favourable due to the quality of the data at the match point and the range over which the data must be interpolated/extrapolated on the other ensemble sets to perform this matching. The ideal point has as small a statistical error as possible and lies within the range of simulated data on all of the matched ensemble sets such that only a small interpolation is required. In practice, the errors on the mass ratios at the match point can be reduced by simultaneously fitting to all partially quenched simulated data on the ensemble set ${\bf M}$ and interpolating to the match point which lies on the unitary curve.
Further details of the procedure are documented in \cite{thirtytwocubed}.

As previously mentioned, the primary ensemble set is chosen to be that with the finest lattice spacing; our 
$32^3\times 64$, $a^{-1}=2.28$ GeV lattice (ensemble \ensA). 
As we have only one other ensemble set, we henceforth drop the superscript on the lattice spacing and quark mass ratios. 

In Table \ref{tab:zlzhzacomparison} we give the values\cite{thirtytwocubed}
for $Z_l$, $Z_h$ and $R_a$ obtained by using several match points on 
both ensemble sets ${\bf M} \in \{\ensA, \ensB\}$. Subject to the condition that we require a match point within the range of simulated data, we can discard the first and last entries. From the remaining, we choose the values $Z_l=0.983(9)$, $Z_h=0.975(7)$ and $R_a=0.759(5)$ from the second entry with ${\bf M}=\ensA$ and $(\wm_l,\wm_h)^{{\bf M}}=(0.006,0.03)$ as our final values. The consistency is excellent, and these are taken as input to our chiral-continuum extrapolation for $B_K$.

\begin{table}[thp]
\centering
\begin{tabular}{ccc|ccccccc}
\hline\hline
$\bf M$ & $(m_l)^{\bf M}$& $(m_h)^{\bf M}$& $(m_l)^{\bf e}$& $(m_h)^{\bf e}$& $Z_l$    & $Z_h$   &$R_a$     \\
\hline
$\bf A$ & 0.004           & 0.03            & 0.00312(13)     & 0.03804(79)     & 0.980(15)& 0.977(11)&0.7623(71)\\
$\bf A$ & 0.006           & 0.03            & 0.00581(12)     & 0.03829(51)     & 0.983(9) & 0.975(7) &0.7591(46)\\
$\bf A$ & 0.008           & 0.03            & 0.00856(19)     & 0.03856(63)     & 0.981(10)& 0.973(8) &0.7556(58)\\
$\bf B$ & 0.005           & 0.04            & 0.00541(10)     & 0.03136(48)     & 0.980(12)& 0.976(8) &0.7604(55)\\
$\bf B$ & 0.01            & 0.04            & 0.00899(18)     & 0.03078(56)     & 0.977(11)& 0.969(9) &0.7520(69)\\
\hline
\end{tabular}
\caption{Values of the quark mass ratios $Z_l$ and $Z_h$ and the lattice spacing ratio $R_a$ determined by matching at five points over both ensemble sets. Quark masses are quoted without the additive $m_\mathrm{res}$ correction.
\label{tab:zlzhzacomparison}}
\end{table}

\subsection{Combined analysis procedure for $B_K$}
\label{subsec:ChiralExtrapRes:Results}

In Reference~\cite{thirtytwocubed}
we obtained the 
the lattice spacings and physical light and strange quark masses
given in Table~\ref{tab:latticeparameters} 
from our two combined analysis procedures.
These are taken as input to our fits to $B_K$ in the present calculation.
This table also 
contains the values of the leading-order $SU(2)$ ChPT LECs $B$ and $f$ 
obtained\cite{thirtytwocubed} from fitting $m_\pi$ and $f_\pi$, and which 
are used as input to our $B_K$ analysis in order to reduce the number of degrees of freedom in the NLO PQChPT fit form. 

In principle, the matrix element fit could be included in our main combined fit analysis, allowing these data to constrain  
the ratio $B/f^2$. In practice however, this constraint is very weak as compared to those from $m_\pi$ and $f_\pi$, 
so the $B_K$ analysis can be decoupled from the main analysis.
On the second line of Table~\ref{tab:latticeparameters} we have given the lattice parameters 
obtained by an NLO PQChPT fit with finite volume effects included by correcting the chiral logarithms using the corresponding finite volume
sum of Bessel functions \cite{Sharpe:1992ft}. These are propagated through to our analysis of the finite volume corrections to $B_K$.

\begin{table}
\centering
\begin{tabular}{c|cc|cc|cc|c|c}
\hline\hline
Fit & $(a^{-1})^\ensA$ & $(a^{-1})^\ensB$ & $(m_l^\mathrm{phys})^\ensA$ & $(m_l^\mathrm{phys})^\ensB$ & $(m_h^\mathrm{phys})^\ensA$ & $(m_h^\mathrm{phys})^\ensB$ & $B$(GeV) & $f$(GeV)\\
\hline
NLO PQChPT & 2.28(3) & 1.73(2) & 0.00099(3) & 0.00133(4) & 0.0278(7) & 0.0376(11) & 4.13(8) & 0.107(2)\\
NLO PQChPT+FV & 2.28(3) & 1.73(2) & 0.00101(3) & 0.00136(4) & 0.0278(7) & 0.0375(11) & 4.04(7) & 0.110(2)\\
LO Analytic & 2.29(3) & 1.74(2) & 0.00105(6) & 0.00140(9) & 0.0277(7) & 0.0374(11) & - & -\\
\end{tabular}
\caption{Parameters of the \ensA and \ensB ensemble sets determined from a combined fit using the fit form given in the first column. We also include the LO ChPT LECs $B$ and $f$ that are used to constrain the fits to $B_K$. 
\label{tab:latticeparameters}} 
\end{table}

Our data are reweighted/interpolated to the physical strange quark 
mass prior to the fit, as discussed above. The data are given in Tables~\ref{tab:32cphysmhmatrixelements} and 
\ref{tab:24cphysmhmatrixelements}. We fit this data with both ChPT and analytic forms,
Equations (\ref{eq:CemNLOBKxh}) and (\ref{eq:CemANBKxh}), 
fitting the NLO PQChPT form of Equation~(\ref{eq:CemNLOBKxh}) both with and without finite volume corrections 
in order to estimate the finite volume systematic error. 

Note that these equations are applied with strange quark mass 
fixed to its physical value having linearly interpolated and reweighted
the data to the physical strange quark mass.

We renormalize the $B_K$ data using the renormalization constants determined in
Section~\ref{subsec:zbk} prior to performing our fit. Thus the fit is performed
seperately for each of the schemes SMOM$(\slashed{q},\slashed{q})$ and SMOM$(\gamma_\mu,\gamma_\mu)$,
and for both $2$ GeV and $3$ GeV matching scales. The central value is taken from the 
SMOM$(\slashed{q},\slashed{q})$ scheme, and the SMOM$(\gamma_\mu,\gamma_\mu)$ contributes to determining
the renormalisation error.

\begin{table}[tb]
\centering
\begin{tabular}{c|cc}
\hline\hline
Fit & $B_K^{\overline{\mathrm{MS}}}(2\,\mathrm{GeV})$ & $B_K^{\overline{\mathrm{MS}}}(3\,\mathrm{GeV})$\\
\hline
NLO PQChPT & 0.544(5) & 0.523(5)\\ 
NLO PQChPT+FV & 0.542(5) & 0.521(5)\\
LO Analytic & 0.557(5) & 0.536(5)
\end{tabular}
\caption{$B_K^{\overline{\mathrm{MS}}}(2\,\mathrm{GeV})$ as obtained by a combined fit to the data at the physical strange quark mass using an NLO PQChPT fit form and a LO analytic fit form. The second line contains the NLO PQChPT fit with finite volume corrections included, from which we estimate the finite volume systematic by comparing to the fit without corrections. Errors are statistical only and do not include the error on the renormalisation coefficient.
\label{tab:bkfitresults}}
\end{table}

\begin{table}
\begin{tabular}{c|c|c|c|c}
\hline
\hline
Parameter & \multicolumn{2}{c|}{NLO PQChPT} & \multicolumn{2}{c}{NLO PQChPT+FV} \\
\hline
          & 2 GeV               & 3 GeV    & 2 GeV            & 3 GeV\\            
\hline
$B^0_K$ & 0.533(5)   & 0.513(5)      & 0.531(5)   & 0.511(5)  \\
$c_a$   & 0.06(4)    & 0.08(4)       & 0.05(4)    & 0.08(4)  \\
$c_0$   & -0.0060(8) & -0.0060(8)    & -0.0062(8) & -0.0062(8)  \\
$c_1$   & 0.0061(3)  & 0.0062(3)     & 0.0071(4)  & 0.0071(4)  \\
\end{tabular}
\caption{Fit parameters of the NLO PQChPT fits to the $B_K$ matrix element, with and without finite volume corrections.
\label{tab:bknlochptparams}} 
\end{table}

\begin{table}[tp]
\begin{tabular}{c|c|c}
\hline
\hline
Parameter & \multicolumn{2}{c}{Result}\\
\hline
          & 2 GeV & 3 GeV\\
\hline
$c_0$ & 0.554(5) & 0.534(5)\\
$c_a$ & 0.06(4) & 0.08(3)\\
$c_l$ & 0.2(3) & 0.2(3)\\
$c_v$ & 0.9(1) & 0.9(1)\\
\end{tabular}
\caption{Fit parameters of the leading order analytic fit to the $B_K$ matrix element.
\label{tab:bkloanalyticparams} }
\end{table}

Performing the fits, we obtain the results given in Table~\ref{tab:bkfitresults}, where the quoted errors are statistical only. 
Here we have also included an NLO PQChPT fit with finite volume corrections, 
which is used below to estimate the finite volume systematic. The fit parameters are 
given in Tables~\ref{tab:bknlochptparams} and ~\ref{tab:bkloanalyticparams}. 

Figure \ref{fig:bk_pq_su2} and \ref{fig:bk_pq_analytic} display the partially quenched light quark valence and
sea mass dependence of both our SU(2) and analytic fit forms to kaon matrix element data with one valence quark 
mass set to the physical strange mass, and the sea heavy quark mass reweighted to the physical strange mass.
Our previous work \cite{Allton:2008pn} contained small indications in the corresponding plot for curvature
consistent with NLO ChPT. These have become less pronounced in our doubled data set and also not supported
by the higher precision data from the second lattice spacing.

Figure \ref{fig:bkfitcomparison} shows the continuum limit chiral extrapolation, 
overlaid by the data corrected to the continuum limit using the 
fit parameters describing $a^2$ dependence. 
Figure \ref{fig:bkfitcomparisonuncorrected} shows the same fits overlaid with the uncorrected data. 
By comparing these plots, the weak lattice spacing dependence of the data is apparent. 

\begin{figure}
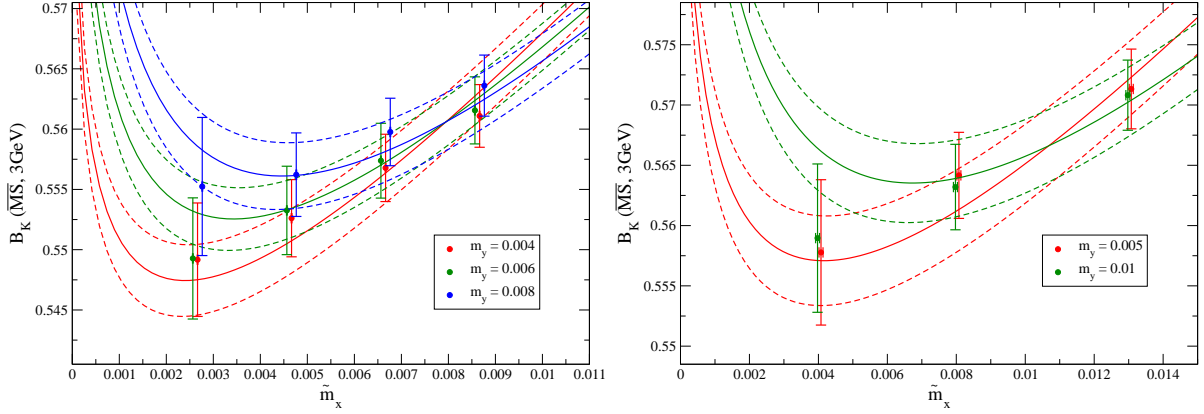

 \centering
\includegraphics*[height=0.33\textwidth]{fig/bk_PQ32_chpt.eps}
\includegraphics*[height=0.33\textwidth]{fig/bk_PQ24_chpt.eps}
\caption{Partially quenched light valence mass dependence of $B_K$ for the three
($32^3$) \ensA ensembles (left panel) and two ($24^3$) \ensB ensembles (right panel) at a valence
strange quark mass fixed to be the physical strange mass, and after reweighting in the
heavier sea quark mass to the physical strange mass. The overlayed curves are the partially
quenched SU(2) chiral perturbation theory expressions used in our fits.
\label{fig:bk_pq_su2}
}
\end{figure}

\begin{figure}
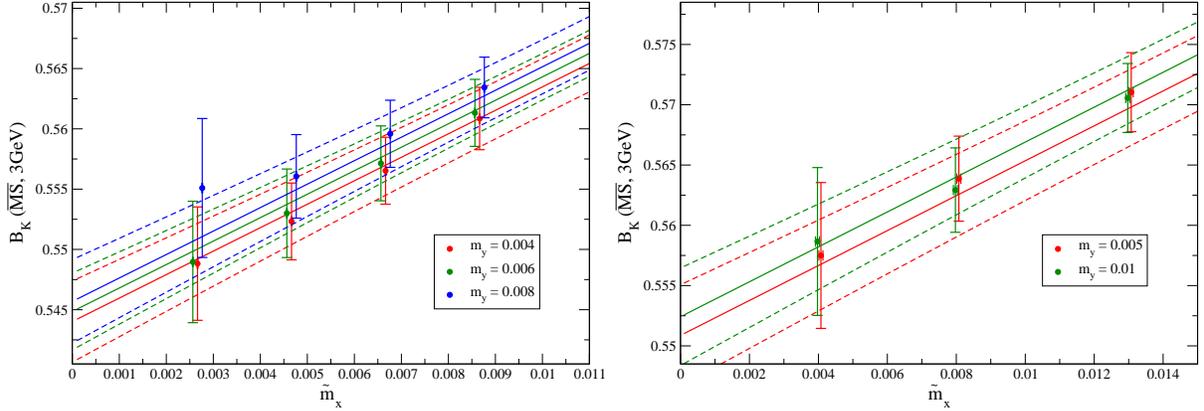

 \centering
\includegraphics*[height=0.33\textwidth]{fig/bk_PQ32_flavourexp.eps}
\includegraphics*[height=0.33\textwidth]{fig/bk_PQ24_flavourexp.eps}
\caption{Partially quenched light valence mass dependence of $B_K$ for the three
($32^3$) \ensA ensembles (left panel) and two ($24^3$) \ensB ensembles (right panel) at a valence
strange quark mass fixed to be the physical strange mass, and after reweighting in the
heavier sea quark mass to the physical strange mass. The overlayed lines represent analytic
fits to this data.
\label{fig:bk_pq_analytic}
}
\end{figure}

\begin{figure}
 \centering
\includegraphics*[width=10cm]{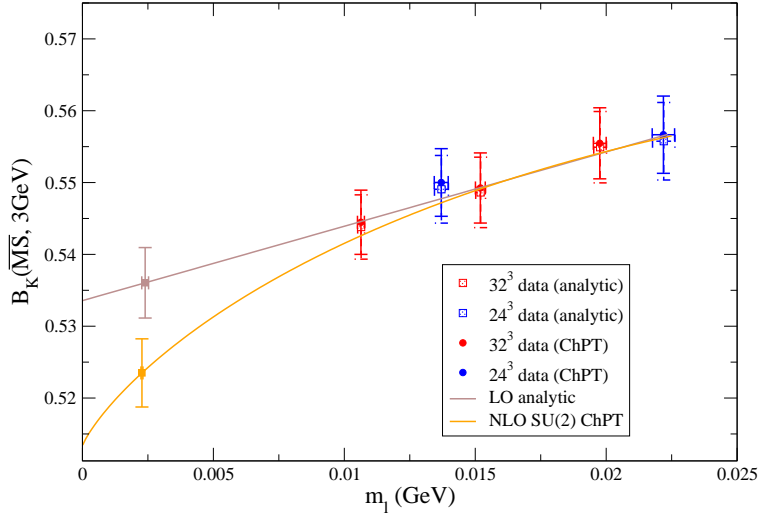}
\caption{The continuum limit chiral extrapolation obtained from our global fits using NLO $SU(2)$ PQChPT and LO analytic fits. The data is shown corrected to the continuum limit using the $\mathcal{O}(a^2)$ corrections obtained from both fit forms.
\label{fig:bkfitcomparison}
}
\end{figure}

\begin{figure}
 \centering
\includegraphics*[width=10cm]{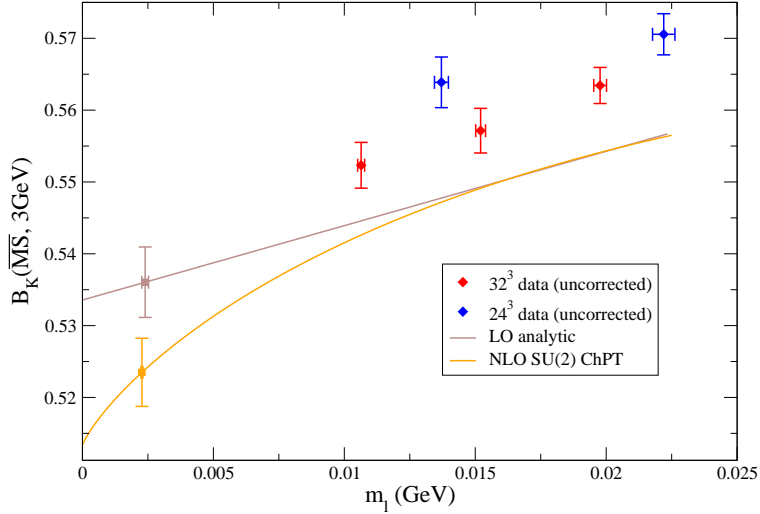}
\caption{The continuum limit chiral extrapolation obtained from our global fits using NLO $SU(2)$ PQChPT and LO analytic fits. As opposed to in Figure~\protect\ref{fig:bkfitcomparison}, the data plotted here has not been corrected to the continuum limit. The fit curves plotted are those performed to the continuum data as before.
\label{fig:bkfitcomparisonuncorrected}
}
\end{figure}

\subsection{Systematic errors on $B_K$} 
\label{sec:bksystematics}
Due to our combined analysis technique, and our use of reweighting in 
the strange sea sector, we eliminate systematic errors associated with discretisation effects and the untuned strange quark mass that were present in our previous analysis~\cite{Antonio:2007pb}. The remaining sources of systematic error are those arising due to the chiral extrapolation, finite volume effects and the renormalization. The systematic errors on the renormalization coefficients were discussed in Section~\ref{sec:NPR}. We discuss the remaining contributions below.

\subsubsection{Chiral fit systematics}
\label{sec:bkchptsystematics}
In Reference~\cite{Allton:2008pn,thirtytwocubed} we showed that a 
continuum fit to our two lattices using NLO $SU(2)$ PQChPT fit forms 
gives a value for $f_\pi$ that is $\sim 10\%$ too low after finite volume 
effects are included. Although this is of the magnitude expected for 
naturally sized NNLO contributions, we show in 
Reference~\cite{Mawhinney:2009jy} that a full NNLO 
fit to our data is heavily dependent on the priors used to constrain 
the fit and thus has little predictive power. We also considered an alternate 
fit form obtained from an analytic expansion at leading order about a 
non-zero unphysical pion mass, as advocated by Lellouch 
\cite{Lellouch:2009fg}. We are able to fit all of our data successfully, 
and obtain a result that is much closer to the known physical value for 
$f_\pi$. We observed that the difference between the analytic and the 
ChPT fit results in this case provides a good estimate of the systematic 
error associated with the chiral fit form\cite{Kelly:2009fp,thirtytwocubed}. 
We concluded that comparing ChPT and LO analytic fits is likely a good, robust method of estimating the 
systematic error for other quantities such as $B_K$. Both approaches must converge upon the physical value as the 
simulated quark masses approach the physical point.

The result of the LO analytic fit to $B_K$ is given alongside the 
NLO PQChPT results and those with NLO PQChPT including finite volume
effects in Table~\ref{tab:bkfitresults}.
To combine these in a final prediction, we follow \cite{thirtytwocubed}
and note that  both the
analytic and finite volume NLO PQChPT fits are reasonable extrapolation 
methods that can be justified in distinct limiting cases: the analytic 
form is certainly the correct
approach when we have data sufficiently close to the physical point regardless
of whether we are in the chiral regime, while
the NLO form including finite volume effects is also 
certainly correct when the data and physical point lie within the
chiral regime.

Given our experience with $f_\pi$, and following the approach taken
in \cite{thirtytwocubed} we take our central value as the average of
those obtained with the analytic extrapolation form, and the
finite volume corrected SU(2) NLO forms. We take the 
difference between these to estimate a chiral fit systematic error
as $(\Delta B_K)_{\chi} = 0.014$ ($2.6\%$). We
take the \emph{full} difference as the systematic and believe this is
a prudent and conservative approach. 

Another reasonable data driven
method would take \emph{half}  the difference as the error
estimate; this would assume that the analytic extrapolation
is a hard upper bound on the mass dependence, and that the NLO
form is a hard lower bound -- given the flexibility in unconstrained
NNLO ChPT forms this would appear to be too optimistic.

We also note that within the mass range
of the data our SU(2) NLO fit estimates the biggest correction to
be around 8\% of the value in the two flavor chiral limit (0.56 vs 0.517).
Squaring this term would suggest a naive estimate of NNLO effects at
around 0.5\%, which is substantially below our more conservative chiral
extrapolation error.

\subsubsection{Finite volume systematics}

We estimate finite volume corrections to our result from finite-volume PQChPT. 
As shown in Reference~\cite{Allton:2008pn} these corrections are obtained from the 
standard PQChPT forms by replacing the NLO chiral logarithms with sums over 
modified Bessel functions of the second kind.   

The result for this fit is given in Table~\ref{tab:bkfitresults}. Comparing this to the uncorrected result we estimate a finite volume error of $(\Delta B_K)_{\mathrm{FV}} = 0.002$ ($0.4\%$).

\subsection{Continuum prediction for $B_K$}

Combining our central value and the systematic uncertainties discussed above, we quote a prediction for $B_K$ using either the
$p^2 = \mu^2 = (2~{\rm GeV})^2$ renormalization scale,
\begin{equation}
\BKRESULTTWOGEV\,.
\label{eqn:bk2GeVresult}
\end{equation}
or the $p^2 = \mu^2 = (3~{\rm GeV})^2$ renormalization scale
\begin{equation}
\BKRESULT\,.
\label{eqn:bkfinalresult}
\end{equation}

The latter is our preferred central value as our systematic error
for the renormalization is halved.

This can be converted to the common RGI scheme for comparison and phenomenological application:
\begin{equation}\label{eq:bkrgiresult}
\boxed{\ \BKRESULTRGI\,,\ }
\end{equation}
and adding all sources of error in quadrature we
obtain 
\begin{equation}\label{eq:bkrgiresultcombined}
\boxed{\ \BKRESULTRGIQUAD\,,\ }\end{equation}
corresponding to an overall error of 3.6\%.

\fi


\section{Conclusions}
\label{sec:Conclusions}

\ifnum\theConclusions=1
In this paper we have calculated $B_K$ to 3.6\% precision with 2+1 flavours of dynamical quarks and,
for the first time, in the continuum limit with a lattice action with
good chiral symmetry. The result is presented in Equation~(\ref{eq:bkrgiresult}) (or equivalently in (\ref{eq:bkrgiresultcombined})).

Our calculation of this important quantity has exploited several significant improvements in lattice techniques which we have been developing for more than a decade. These include:
a) the use of domain wall fermions with good chiral symmetry~\cite{Blum:1997mz,Blum:2001xb},
b) the implementation of domain wall fermions in dynamical simulations with 2 + 1 flavours of light quarks~\cite{Antonio:2008zz,Antonio:2006px,Allton:2007hx,Boyle:2007qe,Antonio:2007pb,Boyle:2008yd,Boyle:2007fn},
and c) the use of SU(2) ChPT for chiral extrapolations of 2+1 flavour simulations, first
exploited by the RBC-UKQCD collaborations~\cite{Allton:2008pn,Antonio:2007pb}.

The present calculation of $B_K$ includes a particularly careful
treatment of the renormalization. We have introduced  several new momentum renormalization
schemes (based on the original works of ~\cite{Martinelli:1994ty}
and of~\cite{Sturm:2009kb} as explained in detail in Section~\ref{sec:NPR}), and our renormalization
also includes, for the first time, the improved scaling procedure of \cite{Arthur:2010ht}.

The small increase in our central value for $B_K$ in
this work and in \cite{Kelly:2009fp} compared to
\cite{Antonio:2007pb,Allton:2008pn} has arisen partly
from significant improvements in our approach to
renormalization as well as from taking the continuum limit.
The difference is within the previously budgeted errors for these sources,
and a large component of this small shift
arises from taking the central value from a new, 
non-exceptional momentum scheme
using the perturbative results derived in this paper.

\begin{table}[t]
\begin{tabular}{ccc}\\
Publication &$f$ & $\hat{B}_K^{\overline{\mathrm{RGI}}}$\\ \hline\hline
\emph{This work}   &2+1 & {\bf 0.749(7)(26)} \\
\hline
Bae'10     \cite{Bae:2010ki}   & 2+1 & 0.724(12)(43)\\
RBC-UKQCD'09\cite{Kelly:2009fp}&2+1 & 0.737(26)\\
Aubin'09   \cite{Aubin:2009jh} &2+1 & 0.724(8)(29)\\
RBC-UKQCD'07\cite{Antonio:2007pb}  &2+1 & 0.720(13)(37)\\
\hline
ETMC'10 \cite{Constantinou:2010qv}  & 2 & 0.729(30)\\
ETMC'09 \cite{Bertone:2009bu}  & 2 & 0.73(3)(3)\\
JLQCD'08  \cite{Aoki:2008ss}   & 2 & 0.758(6)(71)\\
\end{tabular}
\caption{A comparison of our result for $B_K$ with those of other recent calculations with dynamical fermions. Here $f$ denotes the number of dynamical quark flavours. Where separate errors are quoted, the first error is statistical and the second is systematic.
\label{tab:compareres}}
\end{table}

Our result for $B_K$ is compared to other recent calculations in Table~\ref{tab:compareres}.
Since all the results in this table, except for those of Reference~\cite{Bae:2010ki} and the current work, used the
original RI-MOM scheme, there is a substantial correlation
in the perturbative systematics between these five calculations. Thus 
the additional renormalization schemes introduced in this paper give added confidence to the estimates of 
the systematic error from this source.

In the remainder of this section we briefly discuss the significance of the recent lattice results for $B_K$ and the prospects for improving the precision still further.

\subsection{Significance of lattice results of $B_K$}

Flavour physics will continue to be central to the exploration of the limits of the standard model, to searches for new physics and to the eventual understanding of the fundamental theoretical framework of physics beyond the standard model. An important tool in this endeavour is the interpretation of experimental data in terms of the unitarity triangle where, in general, the remarkable consistency of the information from different processes places significant constraints on the possible parameter space of new models. Having said this, a number of \textit{tensions} have arisen in recent years; possible inconsistencies at a $1.5-3\sigma$ level
\cite{Lunghi:2008aa,Bona:2006ah,Charles:2004jd,tensions}
which certainly merit further investigation. The lattice results for $B_K$ contribute to these tensions as we now briefly explain.

Lattice calculations are necessary to evaluate the hadronic effects in tests of the unitarity of the CKM matrix and our results for $B_K$, used in conjunction with the experimental determination of $\epsilon_K$,  the indirect CP violation parameter monitoring $K_L \to \pi \pi$, are a major ingredient in tests of the CKM paradigm (see Equation~(\ref{eq:bk_epsilonk})). We illustrate this here with one example, exploiting lattice inputs not only for $B_K$ but also for the semileptonic $B\to\pi,\rho$ and $B\to D,D^\ast$ formfactors (used to determine $V_{ub}/V_{cb}$) and the
SU(3) breaking ratio, $\xi$, which contains the hadronic effects in the ratio of the mixings of $B_s$
mesons and  $B_d$ mesons. With these three key lattice inputs a nice prediction,
$\sin 2\beta=0.75 \pm 0.04$~\cite{Lunghi:2008aa,Bona:2006ah,Charles:2004jd}, emerges.
This can be compared with {\it direct} experimental measurements from
the time-dependent CP asymmetry in the {\it golden} mode,
$B_d \to J/\psi K_s$ which gives, $\sin 2 \beta^{J/\psi K_s} =0.681 \pm
0.025$~\cite{Amsler:2008zzb},
which is within 2$\sigma$ of the
Standard Model prediction with the lattice input. A similar tension is found in References~\cite{Buras:2009pj,Buras:2008nn,Buras:2010pza} who stress the need to include better approximations to the theoretical expression for $\epsilon_K$ now that $B_K$ is known to such good precision. These improvements include terms proportional to Im$A_0/$Re$A_0$ (where $A_0$ is the $K\to\pi\pi$ amplitude with the two pions in a state with isospin 0) and the recognition that the phase $\arctan (2\Delta M_K/\Delta\Gamma)$ is not precisely equal to $\pi/4$ ($\Delta M_K$ and $\Delta\Gamma$ are the differences of the masses and widths of the $K_L$ and $K_S$ mesons).

From the above discussion it is clear that lattice calculations of weak matrix elements in general, 
and of $B_K$ in particular, in conjunction with experiments, are 
providing ever more precise tests of the CKM explanation for CP violation. Of course our ambitions do not stop here; even if the small tension between the Standard Model prediction for $\sin(2\beta)$ and its direct determination disappears on closer scrutiny, the $O(10\%)$ difference in the central values still leaves ample room for new physics which we wish to squeeze still further. In the next subsection we discuss the prospects for improved precision in the determination of $B_K$ and of course it must be remembered that improvements in the determination of other inputs, including $\xi$ and $V_{cb}$ will also be necessary (recently it was shown that the use of $V_{cb}^4$ with its significant error, can be replaced by information from the leptonic $B\to\tau\nu$ branching ratio and lattice results on the decay constant $f_{B_d}$ and the mixing parameter $B_{B_d}$~\cite{Lunghi:2009ke}).

\subsection{Prospects for $B_K$ with one percent scale precision}

It is interesting to analyse our error budget and to assess what future
gains in precision can be made in the determination of $B_K$. In particular, we consider here what would be required to obtain $B_K$ with one
percent scale precision.

Currently, our dominant uncertainty is the 3\% error
arising from the chiral extrapolation. This will be addressed by simulations at or near
the physical quark masses, some of which are presently being undertaken by RBC and UKQCD.
Although expensive, these are affordable, even with current computer technology. We can therefore envisage these to be under control at the one
percent level in a few years.

The 2\% renormalization error is partly associated with the low scale at which
we presently apply one-loop matching and two-loop running to our operators. This uncertainty can be reduced
in two ways: firstly the scale can be raised at modest expense using a
step scaling technique\cite{Arthur:2010ht}, perhaps raising the matching scale from around 3\,GeV
to approximately 10\,GeV, reducing the $\alpha_s^2$
error on our one-loop matching from 2\% to around 1\%.
A larger gain would be obtained by extending the perturbative calculations presented in this paper
to the next order, leading to an expected $\alpha_s^3$ error
of around 0.7\%. 
The gain from step scaling is of course increased by higher order matching,
and one might expect a step scaled matching to attain 0.2\% renormalization
precision for an $\alpha_s^3$ renormalization error.
Such a two-loop calculation has been performed for the 
determination of light-quark masses~\cite{Gorbahn:2010bf,Almeida:2010ns} contributing to the improved lattice determination 
of these quantities~\cite{thirtytwocubed}. Given the importance of a precise determination of $B_K$, we would hope and expect 
that the two-loop matching calculation will be performed soon.

The remaining statistical and finite volume errors are small, and not 
unduly expensive to reduce still further
as this increases computational cost by only modest factors.

We conclude therefore that we can expect to determine $B_K$ at the one
percent scale over the next few years. What is perhaps more challenging is for lattice simulations to contribute in other ways to the determination of subdominant corrections to the theoretical expression for $\epsilon_K$, for example the long-distance contributions and the direct computation of $K\to\pi\pi$ decay amplitudes; the status of our endeavours in this direction are summarised in ~\cite{kaonsatlattice2010,christLat2010}.











\fi

\section*{Acknowledgments}

The calculations reported here were performed on the QCDOC computers
\cite{Boyle:2005qc,Boyle:2005gf,Boyle:2003mj,QCDOC:sc04} at Columbia University,
Edinburgh University, and at Brookhaven National Laboratory (BNL),
and Argonne Leadership Class Facility (ALCF) BlueGene/P resources at Argonne National Laboratory (ANL).
At BNL, the QCDOC computers of the RIKEN-BNL Research Center and the USQCD Collaboration were used.  
The very large scale capability of the ALCF was critical for carrying out the 
challenging calculations reported here.

The Edinburgh QCDOC system was
funded by PPARC JIF grant PPA/J/S/1998/00756 and operated through
support from the Universities of Edinburgh, Southampton and Wales
Swansea, and from STFC grant PP/E006965/1.

Computations for this work were carried out in part on facilities
of the USQCD Collaboration, which are funded by the Office of Science
of the U.S. Department of Energy.  We thank ANL, RIKEN, BNL and the U.S.
DOE, the University of Edinburgh and STFC
for providing the facilities essential for the completion of
this work.

The software used includes:  the
CPS QCD codes \\
{\tt http://qcdoc.phys.columbia.edu/cps.html},
supported in part by the USDOE SciDAC program; the
BAGEL \cite{Boyle:2009bagel} assembler
kernel generator for many of the high-performance optimized kernels;
and the UKHadron codes.

The work of the Edinburgh authors was supported by PPARC grants
PP/D000238/1 and PP/C503154/1. PAB acknowledges support from RCUK.
TB and RZ were supported by the US DOE under grant DE-FG02-92ER40716.
TI was supported in part by the Grant-in-Aid of the Japanese Ministry
of Education
(Nos. 22540301, 20105002, 20025010).
C.J., T.I., C. St. and A.S. (BNL) were partially supported by the U.S.\ DOE
under contract DE-AC02-98CH10886.
E.E.S is partly supported by DFG SFB/TR 55 and by the Research Executive
Agency of the European Union under grant PITN-GA-2009-238353 (ITN
STRONGnet).
N.C. and R.M. (Columbia University) were partially supported
by the U.S.\ DOE under contract DE-FG02-92ER40699.
D.B. and C.T.S (University of Southampton) were partially
supported by UK STFC Grant PP/D000211/1 and by EU contract
MRTN-CT-2006-035482 (Flavianet).
Y.A. is partially supported by JSPS KAKENHI 21540289. 
We thank Andrzej Buras for useful conversations.




\ifnum\theTables=1
\newpage
\fi


\ifnum\theFigures=1
\newpage
\fi

\end{document}